\documentclass[longauth]{aa}  
\usepackage{graphicx}
\usepackage{xcolor}
\usepackage[varg]{txfonts}
\usepackage{natbib}
\usepackage{amssymb}
\usepackage{sidecap}

\pdfoutput=1
\begin{document}

   \title{ALMA images the many  faces of the \object{NGC\,1068} torus and its surroundings}

 \author{S.~Garc\'{\i}a-Burillo\inst{1}
	   \and
	   F.~Combes\inst{2}	
	   \and	
	   C.~Ramos Almeida\inst{3, 4}
	   \and		
	   A.~Usero\inst{1}
	   \and	
	   A.~Alonso-Herrero\inst{5}		
	  \and			
 	   L.~K.~Hunt\inst{6} 
	   \and
	  D. ~Rouan\inst{7}
	  \and		
	  S.~Aalto\inst{8}
	  \and
          M.~Querejeta\inst{1,9}	  
	  \and
	  S.~Viti \inst{10}	
	  \and
          P.~P.~van der Werf\inst{11}	 
          \and
          H.Vives-Arias \inst{3, 4}
          \and
	  A.~Fuente\inst{1}
	  \and
	  L.~Colina\inst{5}	
	  \and
	  J.~Mart\'{\i}n-Pintado\inst{5}	  	  
	  \and
	  C.~Henkel\inst{12, 13}
	 \and		  	
	  S.~Mart\'{\i}n\inst{14,15} 	
	  \and
	  M.~Krips\inst{16}	  
	 \and	 
	  D.~Gratadour\inst{7}	
	   \and
	  R.~Neri \inst{16}
	  \and
          L.~J.~Tacconi\inst{17}
	    }
   \institute{
          Observatorio Astron\'omico Nacional (OAN-IGN)-Observatorio de Madrid, Alfonso XII, 3, 28014-Madrid, Spain 
			  \email{s.gburillo@oan.es} 		  
        \and 
      Observatoire de Paris, LERMA, CNRS, 61 Av. de l'Observatoire, 75014-Paris, France 
	 \and	 
	Instituto de Astrof\'{\i}sica de Canarias, Calle V\'{\i}a L\'actea, s/n, E-38205 La Laguna, Tenerife, Spain
	\and
	Departamento de Astrof\'{\i}sica, Universidad de La Laguna, E-38205, La Laguna, Tenerife, Spain
        \and
	Centro de Astrobiolog\'{\i}a (CSIC-INTA), ESAC Campus, 28692 Villanueva de la Ca\~nada, Madrid, Spain           
	\and
	INAF-Osservatorio Astrofisico di Arcetri, Largo Enrico Fermi 5, 50125-Firenze, Italy 	 
	 \and	 
	LESIA, Observatoire de Paris, PSL Research University, CNRS, Sorbonne Universit\'es, UPMC Univ. Paris 06, Univ. Paris Diderot, Sorbonne Paris 	Cit\'e, 5 place Jules Janssen, 92195, Meudon, France 	 
	 \and
	 Department of Earth and Space Sciences, Chalmers University of Technology, Onsala Observatory, 439 94-Onsala, Sweden
	\and	
	European Southern Observatory (ESO), Karl-Schwarzschild-Strasse 2, D-85748 Garching bei M\"unchen, Germany	 
	 \and 
         Department of Physics and Astronomy, UCL, Gower Place, London WC1E 6BT, UK
	 \and
	Leiden Observatory, Leiden University, PO Box 9513, 2300 RA Leiden, Netherlands 
	\and	 
	 Max-Planck-Institut f\"ur Radioastronomie, Auf dem H\"ugel 69, 53121, Bonn, Germany	
 	 \and
	 Astronomy Department, King Abdulazizi University, P.~O. Box 80203, Jeddah 21589, Saudi Arabia
		\and	 
	Joint ALMA Observatory, Alonso de C\'ordova, 3107, Vitacura, Santiago 763-0355, Chile
	\and	
	European Southern Observatory (ESO), Alonso de C\'ordova, 3107, Vitacura, Santiago 763-0355, Chile	 
	\and
	 Institut de Radio Astronomie Millim\'etrique (IRAM), 300 rue de la Piscine, Domaine Universitaire de Grenoble, 38406-St.Martin d'H\`eres, France 
	\and	 
        Max-Planck-Institut f\"ur extraterrestrische Physik, Postfach 1312, 85741-Garching, Germany 
	}

   \date{Received: August 30th, 2019; Accepted: October 13th, 2019}

 
  \abstract
  {}
{We investigate the fueling and the feedback of nuclear activity in the nearby ($D =14$~Mpc) Seyfert~2 barred galaxy NGC~1068, by 
studying the distribution and kinematics of molecular gas in the torus and its connections to the host galaxy disk.}
   {We use the Atacama Large Millimeter Array (ALMA ) to image the emission of a set of molecular gas tracers in the  circumnuclear disk (CND) and  the torus of the  galaxy  using the 
   CO(2--1), CO(3--2) and HCO$^+$(4--3) lines and their underlying continuum emission with high spatial resolutions ($0.03\arcsec-0.09\arcsec\simeq2-6$~pc). These 
   transitions, which span a wide range of physical conditions of molecular gas  ($n$(H$_2$)~$\subset10^3-10^7$cm$^{-3}$), are instrumental in revealing the density radial stratification 
   and the complex kinematics of the gas in the torus and its surroundings.}
   {The ALMA images resolve the CND as an asymmetric ringed disk  of $D\simeq400$~pc-size and mass of $\simeq1.4\times10^8 M_{\rm sun}$. The 
   CND shows a marked deficit of molecular gas in its central $\simeq130$~pc-region. The inner edge of the ring is associated with the presence of 
   edge-brightened arcs of NIR polarized emission, which are identified with the current working surface of the ionized wind of the active galactic nucleus (AGN).  ALMA proves the existence  of an 
   elongated molecular disk/torus in NGC~1068 of $M_{\rm torus}^{\rm gas}\simeq3\times10^{5}~M_{\sun}$, which extends over a large range of  spatial scales  $D\simeq10-30$~pc 
   around the central engine.  The new observations evidence the density radial stratification of the torus: the HCO$^+$(4--3) torus, with a full size $D_{\rm HCO^+(4-3)}=11\pm0.6$~pc, is a factor of 2--3 smaller than its CO(2-1) and CO(3-2) counterparts, which have full-sizes $D_{\rm CO(3-2)}=26\pm0.6$~pc and $D_{\rm CO(2-1)}=28\pm0.6$~pc, respectively. This result brings into  light the many faces of the molecular torus. The torus is connected to the CND through a network of 
   molecular gas streamers detected inside the CND ring. The kinematics of  molecular gas show strong departures from circular motions in  the torus,  the  gas streamers, and the CND
   ring. These velocity field distortions are interconnected  and are part of a 3D outflow that reflects the effects of AGN feedback on the  kinematics of molecular gas across a wide 
   range of  spatial scales around the central engine. In particular, we estimate through modeling that a significant fraction of the gas inside the torus 
   ($\simeq0.4-0.6 \times M_{\rm torus}^{\rm gas}$) and a comparable amount of mass along the gas streamers are outflowing. However, the bulk of the mass, momentum and energy of the
    molecular outflow of NGC~1068 is contained at larger radii in the CND region, where the AGN wind and the radio jet are currently pushing the gas assembled at the Inner Lindblad Resonance (ILR) ring of the nuclear stellar bar.}
   {In our favored scenario a wide-angle AGN wind launched from the accretion disk of NGC1068 is currently impacting a sizable fraction of the gas inside the torus.  However, a large gas reservoir ($\simeq 1.2-1.8 \times10^5M_{\sun}$), which lies close to the equatorial plane of the torus, remains unaffected by the feedback of the AGN wind and can therefore continue fueling the AGN for at least $\simeq1-4$~Myr. AGN fueling seems nevertheless  currently thwarted on intermediate scales ($15$~pc~$\leq r \leq50$~pc). }
  \maketitle
%

\section{Introduction}

The fueling of super-massive black holes (SMBH) explains the onset of nuclear activity in galaxies.
High-resolution observations of molecular gas have been instrumental to understand how active galactic nuclei (AGN) are fueled \citep[e.g.,][]
 {Gar05, Gar12, Com13, Com14, Sto19}. These observations have also started to reveal  that the vast amounts of energy released during the feeding process can help regulate gas accretion through the launching of molecular outflows in  different types of active galaxies \citep[e.g.,][]{Fer10, Stu11, Alat11, Alat14, Aal12, Aal16, Com13, Mor13,  Mor15, Cic14, GB14, GB15, Que16, Bar18, Alo19}.  Molecular outflows are a manifestation of AGN feedback and they constitute a key ingredient to understand the co-evolution of galaxies
and SMBH, which is inferred from  the observed scaling laws relating the mass of nuclear black holes ($M_{\rm BH}$) to the properties of the spheroidal components of   galaxies, like the mass  ($M_{\rm bulge}$) \citep[e.g.,][]{Kor95, Mag98, Mar01} or the stellar velocity dispersion ($\sigma$) \citep[e.g.,][]{Geb00, Mer01, Fer05, Gul09}.

Observational properties of AGN are used to classify them into two categories: Type 1 AGN show broad line regions (BLR), while Type 2 AGN only show narrow line regions (NLR). Lines are broadened in BLR close to the central engine, while they are narrow farther out in the NLR. The central engines of Type 2 objects are thought to be hidden behind a screen of obscuring material located in a dusty torus of a few-to-10 parsec size  \citep{Mil83, Ant85, Urr95, Eli12, Net15, Ram17}.  Determining from observations the physical parameters of the putative torus and its surroundings has been challenging due to the foreseen small size of these structures. The list of key parameters to be determined includes the size, the mass and the dynamical state of the gas in the torus (rotating, outflowing and/or inflowing). Furthermore, we need a description of how the torus is connected to the host galaxy disk.

In the absence of observational constraints, the first estimates  of the  torus size came  from theoretical  models  of the  spectral energy distributions (SED). \citet{Pie92}  favored compact tori with an outer radius  $\leq5-10$~pc. Later works extended these compact
structures out to radii $\simeq30-100$~pc  \citep[e.g.,][]{Pie93, Gra94}. A recent compilation of high-resolution observations in the near and mid-infrared (NIR and MIR) of 23 nearby Seyfert galaxies found relatively compact sizes for their AGN tori, yet with a remarkable range of wavelength-dependent values, from $r\simeq0.1-1$~pc in the NIR, to  $r\simeq1-$a few pc in the MIR \citep[e.g.,][and references therein]{Bur13}. 
Furthermore, the observed nuclear NIR-to-MIR SED (including MIR spectroscopy) of Seyferts were reproduced with clumpy tori of sizes $r\simeq1-14$~pc \citep{Ram11, Alo11, Ich15, Gar19}.

The question of the dynamical status of the torus has also evolved from the first proposed scenario, depicting the torus as a geometrically thick  rotating doughnut-like disk in hydrostatic equilibrium \citep{Kro88}, towards a more dynamical picture, describing the torus as an outflowing structure, where clouds are embedded in a hydro-magnetically or radiatively driven wind   \citep{Bla82, Emm92, Bot97, Kar99, Eli06, Wad12, Wad15,Wad16, Hoe17, Cha16, Cha17, Will19, Hoe19}. This outflowing torus may be part of the obscuring structures that have been identified as polar-like dust components in a notable fraction of Seyferts observed with interferometers in the MIR  \citep{Tri09, Bur13, Hoe13, Lop14, Lop16}.

Due to their limited  $(u,v)$-coverage,  MIR interferometers have not been optimized thus far to provide a direct image of the torus and its surroundings. This limitation can be  overcome by mm/submm interferometers like  the Atacama Large Millimeter Array (ALMA).
Building on an exquisite coverage of the $(u,v)$-plane, ALMA has started to image with high fidelity the distribution and kinematics of molecular gas in the tori of nearby Seyfert galaxies (D$\leq$10-30~Mpc), thanks to its ability to reach angular resolutions $\leq0.05\arcsec-0.1\arcsec$ \citep{GB16, Gal16, Ima16, Ima18, Alo18, Alo19, Izu18, Com19, Imp19, Aud19}.

The nearby ($D\sim14$~Mpc; \citealt{Bla97}) Seyfert~2 galaxy \object{NGC~1068} has been a testbed for unifying theories of AGNs after the discovery of polarized optical continuum and broad-line emission in this source \citep{Mil83, Ant85}.  Single-dish observations in the NIR and MIR showed the existence of emission extending over a wide range of spatial scales ($\simeq$~5--70~pc) around the AGN \citep{Boc00, Mar00, Tom01, Rou04, Gal05, Gra06, Gra15, LopR16}.  Interferometric observations in the NIR and MIR confirmed the existence of two components \citep{Jaf04,Wei04, Rab09, Bur13,Lop14}: a compact 0.5--1.4~pc-sized disk centered at the AGN with a position angle $PA\simeq140^\circ$, and a  3--10~pc-sized elongated structure oriented along the north-south axis.  The compact disk contains hot ($\simeq800$~K) dust co-spatial with the H$_2$O megamaser disk \citep{Gre96, Gre97, Gal01, Gal04}, while the elongated source contains warm ($\simeq300$~K) dust likely associated with the ionization cone.

\citet{GB16} and \citet{Gal16}  used ALMA  to map the emission of the CO(6--5) line and its underlying continuum  in \object{NGC 1068} with a spatial 
resolution of $\sim4$~pc. These observations  imaged the dust  and molecular line emission  from  the  torus of \object{NGC\,1068}. The  $\simeq7-10$~pc CO/dusty torus extends along $PA\simeq112\pm20^{\circ}$ over spatial scales a factor of four larger than the MIR sources detected at the AGN \citep{Bur13, Lop14}. The CO torus shows a lopsided disk morphology, an orientation roughly perpendicular to the AGN wind/jet axis, and a surprisingly perturbed kinematics, whose origin remains a matter of debate. Firstly, \citet{GB16} reported the apparent counter-rotation of the outer disk $r\simeq3.5$~pc (imaged in CO(6--5)), relative to the 
inner $r\simeq1$~pc (imaged in  H$_2$O maser emission). This apparent counter-rotation has also been evidenced by the recent HCN(3--2) and HCO$^+$(3--2) ALMA maps of the torus \citep{Ima18, Imp19}. Secondly, the maps of \citet{GB16} and \citet{Gal16} show a velocity gradient along the morphological minor 
axis of the torus. This velocity gradient could be explained by gas being entrained in the outflow \citep{GB16, Gal16, Imp19}.  This interpretation is nevertheless challenged by the fact  that the reported radial shift of velocities in the torus is reversed farther out ($r\geq100-200$~pc) at the circumnuclear disk (CND), where the CO outflow imaged by \citet{GB14} follows the pattern of  low-ionization lines \citep[see discussion in][]{GB16}. Alternatively, the  disturbed  kinematics of the torus could be the signature of the Papaloizou-Pringle  instability (PPI),  predicted to drive the dynamical evolution of AGN tori on scales $r\leq0.1$~pc \citep{Pap84}. This instability  has been studied  both in isolated  tori \citep{Kiu11, Kor13} and also in tori perturbed due to the accretion of gas \citep{Don14, Don17}. These simulations show the growth of a long-lasting non-axisymmetric $m=1$ mode. The PPI is able to sustain a lopsided distribution and non-circular motions in the gas. However, whether this instability  can propagate efficiently and in due time to the significantly larger radii of the CO torus is still an open question.

Furthermore, several groups found evidence of the existence of gas lanes connecting  the torus with the CND out to $r\simeq30$~pc. These gas lanes emit in the 2.12~$\mu m$ H$_2$ line, a tracer mostly sensitive to hot molecular gas ($T_{\rm k}\geq10^3$~K).  However, these works have led to contradicting conclusions regarding the interpretation of what is the kinematics of  the gas in these structures, interpreted in terms of either inflow \citep{Mue09} or outflow motions \citep{Bar14, May17}. \citet{GB16} detected no clear CO(6--5) counterpart of the NIR lanes/tongues.
The nature of the CND-torus connection  and the question of what is the fueling budget in this Seyfert are therefore mostly unsettled.

The new ALMA observations presented in this paper image the emission of molecular gas in the CND and the torus of \object{NGC~1068} using the CO(2--1), CO(3--2) and HCO$^+$(4--3) lines, with a set of spatial resolutions $\simeq 0.03-0.09\arcsec$~(2--6~pc) that  improve by a significant factor $\geq6-10$ the best spatial resolutions achieved in these lines in the previous maps of \citet{Sch00} and \citet{GB14}. The use of these transitions, which purposely span a wide range of physical conditions of molecular gas  ($n$(H$_2$)~$\subset10^3-10^7$cm$^{-3}$), is instrumental to reveal the density radial stratification of the gas in the torus.  The non-detection by \citet{GB16} of a CO(6--5) counterpart of the NIR lanes/tongues may be due to the fact that the  gas in the connecting bridges has volume densities below the critical density of the CO(6--5) line ($\geq 10^5$~cm$^{-3}$). In this context, the choice of the CO(2--1) line, which is sensitive to densities
$n$(H$_2$)~$\simeq10^3$cm$^{-3}$, is motivated by the need to trace the bulk of the H$_2$ gas in the CND, in the torus and, also, in the structures connecting the CND with the torus. Furthermore, with the chosen requirement on what should be the minimum largest angular scale (LAS) recovered ($\simeq1.3"-1.8"=90-130$~pc), the aimed spatial resolution goal in both ALMA Bands  (\#~6 and \#~7)  is still compatible with the ability to recover the emission in the CND and, also, in the CND-torus connections.

We describe in Sect.~\ref{observations} the  new ALMA observations  used in this work. Section~\ref{continuum} presents the continuum maps derived at 229.7\,GHz and 344.5\,GHz. The distribution and kinematics of molecular gas derived from the CO and HCO$^+$ line maps in the CND and the torus are discussed, respectively,  in Sects.~\ref{CND} and ~\ref{torus}. We describe in Sect.~\ref{torus-conn} the gas connections between  the torus and the CND and attempt to find a global description for the kinematics of the gas in the torus and the large-scale molecular outflow
in Sect.~\ref{BBarolo}. The main conclusions of this work are summarized in Sect.~\ref{summary}.

 
  \begin{table}[bt!]
\caption{\label{Tab1} Observational parameters}
\centering
\resizebox{0.5\textwidth}{!}{
\begin{tabular}{lcccc}
\noalign{\smallskip} 
\hline
\hline
\noalign{\smallskip} 
Tracer & beam  & LAS & FOV & sensitivity \\
\noalign{\smallskip}
 & $\arcsec \times \arcsec$  &  $\arcsec$ & $\arcsec$ & mJy~beam$^{-1}$ \\
\noalign{\smallskip} 	      	      
\hline
\noalign{\smallskip} 	
continuum@229.7GHz &  $0\farcs089\times0\farcs087$ at  $PA=40^{\circ}$ & 1.3  & 25.4 & 0.030 \\
continuum@344.5GHz &  $0\farcs11\times0\farcs072$ at  $PA=90^{\circ}$ &  1.8 & 16.9 & 0.046 \\
CO(2--1)-MSR &  $0\farcs11\times0\farcs068$ at  $PA=80^{\circ}$ & 1.3  & 25.4 &   0.11    \\
CO(2--1)-HSR &    $0\farcs04\times0\farcs03$ at  $PA=25^{\circ}$   & 1.3  & 25.4 & 0.14 \\ 
CO(3--2)-MSR & $0\farcs13\times0\farcs065$ at  $PA=100^{\circ}$ & 1.8 & 16.9 & 0.23 \\
CO(3--2)-HSR &    $0\farcs04\times0\farcs03$ at  $PA=74^{\circ}$      & 1.8  & 16.9 & 0.23 \\
HCO$^+$(4--3)-MSR & $0\farcs081\times0\farcs046$ at  $PA=78^{\circ}$  & 1.7  & 16.4 & 0.28 \\
HCO$^+$(4--3)-HSR &    $0\farcs04\times0\farcs03$ at  $PA=73^{\circ}$     & 1.7  & 16.4 & 0.28 \\
\noalign{\smallskip}    
\hline
\hline 
\end{tabular}} 
\tablefoot{Column (2) lists the spatial resolutions reached for the different line and continuum observations used in this work. We distinguish
the moderate spatial resolution data set (MSR) from the high spatial resolution data set (HSR) obtained by using a different robust weighting parameter
of the visibilities in the UV plane: $robust=1$ (0.1) for the MSR (HSR) configuration.   Columns (3) and (4) list, respectively,  the largest angular scale recovered  (LAS) and the field-of-view (FOV)  of the observations, in arc second units. Column (5) lists the sensitivity (1$\sigma$) for the continuum and line data cubes. The line sensitivities are
derived for channels of 20~km~s$^{-1}$-width.}
\\
\end{table}

\section{Observations}\label{observations}

\subsection{ALMA data}\label{ALMA}

 We observed the emission of the CO(2--1), CO(3--2) and HCO$^+$(4--3) lines and their underlying continuum emission in the CND of NGC~1068 with ALMA during Cycle~4 using Band~6 and Band~7 receivers (project-ID: $\#$2016.1.0232.S; PI: S.~Garc\'{\i}a-Burillo). The phase tracking center was set to $\alpha_{2000}=02^{h}42^{m}40.771^{s}$,   $\delta_{2000}=-00^{\circ}00^{\prime}47.84\arcsec$, which is the galaxy's center according to SIMBAD taken from the Two Micron 
  All Sky Survey--2MASS survey \citep{Skr06}.  The tracking center is offset by $\leq$1$\arcsec$ relative to the AGN position:   
  $\alpha_{2000}=02^{h}42^{m}40.71^{s}$, $\delta_{2000}=-00^{\circ}00^{\prime}47.94\arcsec$ \citep{Gal96,Gal04,GB14,GB16, Gal16, Ima16, Ima18}.
The data in the two bands were calibrated using the ALMA reduction package  {\tt CASA\footnote{http//casa.nrao.edu/}}. The calibrated uv-tables were exported to {\tt GILDAS\footnote{http://www.iram.fr/IRAMFR/GILDAS}}-readable format  in order to perform the mapping and cleaning steps as detailed below. For both bands we estimate that the absolute flux accuracy is
 about 10$\%$, which is in line with the goal of standard ALMA observations at these frequencies.  Rest frequencies for all the lines used in this work were corrected for the 
  recession velocity initially assumed to be $v_{\rm o}$(LSR)$~=1129$~km~s$^{-1}=v_{\rm o}$(HEL)$=1140$~km~s$^{-1}$. 
  Systemic velocity is nevertheless re-determined in this work  as $v_{\rm sys}$(LSR)$~=1120$~km~s$^{-1}=v_{\rm sys}$(HEL)$=1131$~km~s$^{-1}$.  When appropriate, relative velocities throughout the paper refer to  $v_{\rm sys}$.  
  We also use the continuum map  of NGC~1068 obtained with ALMA during Cycle 0 using Band~9 receivers (project-ID: $\#$2011.0.00083.S; PI: S.~Garc\'{\i}a-Burillo) and published by \citet{GB14} and \citet{Vit14}. Hereafter we assume a distance to NGC~1068 of
$D\simeq14$~Mpc \citep{Bla97}; the latter implies a spatial scale of $\simeq70$~pc/$\arcsec$.

   \begin{figure*}
   \centering
    \includegraphics[width=1\textwidth]{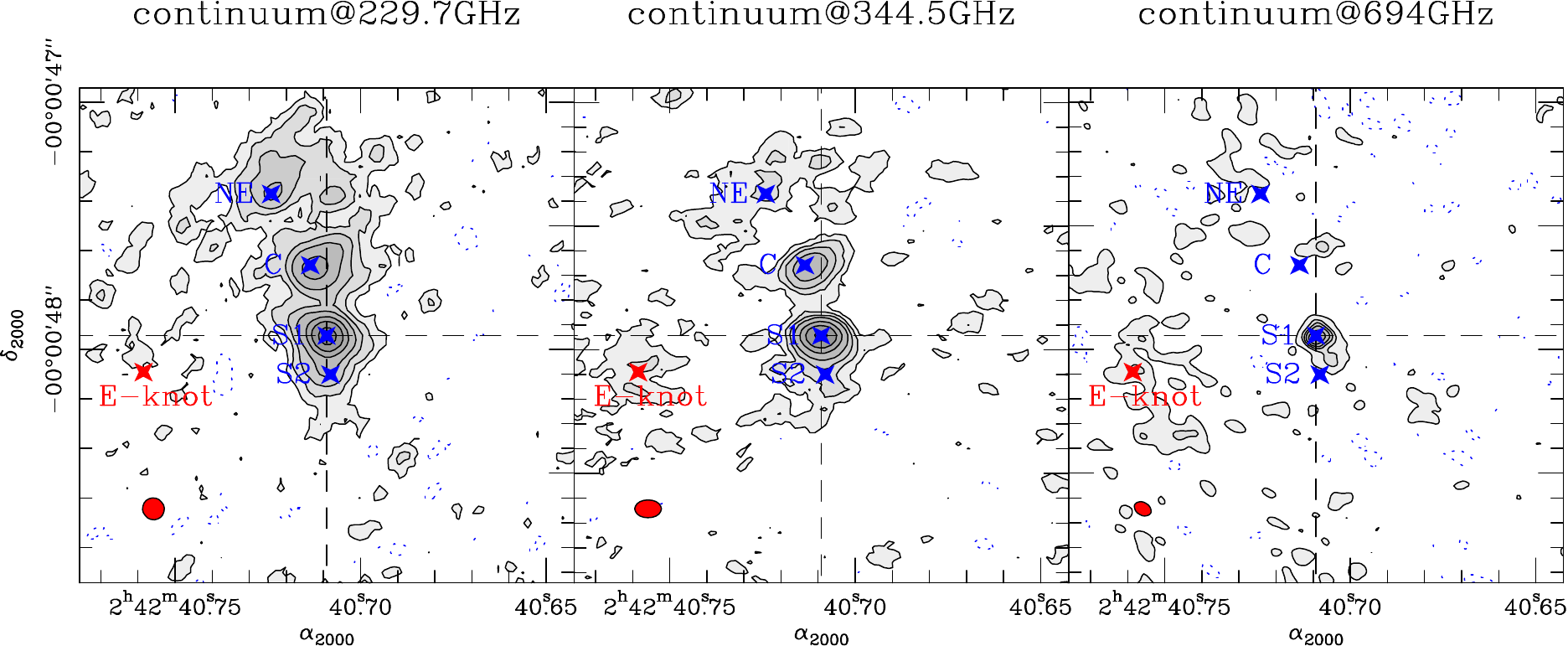}
   \caption{{\it Left panel:}~The continuum emission map of the central $r\leq1\arcsec$ (70~pc) region of NGC~1068 obtained with ALMA at 229.7~GHz (1306~$\mu$m).  The map is shown in grey scale with contour levels -2$\sigma$ (dashed contour), 2$\sigma$, 5$\sigma$, 10$\sigma$, 20$\sigma$, 50$\sigma$, 100$\sigma$, and 200$\sigma$, where 1$\sigma$~=~30$~\mu$Jy~beam$^{-1}$. ~{\it Middle panel:}~Same as {\it left panel} but  showing the continuum emission at 344.5~GHz (871~$\mu$m) . Contour spacing is as in {\it left panel}, but with 1$\sigma$~=~50$~\mu$Jy~beam$^{-1}$.  ~{\it Right panel:} Same as {\it left panel} but  showing the continuum emission at  694~GHz (432~$\mu$m) published by \citet{GB16}. Contour levels are -2$\sigma$ (dashed contour), 2$\sigma$, 5$\sigma$, 7$\sigma$, 9$\sigma$, 12$\sigma$, and 16$\sigma$, where 1$\sigma$~=~0.5mJy~beam$^{-1}$. The (red) filled
ellipses at the bottom left corners of the panels represent the beam sizes of ALMA at  229.7~GHz  ($0\farcs089\times0\farcs087$ at  $PA=40^{\circ}$), 344.5~GHz ($0\farcs11\times0\farcs072$ at  $PA=90^{\circ}$) and  694~GHz ($0\farcs07\times0\farcs05$ at  $PA=60^{\circ}$). The AGN lies at the intersection of the dashed lines: $\alpha_{2000}=02^{h}42^{m}40.709^{s}$, $\delta_{2000}=-00^{\circ}00^{\prime}47.94\arcsec$ (S1 knot). We highlight the position of other radio continuum knots (S2, C and NE) as given in the VLBI maps of \citet{Gal96,Gal04}, and the E-knot, characterized by strong dust continuum and molecular line emission \citep{GB14, Vit14}.}
   \label{cont-maps}
    \end{figure*}
   \begin{figure*}
   \centering
    \includegraphics[width=1\textwidth]{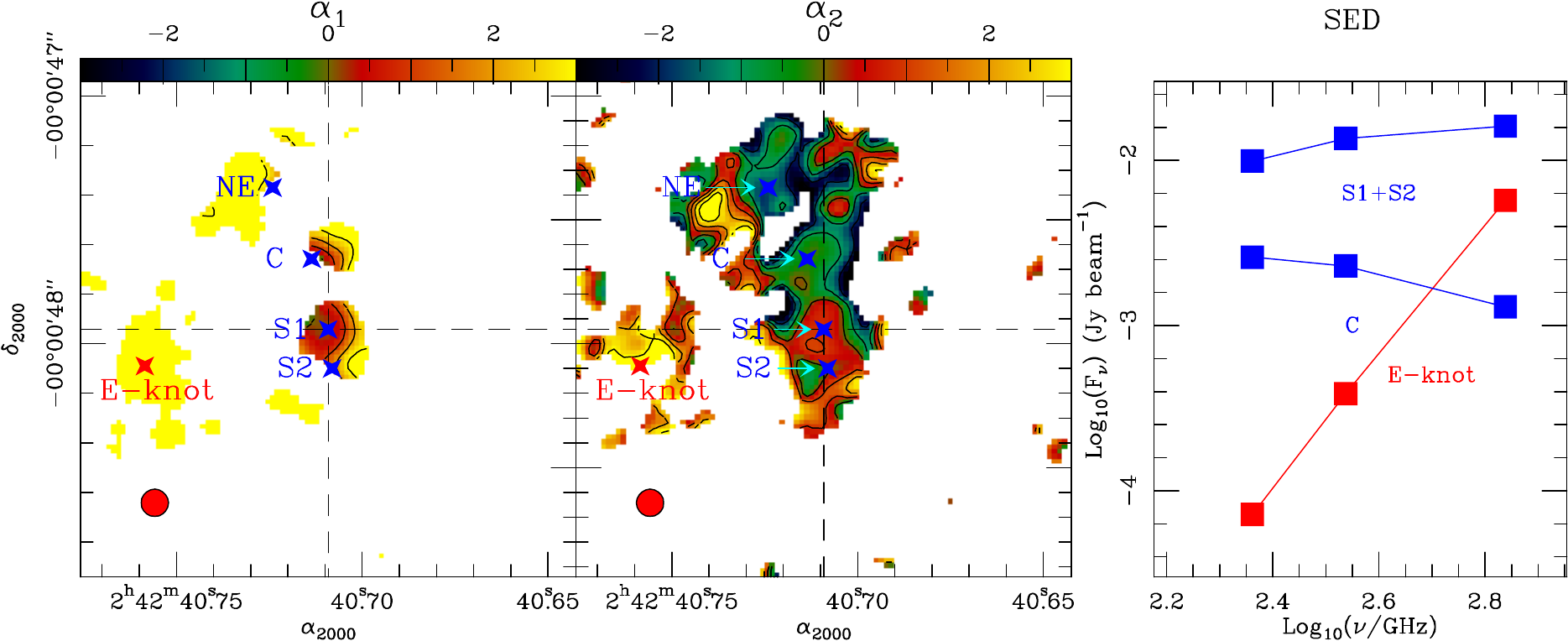}
   \caption{We show the spectral index maps ($\alpha$, with $S_{\nu}\propto\nu^{\alpha}$)  derived from the continuum emissions at $\nu$=344.5~GHz and 694~GHz ($\alpha_1$: {\it left panel}), and from $\nu$=229.7~GHz and 344.5~GHz ($\alpha_2$: {\it right panel}).  Contours  span the range $\alpha=[-3,3]$ in steps of 1. The common aperture adopted to derive the spectral index map is $0\farcs11\times0\farcs11$ (red circle).  The SED of continuum emission from 229.7~GHz to 694~GHz for the E-knot, the C-knot and the combined S1+S2 region are shown in the {\it right panel}.}
   \label{cont-alpha}
    \end{figure*}

\subsubsection{Band~7 data} 

We used a single pointing with 
a field-of-view (FOV) of 17$\arcsec$. In total two tracks were 
observed between October 2016 and September 2017 resulting in a total of 
411278~visibilities. We had 46 antennas available during the 
observations with projected baselines ranging from 18~m to 7.4~km. Four spectral windows were placed, two in the 
lower side band (LSB) and two in the upper sideband (USB). The four windows were centered on 
the following sky frequencies: 344.484~GHz and 345.681~GHz in the LSB and 
355.380~GHz and 357.339~GHz in the USB. All the sub-bands have a spectral bandwidth of 1.875~GHz, except for the one centered around
345.681~GHz, which is  factor of two narrower. This setup allowed us to simultaneously observe 
CO($J=3-2$) (345.796~GHz at rest) and H$^{13}$CO$^+$($J=4-3$) (346.998~GHz at rest) in the 
LSB bands, and  HCO$^+$($J=4-3$) (355.380~GHz at rest) and the continuum emission 
in the USB bands. Images of the continuum emission were obtained by averaging in each of the three sub-bands centered around
spectral lines those channels free of line emission. These maps were used to subtract in the $(u,v)$-plane the underlying continuum 
emission from the visibilities and subsequently obtain continuum-free spectral line images for all the lines.  In this paper we present the  CO(3--2) and HCO$^+$(4--3) maps but omit any discussion on the H$^{13}$CO$^+$(4--3) data due to its low S/N ratio.

Table~\ref{Tab1} summarizes the main observational parameters. We obtain two sets of angular resolutions by changing in the {\tt GILDAS} task {\tt UV-MAP} the robust parameter ($b$) from 1 (in the moderate spatial resolution data set, hereafter MSR) to 0.1 (in the high spatial resolution data set, hereafter HSR). The range of angular resolutions obtained is $\simeq0\farcs03-0\farcs1$  (2--7~pc) in the CO(3--2) line, and $\simeq0\farcs03-0\farcs06$  (2--4~pc) in the HCO$^+$(4--3) line.  The conversion factors between 
mJy~beam$^{-1}$ and K are 1.2~K~mJy$^{-1}$~beam (MSR) and 8.4~K~mJy$^{-1}$~beam (HSR) for the CO(3--2) line, and 2.6~K~mJy$^{-1}$~beam (MSR) and 8.5~K~mJy$^{-1}$~beam (HSR) for the HCO$^+$(4--3) line.  The line data cubes were binned to a common frequency 
resolution of 23.1~MHz (equivalent to $\sim$20~km~s$^{-1}$ in Band~7). The point source sensitivities in the line data cubes 
were derived selecting areas free from emission in all channels, resulting in a common value of 0.23~mJy~beam$^{-1}$  for the CO(3--2) line and 0.28~mJy~beam$^{-1}$ for the HCO$^+$(4--3) line, in channels of 20~km~s$^{-1}$ width. The corresponding point source sensitivity for the continuum is 46~$\mu$Jy~beam$^{-1}$.   

 As our observations do not contain short-spacing correction, we expect that the flux will start to be filtered out on scales $\geq1.8\arcsec$ (130~pc). We nevertheless 
 foresee that the highly clumpy distribution of the gas and the strong velocity gradients of the emission in the nuclear regions will help us
recover the bulk of the flux on the spatial scales of the CND relevant to this paper. 
As a sanity check, we have compared the total flux of the CO(3--2) line integrated over the CND region, shown in Fig.~\ref{co-full}, with the corresponding value derived 
using the ALMA map of \citet{GB14}, which is expected to recover all the flux out to scales $\simeq6\arcsec\simeq420$~pc, which are similar to the total size of the CND ($\simeq300-400$~pc).  Within the calibration errors, we obtain virtually identical fluxes from the two data sets: 1500~Jy~km~s$^{-1}$ \citep{GB14} $\simeq$ 1450~Jy~km~s$^{-1}$ (this work), an indication that the emission stemming from spatial scales $\simeq130-420$~pc in the new CND map is kept to a minimum. We can therefore conclude that the new CO(3--2) map  contains the bulk of the flux  for the whole range of spatial scales of the CND analyzed in this work. Furthermore, we can assume that a similar scenario is valid for the HCO$^+$(4--3) line, considering that this transition is likely tracing comparatively denser and in all probability clumpier molecular gas complexes in the CND.

\subsubsection{Band~6 data}
 
Similarly to Band~7, we used a single pointing that translates at the lower frequencies of Band~6 into
a FOV of 25$\arcsec$. We observed three tracks between October 2016 and September 2017 resulting in a total of 
1718451~visibilities. We had 50 antennas available during the 
observations with projected baselines ranging from 18~m to 12~km. We placed five spectral windows, two in the LSB and three in the USB. The five windows were centered on 
the sky frequencies: 217.647~GHz and 216.281~GHz in the LSB and 
229.663~GHz, 230.343~GHz, and 231.021~GHz in the USB. This setup allowed us to  observe 
H$_2$CO[3(2,2)-2(2,1)] (218.476~GHz at rest) and SiO($J=5-4$) (217.105.~GHz at rest) in the 
LSB bands, and  CO($J=2-1$) (230.538~GHz at rest),  $^{13}$CS($J=5-4$) (231.221~GHz at rest), and H$_{\rm 30\alpha}$ (231.901~GHz at rest)
in the USB bands. 
Images of the continuum emission were obtained by averaging in all sub-bands those channels free of line emission. As for Band~7, these maps were used to obtain continuum-free spectral line images for all the lines.  In this work we discuss the results  obtained in the CO(2--1) line and leave out any discussion on the remaining lines, to be presented in a forthcoming paper.

The main observational parameters are listed in Table~\ref{Tab1}.  Similarly to Band~7, we use two sets of angular resolutions for the CO(2--1) line that lie in the range 
 $\simeq0\farcs03-0\farcs09$  (2--6~pc).  The conversion factors between 
mJy~beam$^{-1}$ and K are 3~K~mJy$^{-1}$~beam (MSR) and 18~K~mJy$^{-1}$~beam (HSR).  For the sake of comparison with the Band~7 data, we binned the Band~6 data to a frequency  resolution of 15.4~MHz (equivalent to $\sim$20~km~s$^{-1}$). The corresponding point source sensitivities in the line data cubes range from 0.11~mJy~beam$^{-1}$ (MSR) to 0.14~mJy~beam$^{-1}$ (HSR) in channels of 20~km~s$^{-1}$ width.  The corresponding point source sensitivity for the continuum is 30~$\mu$Jy~beam$^{-1}$.   

\begin{figure*}[th!]
      \centering
    \includegraphics[width=0.95\textwidth]{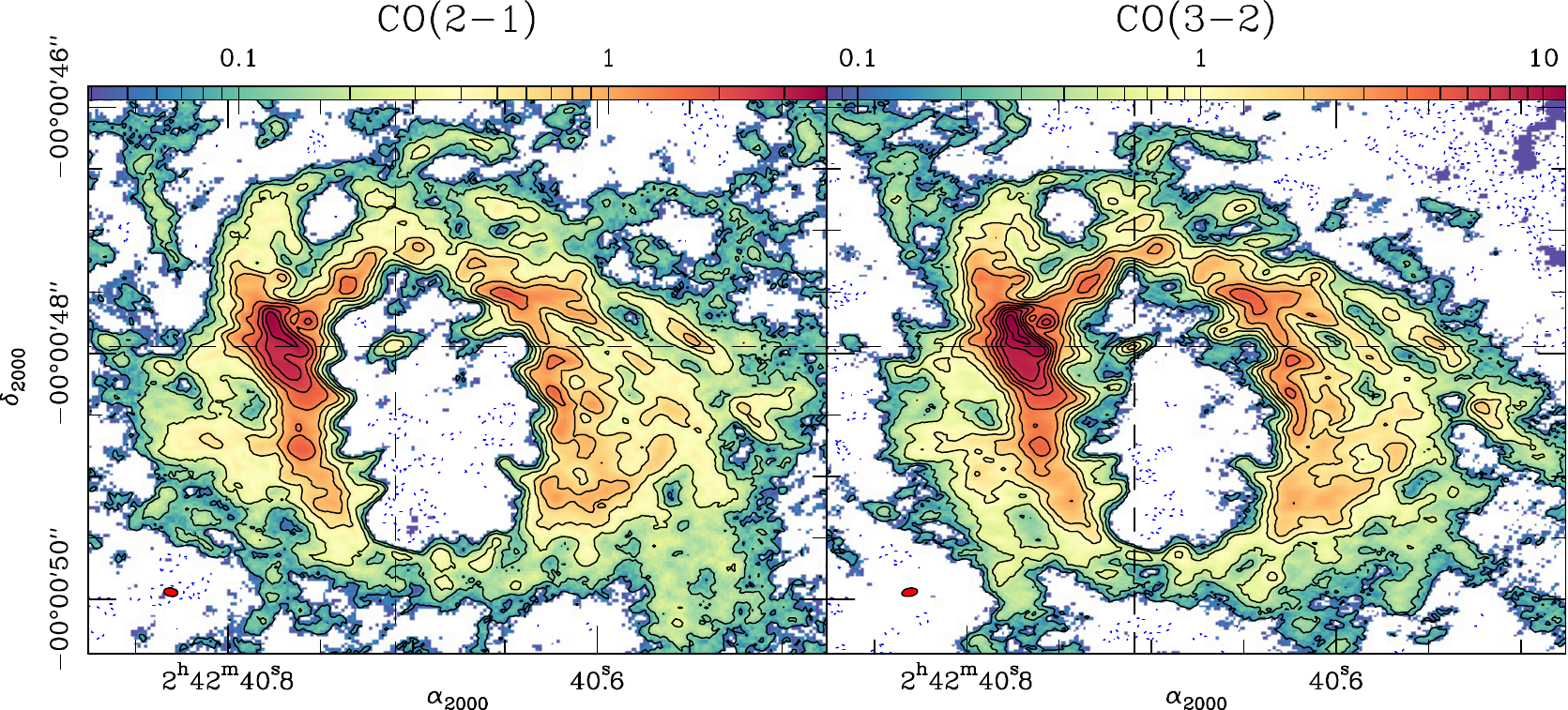}
        \caption{{\it Left panel:}~The CO(2--1) integrated intensity map obtained with ALMA in the CND of NGC~1068 using the MSR data set as defined in Table~\ref{Tab1}. 
   The map is shown in 
color scale spanning the range [3$\sigma$, 300$\sigma$] in logarithmic scale with contour levels  -5$\sigma$ (dashed contour), 5$\sigma$, 10$\sigma$, 20$\sigma$, 40$\sigma$, 60$\sigma$, 100$\sigma$ to
250$\sigma$ in steps of 50$\sigma$ where 1$\sigma=13$~mJy~km~s$^{-1}$beam$^{-1}$.~{\it Right panel:}~Same as {\it left panel} but showing the CO(3--2) integrated intensity map. The color scale
spans the range  [3$\sigma$, 430$\sigma$] in logarithmic scale with contour levels  -5$\sigma$ (dashed contour), 5$\sigma$, 10$\sigma$, 20$\sigma$, 40$\sigma$, 60$\sigma$, 100$\sigma$ to
400$\sigma$ in steps of 50$\sigma$ where 1$\sigma=27$~mJy~km~s$^{-1}$beam$^{-1}$.  The 
AGN locus lies at the intersection of the dashed lines in both panels.  The (red) filled
ellipses at the bottom left corners in both panels represent the beam sizes in CO(2--1) ($0\farcs11\times0\farcs068$ at  $PA=80^{\circ}$) and CO(3--2) ($0\farcs13\times0\farcs065$ at  $PA=100^{\circ}$).}
   \label{co-full}
    \end{figure*}

As for Band~7, we have checked if the lack of short spacing correction in the CO(2--1) map, which implies that the flux will start to be filtered out on scales $\geq1.3\arcsec$ (90~pc),
can affect the results discussed in this work. To this aim, we have compared the flux integrated over the CND region estimated from the IRAM Plateau de Bure interferometer (PdBI) array image of \citet{Kri11}, which recovers the flux out to  scales\footnote{In particular,  according to the results of simulations performed for this work, we expect a flux loss of 20$\%$ for scales $>3\arcsec$.}  $\simeq3\arcsec\simeq210$~pc,  with that derived from the new ALMA map of Fig.~\ref{co-full}. This comparison yielded similar values for the flux recovered within the errors:  530~Jy~km~s$^{-1}$ \citep{Kri11} $\simeq$ 580~Jy~km~s$^{-1}$ (this work). This suggests that the emission stemming from the missing spatial scales in the range $\simeq90-210$~pc of the new CND map is negligible. In conclusion, our comparisons show that the new CO(2--1) map  of the CND contains most, if not all, of the flux  for the range of spatial scales that are relevant to this paper.

%


   \begin{figure*}
   \centering
      \includegraphics[width=0.85\textwidth]{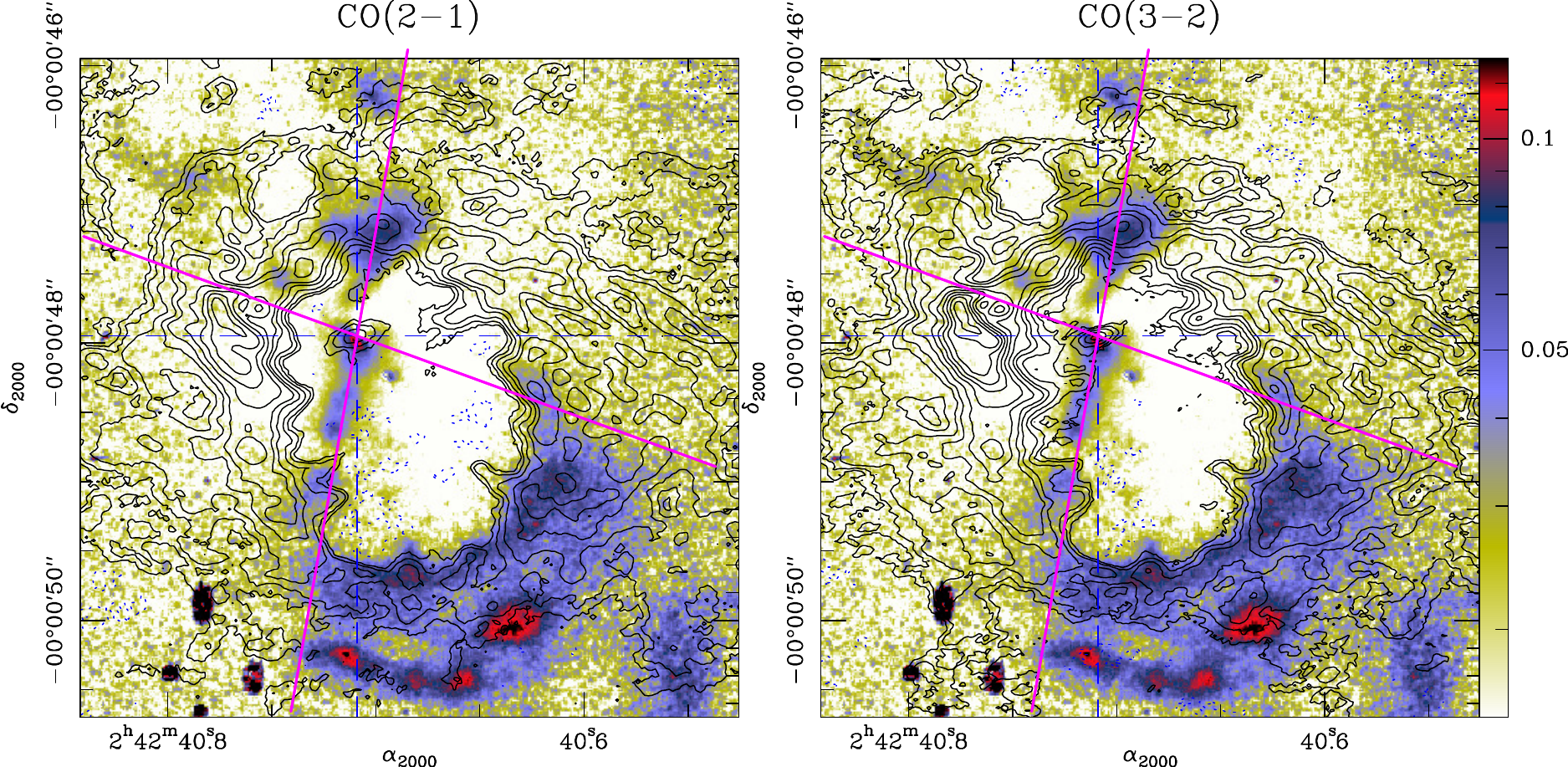} 
       \caption{{\it Left panel:}~We overlay the CO(2--1) contour map (levels as in Fig.~\ref{co-full}) on the image of linear polarization of dust emission obtained in the H band by \citet{Gra15} (color logarithmic scale from 1.5$\%$ to 13$\%$). {\it Right panel:}~Same as {\it left panel} but showing the comparison between the CO(3--2) map and the polarization degree. The magenta lines identify the region occupied by the AGN wind bicone modeled by \citet{Das06}: $PA_{\rm outflow}=30^{\circ}$, FWHM$_{\rm outer}=80^{\circ}$\citep[see also][]{Bar14}.}       
   \label{torus-polar}
    \end{figure*}

   \begin{figure}[th!]
      \centering
      \includegraphics[width=0.47\textwidth]{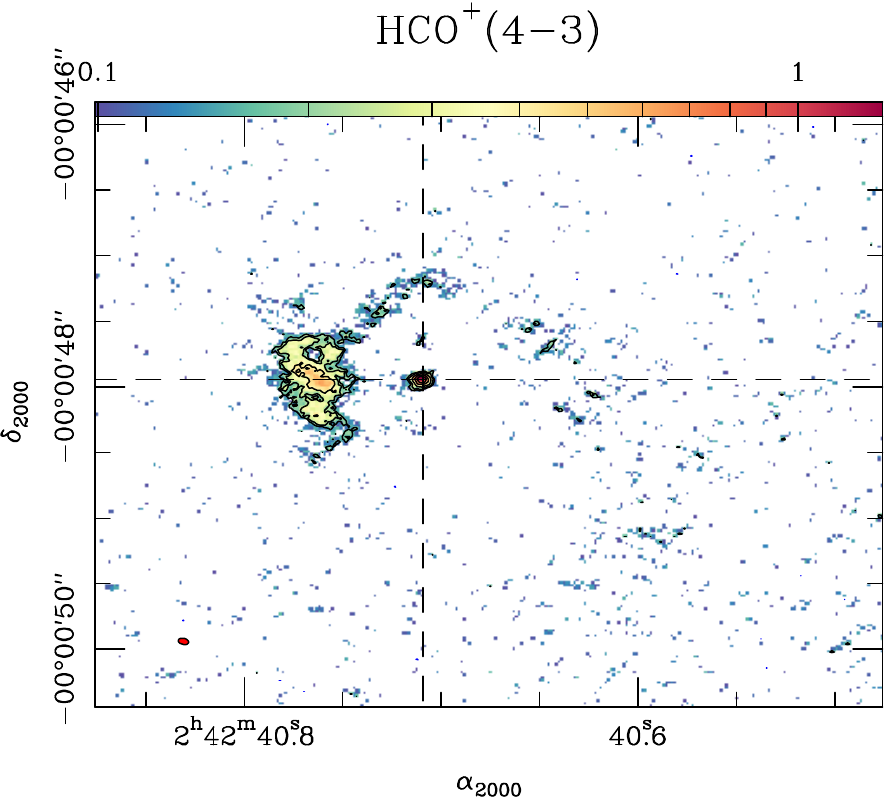}    
   \caption{The HCO$^+$(4--3) integrated intensity map obtained with ALMA in the CND of NGC~1068 using the MSR data set as defined in Table~\ref{Tab1}. 
   The map is shown in 
color scale spanning the range [3$\sigma$, 40$\sigma$] with contour levels  -5$\sigma$ (dashed contour), 5$\sigma$, 7$\sigma$, 12$\sigma$, 20$\sigma$, and 30$\sigma$ where 1$\sigma=33$~mJy~km~s$^{-1}$beam$^{-1}$.  Symbols and markers as in Fig.~\ref{co-full}. 
The (red) filled ellipse at the bottom left corner represents the beam size in HCO$^+$(4--3) ($0\farcs081\times0\farcs046$ at  $PA=78^{\circ}$).}
   \label{hcop-full}
    \end{figure}



\section{Continuum maps}\label{continuum}

Figure~\ref{cont-maps} shows the continuum maps derived at 229.7\,GHz (1306~$\mu$m) and 344.5\,GHz (871~$\mu$m) in the inner $r\leq1\arcsec$(70~pc) region of NGC~1068. We also show the
continuum map obtained by \citet{GB16} at 694\,GHz (432~$\mu$m). 
At the three frequencies examined, the strongest emission corresponds to a point-like source. We used the  {\tt GILDAS} task {\tt UV-FIT} to locate this source at $\alpha_{2000}=02^{h}42^{m}40.709^{s}$, $\delta_{2000}=-00^{\circ}00^{\prime}47.94\arcsec$. This is in agreement with the position of the source S1, identified as the  AGN core in the previous subarcsecond radio continuum maps of \citet{Gal96,Gal04}. We also detect significant emission extending over $\simeq70$~pc  in a highly clumpy jet-like structure at the two lowest frequencies. As illustrated in Fig.~\ref{cont-maps}, the millimeter continuum clumps  at  229.7\,GHz and 344.5\,GHz show a good correspondence with sources S2, C, and NE, which form the inner section of the collimated jet emitting synchrotron emission \citep{Gal96,Gal04}.
This correspondence is much weaker at 694\,GHz.  The jet is seen to be diverted at C due to a jet-ISM interaction \citep{Bic98, Gal96, Gal01}. The C knot is also characterized by highly polarized MIR emission \citep{LopR16}. Reflecting this interaction, the jet changes its orientation from $PA\simeq10^{\circ}\pm5^{\circ}$ along its inner section defined by the C-S1-S2 clumps, towards  $PA\simeq30^{\circ}\pm5^{\circ}$, as defined by the NE-C axis. The latter aligns with the general orientation of the AGN wind on larger scales  \cite[$PA_{\rm outflow}\simeq30^{\circ}\pm5^{\circ}$;][]{Mac94, Arr96, Cre00, Tec01, Cec02, Mue11, Gra15}.

Figure~\ref{cont-alpha} shows  the spectral index maps $\alpha$ (with $S_{\nu}\propto\nu^{\alpha}$)  derived from the continuum emissions at $\nu$=344.5~GHz and 694~GHz  ($\alpha_1$), and from $\nu$=229.7~GHz and 344.5~GHz ($\alpha_2$) derived for a common aperture size of $0\farcs11\times0\farcs11$ ($8$~pc$~\times$~8~pc). We also show in Fig.~\ref{cont-alpha} the corresponding SED in the (sub)millimeter range at different regions of the CND.  Large and positive values of the two spectral indexes $\alpha\simeq$~3--4 can be attributed to the prevalence of dust emission in those regions of the CND that are located far from the radio jet trajectory (e.g., the E-knot) or at the edges of the  jet trail itself. Spectral indexes are flat or marginally positive at the location of the torus and close to its southern polar elongation (knots S1 and S2, respectively). In contrast, spectral indexes become clearly negative ($\alpha\leq$~--1) throughout the radio jet (e.g., the NE-C axis), as expected for synchrotron-like emission.

We used the spectral index values measured in the CND regions dominated by dust emission ($\alpha\simeq$~3--4) to derive the fraction of emission at 871~$\mu$m that can be attributed to dust at the S1 knot and estimated this to be $\leq10\%$, in close agreement with the estimate derived at lower spatial resolution ($\simeq35$~pc) by \citet{GB14,GB16}. This result warns against the use of submillimeter continuum emission  at  $\sim850-870$~$\mu$m as a straightforward tracer of the dust content of AGN tori and/or of their immediate surroundings, in particular in active galaxies that are not radio-silent such as NGC~1068 \citep[see also the recent discussion in Alonso-Herrero et al. 2019 and][]{Pas19}.

   \begin{figure*}[tbh!]
      \centering
      \includegraphics[width=0.95\textwidth]{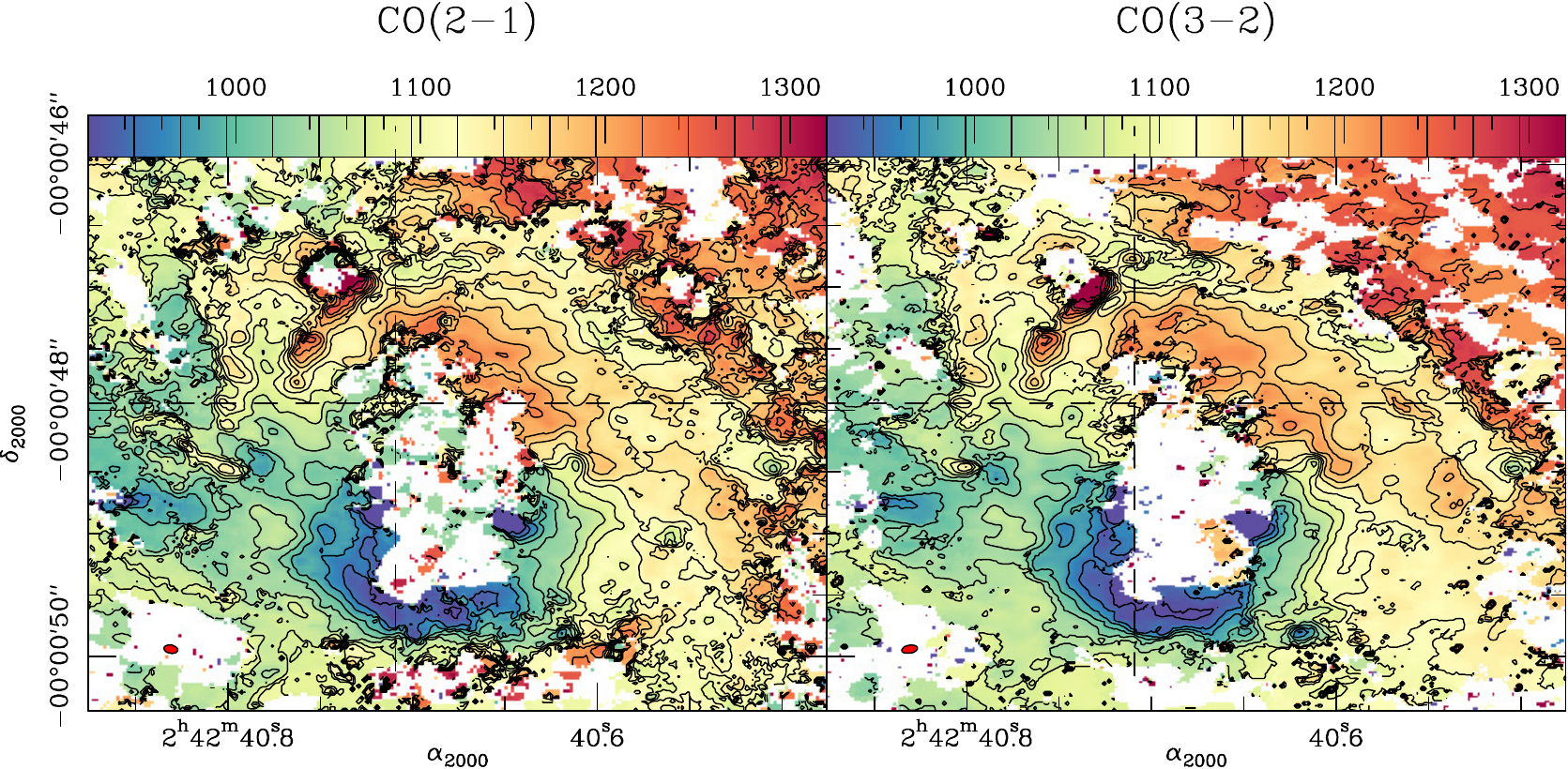}
      \caption{Mean-velocity fields derived for the CO(2--1) ({\it left panel}) and  CO(3--2) ({\it right panel}) lines in the CND of NGC~1068, as obtained from the MSR data set. Isovelocity contours and color scale span the range [--200+$v_{\rm sys}$, 200+$v_{\rm sys}]$~km~s$^{-1}$, where $v_{\rm sys}^{\rm LSR}=1120$km~s$^{-1}$ in steps of 25~km~s$^{-1}$.  Symbols and markers as in Fig.~\ref{co-full}.}
   \label{full-velo}
    \end{figure*}


\section{Molecular line emission: the circumnuclear disk (CND)}\label{CND}

\subsection{The CO maps}\label{CND-CO}

Figure~\ref{co-full} shows the CO(2--1) and CO(3--2) velocity-integrated intensity maps of the CND. The line fluxes have been integrated to include any significant emission that arises over the full velocity span due to rotation and outflow motions in NGC~1068: $v-v_{\rm sys}^{\rm HEL}\subset[ -350, 350]$~km~s$^{-1}$ \citep{GB14,GB16, Gal16}. The maps have also been corrected for 
primary beam attenuation at the respective frequencies.

Based on the measured total flux of the 2--1 line ($\simeq580$~Jy~km~s$^{-1}$), we used Equation~(3) of \citet{Bol13} and estimated the molecular gas mass of the CND to be $M_{\rm gas}\simeq1.4\times10^8~M_{\sun}$, including the mass of Helium. In our estimate we assumed an average 2--1/1--0 brightness temperature ratio $\simeq2.2$, as measured by \citet{Vit14} in the CND, and a galactic CO--to--H$_2$ conversion factor  ($X_{\rm CO}=2\times10^{20}$mol~cm$^{-2}$~(K~km~s$^{-1}$)$^{-1}$).

The high spatial resolution of the ALMA data fully resolves the CND, which displays in both lines of CO the shape of a highly-structured asymmetric ringed disk. The CND has a total size  $\simeq 5\farcs5 \times  4\farcs$0 (400~pc~$\times$~300~pc) and shows an east-west orientation of its principal axes ($PA\simeq90^{\circ}$). If we de-project the CND size onto the plane of the galaxy, assuming  $PA=286^{\circ}\pm5^{\circ}$ and $i=41^{\circ}\pm2^{\circ}$\citep{GB14}, we derive a nearly circular shape for the CND, which would have a  $\simeq400$~pc-diameter. The CND ring feature likely corresponds to the molecular gas assembly at the Inner Lindblad Resonance (ILR) region of the nuclear bar detected  in the NIR \citep{Sco88, Bla97, Sch00, Ems06, GB14}.
 The center of the CND is noticeably shifted $\simeq70$~pc south-west relative to the position of the central engine, an indication that the morphology of the ring may have been shaped to a large extent by the feedback of nuclear activity, as we shall discuss in Sect.~\ref{CND-ring}.
 
Although the two CO line maps show notable differences at small radii, $r\leq50$~pc, their overall morphology elsewhere at larger radii is remarkably similar  down to spatial scales of a few parsecs.   Molecular line emission is detected in both CO lines at and around the AGN locus stemming from  a spatially-resolved elongated disk, which has a diameter $\geq$20~pc. Hereafter we will refer to this disk as the {\it torus}. The torus is partly connected to the CND through gas lanes/streamers that display different morphologies in the 2--1 and 3--2 lines of CO. The torus and its connections to the host are described in detail in Sects.~\ref{torus} and \ref{torus-conn}, respectively.

   \begin{figure*}[tbh!]
      \centering
      \includegraphics[width=0.95\textwidth]{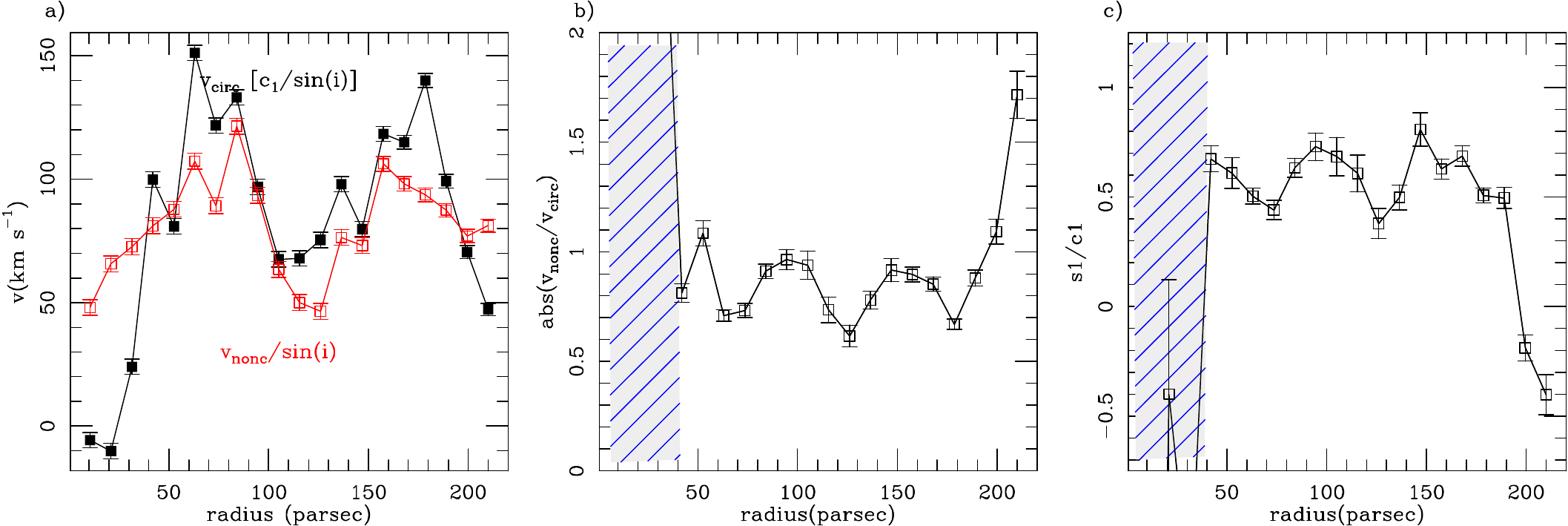}
          \caption{{\bf a)}~({\it Left panel})~The radial profile of the $c_1/$sin$(i)$ ratio,  which represents the best fit obtained by {\tt kinemetry} for the deprojected axisymmetric circular component of the velocity field ($v_{\rm circ}$; black curve and filled symbols).  We also plot the radial profile of the deprojected non-circular motions ($v_{\rm nonc}$/sin$(i)$) derived from the Fourier decomposition until third order (red curve and open symbols). {\bf b)}~({\it  Middle panel}). The variation of the absolute value of $v_{\rm nonc}/v_{\rm circ}$ ratio as function of radius (black curve and open symbols). {\bf c)}~({\it Right panel})~The variation of the  $s_{1}/c_{1}$ ratio as function of radius. The $s_{1}$ coefficient represents the best fit of the (projected) axisymmetric radial component of the velocity field (black curve and open symbols). The hatched grey-filled rectangles in panels {\bf b} and {\bf c} identify the region where the solution found by {\tt kinemetry} is judged less reliable.}   
          \label{harm-I}
       \end{figure*}


   \begin{figure}[tbh!]
      \centering
      \includegraphics[width=0.45\textwidth]{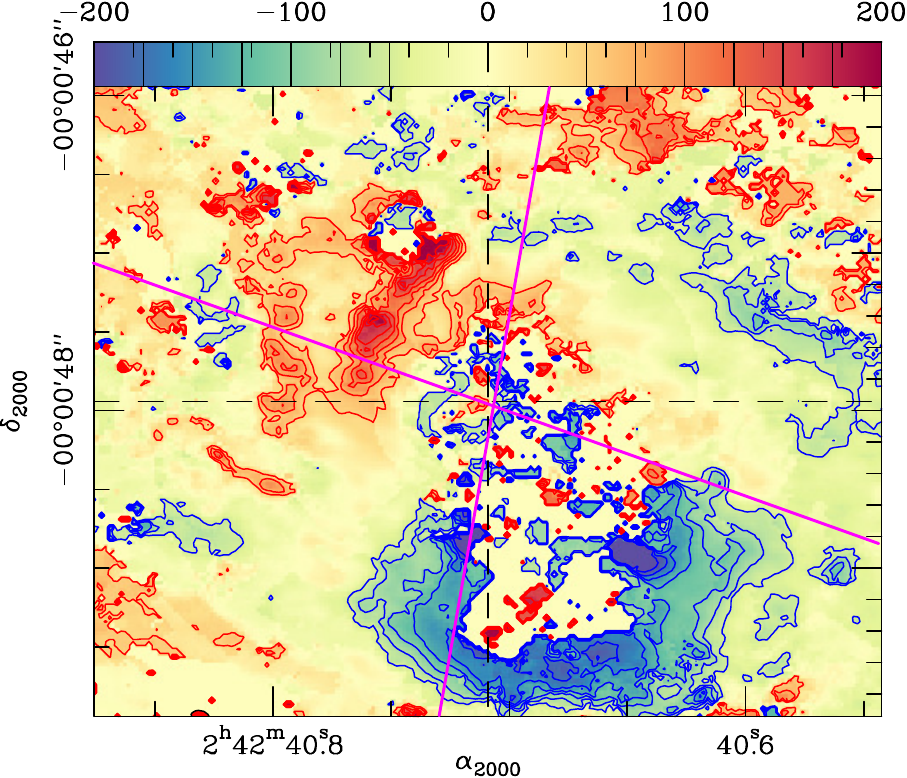}   
       \caption{The residual mean-velocity field  of the CND obtained after subtraction of the best-fit rotation component from the CO(2-1) MSR data, as described in Sect.~\ref{kinemetry}. The color scale spans the range [-200, 200]~km~s$^{-1}$ relative to $v_{\rm sys}^{\rm LSR}=1120$~km~s$^{-1}$. Blue (red) contours go from --200 (+50) to --50 (+200) ~km~s$^{-1}$ in steps of 25~km~s$^{-1}$  relative to $v_{\rm sys}$.  The magenta lines identify the region occupied by the AGN wind bicone.} 
   \label{residuals}
    \end{figure}



   \begin{figure*}[th!]
      \centering
    \includegraphics[width=0.95\textwidth]{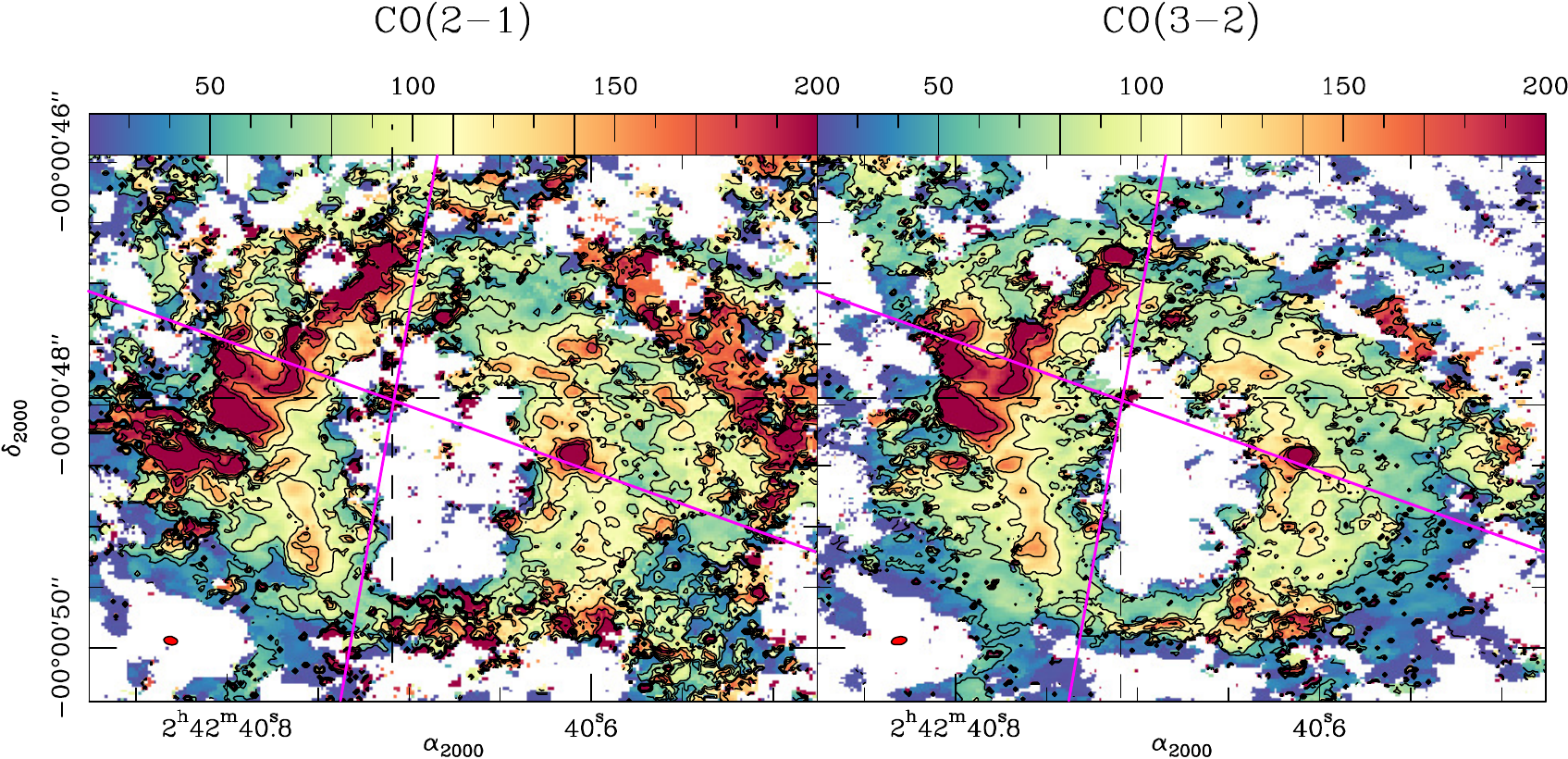}
   \caption{Velocity-width maps in units of full width at half maximum (FWHM) derived for the CO(2--1) ({\it left panel}) and  CO(3--2) ({\it right panel}) lines in the CND of NGC~1068, as obtained from the MSR data set. Contours and color scale span the range [20, 200]~km~s$^{-1}$ in steps of 30~km~s$^{-1}$.  Symbols and markers as in Figs.~\ref{co-full} and \ref{torus-polar}.}
   \label{full-width}
    \end{figure*}


\subsubsection{The CND morphology: evidence of an outflowing ring ?}\label{CND-ring}

Leaving aside the gas concentration at the torus and its surroundings mentioned above, the CND shows a deficit of molecular gas in its central $\simeq130$~pc-region. This deficit leaves the imprint of a highly contrasted ring structure on the CND.  The inner edge of the ring shows a steep radial profile that delimits a gas-deficient region, which has an elliptical shape.  The principal axes of the gas-deficient region ($2\farcs3 \times  1\farcs4\simeq160$~pc$~\times~100$~pc) are oriented roughly at right angles relative to the overall  orientation of the CND.  Compared to the 2--1 line,  this {\it hole} in the 3--2 line seems to shrink in size closer to the central engine. The outer edge of the CND is connected on larger scales to the bar region through a network of spiral gas lanes from $r\geq100-150$~pc. These bar-CND connections were previously identified in the map of \citet{GB14}.

While the central $\simeq130$~pc of the  CND is deficient in molecular gas, this region is known to be pervaded by strong emission from a wide-angle bicone of ionized gas oriented along $PA\simeq30^{\circ}$ \citep{Mac94, Arr96, Cre00, Cec02, Das06, Mue11, Bar14}.  The morphology and the kinematics of the gas
in this component have been interpreted in terms of an outflowing AGN-driven wind.  In particular, the presence of blueshifts and redshifts  on  either  side  of  the  AGN locus along the outflow axis is  consistent  with  a model where the AGN wind axis is oriented close the plane of the galaxy disk \citep[e.g.,][]{Cec02}. This geometry would favor a strong coupling between the AGN wind and the molecular ISM in the disk. Based on the analysis of the distortions of the molecular gas velocity field, studied with ALMA at a spatial resolution of $\simeq35$~pc, \citet{GB14}  concluded that the ionized outflow  is responsible for launching a molecular outflow in the disk over a region extending from $r\simeq50$~pc out to $r\simeq500$~pc.
The highly contrasted ring morphology of the CND imaged by the new ALMA data shown in Fig.~\ref{co-full}, which reveals a remarkably steep radial profile on its inner edge, likely reflects the accumulation of molecular gas at the working surface of the AGN wind. The outflowing CND scenario of \citet{GB14} is re-examined in Sects.~\ref{CND-mean} and \ref{CND-width}, based on a first analysis of the gas kinematics derived from the new ALMA maps. 

Further evidence showing the existence of a large-scale shock in the CND is provided by the comparison of the ALMA CO maps with the hourglass structure displayed by the NIR polarized emission imaged by \citet{Gra15}. This comparison,  illustrated in Fig.~\ref{torus-polar}, shows a fair correspondence between the inner edge of the CND and the edge-brightened arcs of NIR polarized emission, which are prominent at the northern and, most particularly, in the southern regions of the CND\footnote{The southern arcs are brighter in NIR polarized emission due to the higher extinction of the optical emission coming from the southern lobe of the AGN wind, which is partly located below the plane of the galaxy (see Fig.~\ref{outflow-scheme}).}. In the model proposed by \citet{Das06} \citep[see also][for a more complete set of model parameters]{Bar14}, the AGN wind occupies a hollow bicone  that is oriented along $PA\simeq30^{\circ}$ and is characterized by an opening angle, measured from the inner to the outer edge of the bicone, ranging from FWHM$_{\rm inner}\simeq40^{\circ}$  to  FWHM$_{\rm outer}\simeq80^{\circ}$. Based on this model   the interaction of the AGN/wind with the molecular ISM of the disk is expected to be at work mainly at the northern and the southern extremes of the CND. The increased degree of polarization shown by the dust-scattered NIR emission seems to be taking place at the working surface of the AGN wind  as a result of a significant gas/dust pile-up. We also find an increased degree of polarization of NIR emission along the gas
streamers connecting the torus with the CND (see discussion in Sect.~\ref{torus-conn}).

\subsection{The HCO$^+$ map}

Figure~\ref{hcop-full} shows the HCO$^+$(4--3) integrated intensity map of the CND. In stark contrast to the CO line maps shown in Figure~\ref{co-full}, the brightest emission of  the HCO$^+$(4--3) line in the CND corresponds to the  molecular torus. The HCO$^+$ torus is notably smaller in size ($D<10$~pc) compared to the CO molecular tori (see discussion in Sect.~\ref{torus}).  Significant HCO$^+$  emission is also detected at the E-knot, a spatially-resolved region located $\simeq70$~pc east of the AGN, as well as  in a number of isolated clumps throughout the CND. The different distributions of HCO$^+$ and CO line emissions in the CND reflect the wide range of critical densities probed by these transitions, which span nearly  four orders of magnitude: $n$(H$_2$)~$\subset10^3-10^7$cm$^{-3}$.  The location of the brightest   HCO$^+$ emission in Fig.~\ref{hcop-full}, namely the molecular torus and the E-knot, is a signpost of the location of the densest molecular gas: $n$(H$_{\rm 2}$)~$\simeq$~a few 10$^5$--a few 10$^6$cm$^{-3}$, according to the analysis of the different line ratios studied by \citet{Vit14}  at the common $\simeq$35~pc-spatial resolution achieved in the maps of \citet{GB14}.


   \begin{figure}[tbh!]
   \centering
    \includegraphics[width=0.45\textwidth]{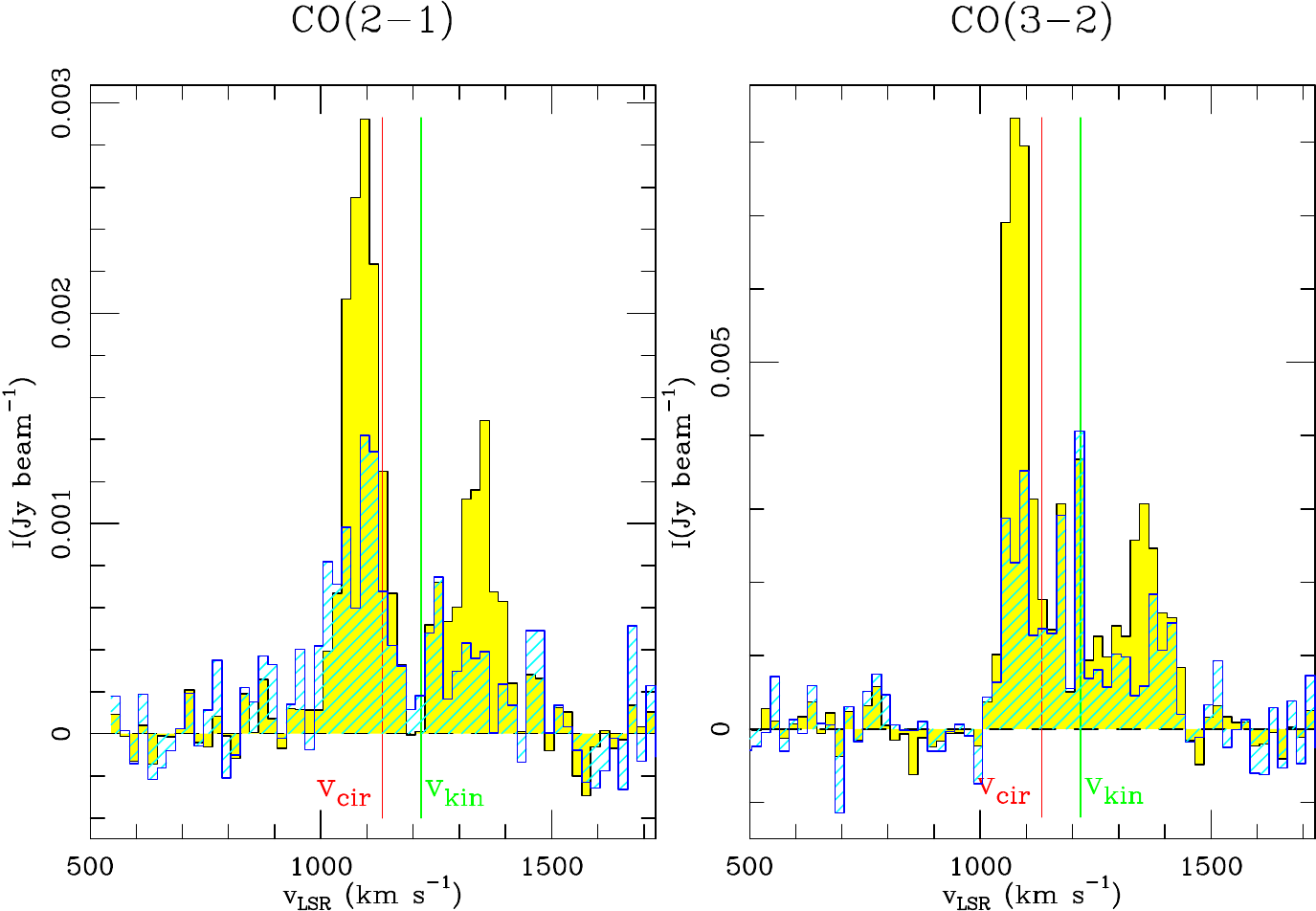} 
    \caption{We show the spectra extracted  at [$\Delta \alpha$, $\Delta \delta$]~=~[$+0\farcs15,  $+0\farcs94] from the MSR data sets for the $J=2-1$ line of CO  ({\it left panel}) and the $J=3-2$ line of CO ({\it right panel}) in yellow-filled histograms; these spectra were extracted using a common aperture of $\simeq$6--7~pc-size. We also show in blue-hatched histograms the corresponding spectra obtained from  the HSR data sets using a common aperture of $\simeq$2--3~pc-size. The green line highlights the radial velocity predicted by {\em kinemetry}, which accounts for the contribution of a co-planar outflowing component ($v_{\rm kin}$). We also indicate by the red line the radial velocity due to circular motions  predicted by the {\em kinemetry} best-fit model ($v_{\rm cir}$).}
   \label{spectra-width} 
    \end{figure}

\subsection{Kinematics of the CND: mean-velocities}\label{CND-mean}

Figure~\ref{full-velo} shows the mean-velocity fields derived from the CO(2--1)  and  CO(3--2)  lines in the CND of NGC~1068. To maximize the reliability of the maps, isovelocities have been obtained by integrating the emission with a threshold of 4$\sigma$ in the two lines. 

The  apparent kinematic major axis  of the CND at small radii is roughly oriented along the North-South direction ($PA\simeq360^{\circ}$). However, the orientation of the major axis changes at the outer edge of the CND towards  $PA\simeq290^{\circ}$, i.e., it tends to align with the large-scale disk of the galaxy, defined by  $PA=286^{\circ}\pm5^{\circ}$ and $i=40^{\circ}\pm3^{\circ}$, where the south-west (north-east) side of the disk corresponds to the near (far) side \citep{Bla97}. \citet{GB14} used the software package {\tt kinemetry} \citep{Kra06} to account for a similar yet less extreme tilt of the major axis identified in previous ALMA observations of  NGC~1068\footnote{Differences with respect to the magnitude of the distortions of the velocity field reported here can be attributed to the comparatively lower spatial resolution of the observations used by \citet{GB14}.}. \citet{GB14} concluded that the distortions  of the mean-velocity field of molecular gas in the CND  could be reproduced by the inclusion of an outflowing radial component \citep[see also][]{Kri11}.  In the following section  we use the package {\tt kinemetry} in an attempt to improve the fit of the mean-velocity field distortions derived from the new higher resolution ALMA data presented in this work. 

\subsubsection{Modeling the mean-velocity field with  {\tt kinemetry}: a coplanar outflow?} \label{kinemetry}

The observed mean-velocity field of a galaxy disk, $v_{\rm mean}(r, \psi)$, expressed a function of the radius ($r$) and the phase angle ($\psi$, measured from the receding side
of the line of nodes), can be divided into a number of elliptical ring profiles with a geometry defined by the position angle and inclination ($PA$, $i$). We can decompose $v_{\rm mean}(r, \psi)$ as a Fourier series with harmonic coefficients $c_j(r)$ and $s_j(r)$, where 
\begin{equation}
v_{\rm mean} (r, \psi)= c_0 + \sum_{j=1}^n [c_j(r) \cos (j\psi) + s_j(r) \sin (j\psi)]
\label{eq-1}
\end{equation}

While $c_1$ accounts for  the contribution from circular rotation, the other terms of the series contain the contributions from noncircular 
motions, $v_{\rm nonc}$, which by construction has an axisymmetric dependence  \citep{Sch97, Sch99}. In particular, if we expand the series out to $n=3$:

 \begin{equation}
v_{\rm nonc} (r)= \sqrt{s_1^2(r)+s_2^2(r)+c_2^2(r)+s_3^2(r)+c_3^2(r)}  
\label{eq-2}
\end{equation}

We have used the software package {\tt kinemetry}, developed by \citet{Kra06}, to fit the CO(2--1) mean-velocity field of the CND of Fig.~\ref{full-velo} purposely using a minimum number of free parameters. In particular, we first assumed that the dynamical center coincides with the AGN position derived in Sect.~\ref{continuum}. Secondly, we adopt the solution found by {\tt kinemetry} for the basic parameters of the large-scale $r\simeq1.5$~kpc --disk covered by the first ALMA observations of \citet{GB14}, defined by $PA=289\pm5^\circ$ and $i=41\pm2^\circ$, and apply it to the relevant radii of the CND. These parameters were obtained to fit the mean-velocity field derived from the previous lower-resolution CO(3--2) ALMA map of the galaxy \citep[see for a detailed discussion ][]{GB14}. This approach implicitly assumes that the molecular gas kinematics in the CND can be modeled by coplanar orbits in a thin disk\footnote{More realistic 3D scenarios are explored in Sect.~\ref{BBarolo}}. From this first iteration we obtained  the best fit for the systemic velocity:  $v_{\rm sys}$(LSR)~$=1120\pm3$~km~s$^{-1} = v_{\rm sys}$(HEL)~$=1131\pm3$~km~s$^{-1}$;  this is compatible within the errors with the values reported in the literature inferred using different diagnostics \citep[e.g.,][]{Gre97,GB14}. We then subtract $v_{\rm sys}$ from 
$v_{\rm mean}$ and re-derive the Fourier decomposition of Eq.~\ref{eq-1} for a set of 20 radii sampling the CND from $r=0\farcs15~(\simeq10$~pc) to   $r=3\arcsec~(\simeq210$~pc).

 Figure~\ref{harm-I}{\it a} represents the radial profiles of the deprojected circular and non-circular components of the CO(2--1) velocity field obtained in the fit ($v_{\rm circ} = c_{1}$/sin$(i)$ and $v_{\rm nonc}/$sin$(i)$, respectively). The fit fails to converge to a physically sound solution for  $v_{\rm circ}$ at small radii  $r<40$~pc, where {\tt kinemetry} finds anomalous
 retrograde rotation and/or very  high values of the $v_{\rm nonc}$/$v_{\rm circ}$ ratio. This reflects the random pattern of the observed  $v_{\rm mean}$ at these small radii (Fig.~\ref{full-velo}). As we shall argue in Sects.~\ref{CND-width} and \ref{torus-conn},  $v_{\rm mean}$ gives an incomplete representation of the complex kinematics of molecular gas in this region, which is characterized by different velocity components around $v_{\rm sys}$.  Outside $r\simeq40$~pc,  the value of $v_{\rm nonc}/v_{\rm circ}$ stays within the range $\sim 0.6-1$ over a sizable fraction of the CND (40~pc~$ < r < $~190~pc; Fig.~\ref{harm-I}{\it b}). As shown in  Fig.~\ref{harm-I}{\it c}, the 
main contribution to $v_{\rm nonc}$ comes from the $s_1$ term. In particular, the sign ($>0$) and normalized  strength of $s_1$  ($s_1/c_1\simeq  0.4-0.8$) indicates significant 
outflow radial motions in the CND, which reach deprojected values, evaluated by the $s_1/$sin$(i)$ term, of up to $\simeq85$~km~s$^{-1}$. This result is compatible with the average value reported by \citet{GB14} for the CND, obtained using a much coarser radial sampling and a significantly lower spatial resolution. Furthermore, at the outer edge of the CND ($r \simeq$~190-210~pc), strong non-circular motions ($v_{\rm nonc}$/$v_{\rm circ}>$~1) are also required to fit the complexity of the observed velocity field, which partly reflects the bar-driven gas inflow ($s_1/c_1\simeq -0.4$). The gas inflow due to the bar extends farther out in the disk up to $r\simeq1.5$~kpc, as discussed by \citet{GB14}.

Figure~\ref{residuals} shows the residual mean-velocity field obtained after subtraction of the rotation component derived in the analysis above\footnote{For the inner 40-pc region, where the {\tt kinemetry} solution is judged doubtful we adopt the solution found by {\tt $^{3D}$Barolo} instead, as discussed in Sect.~\ref{BBarolo}.}.  The residual velocity field of the CND shows the largest departures from rotation (locally up to $\pm$200~km~s$^{-1}$) for the molecular gas located in the region being swept by the AGN wind bicone. The signs of the residuals, mostly redshifted (blueshifted) on the northern (southern) side of the CND across the AGN wind trail, indicate that the kinematics of the gas is shaped by outward radial motions, as captured by the axisymmetric fit of {\tt kinemetry}.  In conclusion, these results confirm that the imprint of the outflow on the mean-velocity field of the molecular gas can be accounted for to first order by a scenario where the gas is being pushed by the AGN wind inside the disk creating a radial expansion of the CND with velocities that are on average $\simeq85$~km~s$^{-1}$ from  $r\simeq50$~pc out to $r\simeq200$~pc, with maximum deprojected values that can reach locally $\simeq200$~km~s$^{-1}$. As expected, the residual velocity field of the CND, shown in Fig.~\ref{residuals}, exhibits significantly larger departures from circular rotation compared to the lower-resolution CO map of \citet{GB14} (See their Fig.~{15}).  

The mass of gas in the outflow has been calculated from the CO(2--1) data cube, after subtraction of the projected rotation 
curve derived by {\tt kinemetry}, by integrating the emission of the line outside a velocity interval $\langle  v_{\rm res} \rangle=[v_{\rm sys}-50,v_{\rm sys}+50]$~km~s$^{-1}$ that encompasses by construction most of the expected virial range attributable to turbulence.  We determine that $\simeq 55\%$ of the total CO(2--1) flux in the CND stems from the outflow component, which implies that around half of the mass of the CND participates in the outflow. A similar percentage ($\simeq 50\%$) was derived by \citet{GB14}.

   \begin{figure*}
   \centering
    \includegraphics[width=0.95\textwidth]{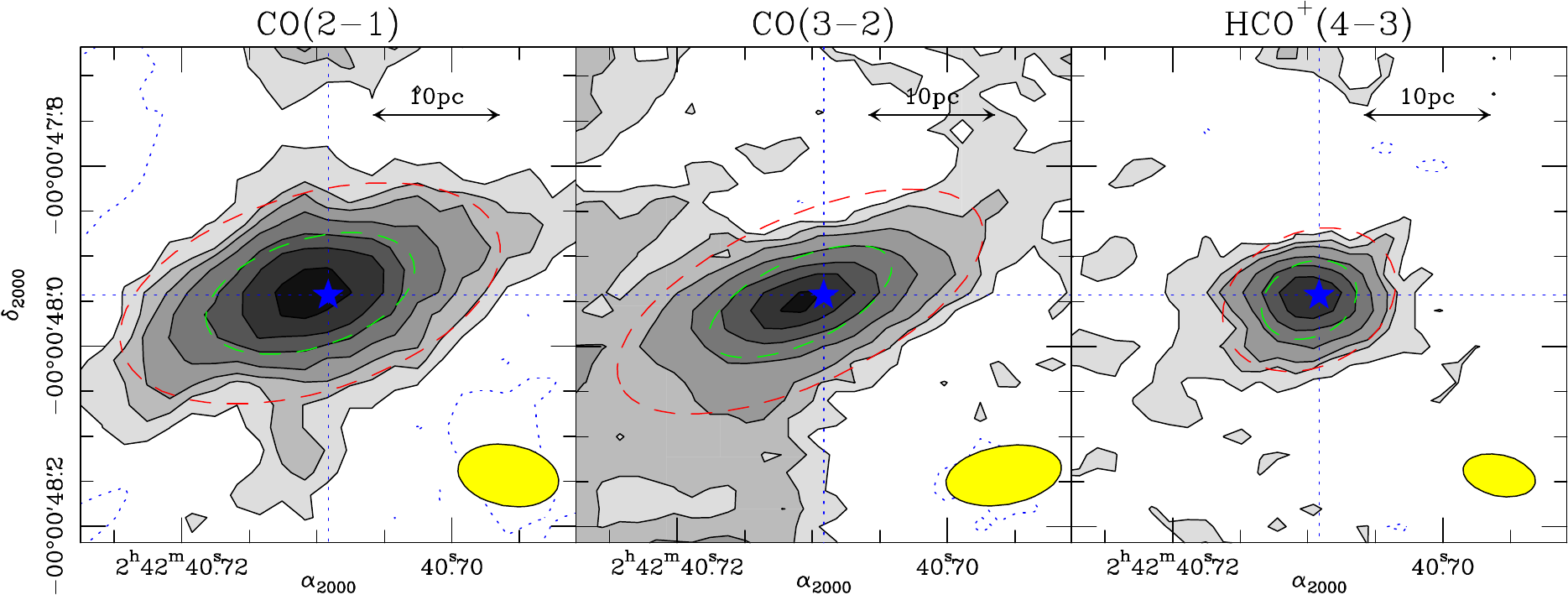} 
    \caption{{\it Left panel:}~The CO(2--1) map of the central $r\leq0\farcs28$ ($\simeq$20~pc) region around the central engine of NGC~1068 obtained using the MSR data set, as defined in Table~\ref{Tab1}. 
    The image shows the molecular torus/disk and its connections.  The map is shown in grey-filled contour levels:  -2.5$\sigma$ (dashed), 2.5$\sigma$, 5$\sigma$, 7$\sigma$, 10$\sigma$, 15$\sigma$, 20$\sigma$, and 30$\sigma$, where 1$\sigma$~=~13~mJy~km~s$^{-1}$beam$^{-1}$.  ~{\it Middle panel:}~Same as {\it left panel} but  showing the CO(3--2) map. Contour spacing is  -2.5$\sigma$ (dashed), 2.5$\sigma$, 5$\sigma$, 10$\sigma$, 10$\sigma$, 20$\sigma$, 35$\sigma$,  50$\sigma$, and 65$\sigma$, where 1$\sigma$~=~27~mJy~km~s$^{-1}$beam$^{-1}$.  ~{\it Right panel:} ~Same as {\it left panel} but  showing the HCO$^+$(4--3) map. Contour spacing is  -2.5$\sigma$ (dashed), 2.5$\sigma$, 5$\sigma$, 7$\sigma$, 12$\sigma$, 20$\sigma$, and 30$\sigma$, where 1$\sigma$~=~33~mJy~km~s$^{-1}$beam$^{-1}$. The dashed green (red) ellipses represent the FWHM-sizes (full-sizes at a 3$\sigma$-level) of the Gaussian fits to the distribution of intensities of the three transitions imaged in the torus prior to deconvolution, as listed in Table~\ref{Tab2}. The position of the AGN is identified by the (blue) star marker. The (yellow) filled
ellipses at the bottom right corners of the panels represent the beam sizes of ALMA.}
   \label{torus-maps-lr}
    \end{figure*}
   \begin{figure*}[bth!]
   \centering
    \includegraphics[width=0.95\textwidth]{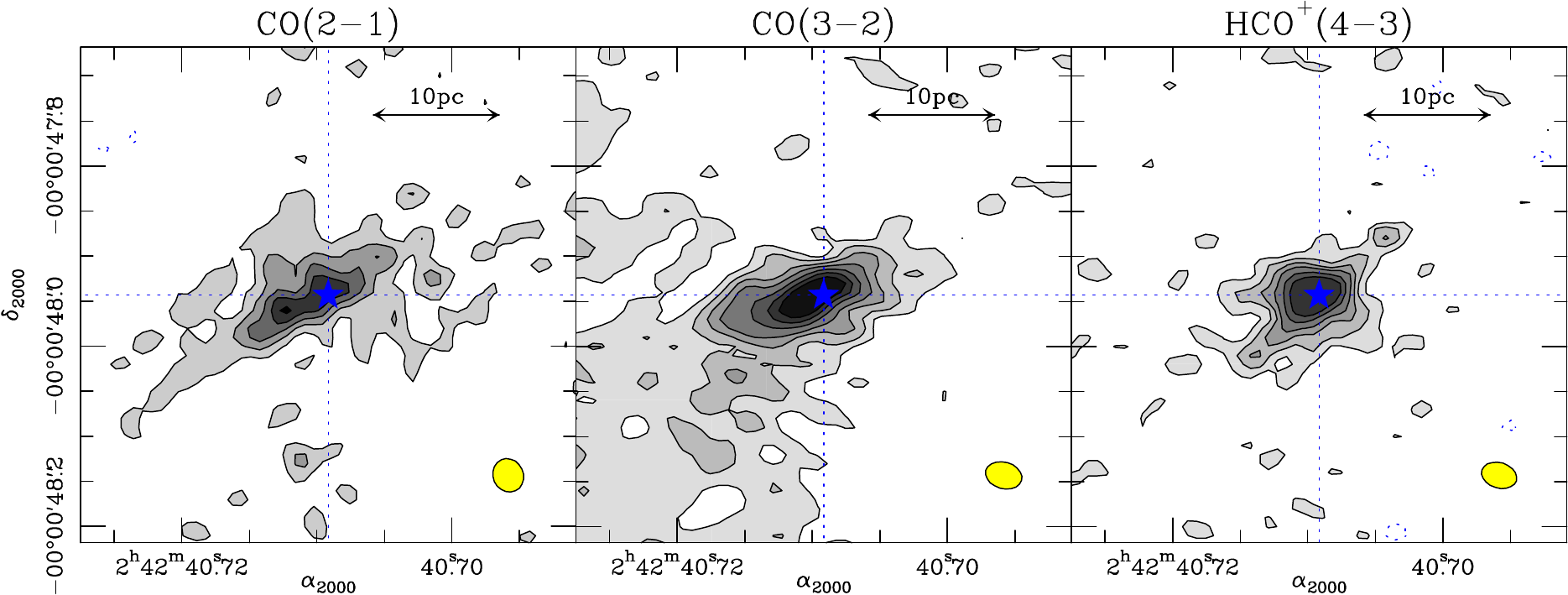} 
    \caption{Same as Fig.~\ref{torus-maps-lr} but showing the maps obtained using the HSR data set, as defined in Table~1. The CO(2--1) contours are:  -3$\sigma$ (dashed), 3$\sigma$, 5$\sigma$, 7$\sigma$, 9$\sigma$, and 11$\sigma$, where 1$\sigma$~=~22~mJy~km~s$^{-1}$beam$^{-1}$. The CO(3--2) contours are   -3$\sigma$ (dashed), 3$\sigma$, 5$\sigma$, 7$\sigma$, 9$\sigma$, 12$\sigma$, 15$\sigma$, and 18$\sigma$, where 1$\sigma$~=~35~mJy~km~s$^{-1}$beam$^{-1}$. The HCO$^+$(4--3) are -3$\sigma$ (dashed), 3$\sigma$, 5$\sigma$, 7$\sigma$, 9$\sigma$, 12$\sigma$ and 15$\sigma$, where 1$\sigma$~=~34~mJy~km~s$^{-1}$beam$^{-1}$.}
   \label{torus-maps-hr}
    \end{figure*}

\subsection{Kinematics of the CND: line-widths}\label{CND-width}

Figure~\ref{full-width}  shows the line width maps (FWHM) of the CND derived for the 2--1 and 3--2 lines using a 4$\sigma$-clipping. The picture drawn from both lines is similar: line widths lie in the range $\leq25-50$~km~s$^{-1}$ only at the outer edge of the CND, while they increase up to significantly higher values  $\geq100$~km~s$^{-1}$ at the smaller radii of the inner edge of the CND, coinciding with the region associated with the working surface of the AGN wind. Line widths can reach very high values ($\simeq200-250$~km~s$^{-1}$) at a fairly large number of positions located 
on the northern and southern extremes  of the CND, i.e., in the way of the jet/AGN wind trajectories. The extreme broadening of the spectra identified in these regions cannot be easily explained by the beam smearing of velocity gradients produced by a combination of rotation and non-circular motions in a coplanar geometry, considering, firstly, the high spatial resolution of the observations ($\simeq 2-7$~pc) and, secondly,  the moderate inclination of the galaxy disk \citep[$i=41^{\circ}\pm2^{\circ}$;][]{GB14}.

 Figure~\ref{spectra-width} shows the CO 2--1 and 3--2 spectra obtained at a representative offset where we have identified extreme line broadening ([$\Delta \alpha$, $\Delta \delta$]~=~[$+0\farcs15,  $+0\farcs94], where offsets are measured relative to the AGN; see Sect.~\ref{ALMA}).  At this position,  the CO profiles show line splitting into two velocity components that are shifted  by $\geq250$~km~s$^{-1}$ relative to each other. As illustrated by Fig.~\ref{spectra-width}, this velocity span  is mostly independent of the range of aperture sizes used to extract the spectra, which coincide with the spatial resolutions of the observations, namely $\simeq~2-7$~pc.  This is an indication that the main contributor to the widening of the spectra must necessarily come from the projection of gas motions perpendicular to the plane of the galaxy.

 As argued in Sect.~\ref{CND-mean}, the coplanar outflow scenario is a first-order explanation for the overall redshift of the line emission at the northern region of the CND: Fig.~\ref{spectra-width} shows that the mean-velocity predicted by the best-fit model found by {\tt kinemetry} ($v_{\rm kin}$) is shifted by $\simeq+100$~km~s$^{-1}$ relative to the radial velocity attributable to circular motions  ($v_{\rm cir}$).  A similar blueshift of CO lines relative to the predicted motions due to circular rotation at the southern region CND is also explained. However, the observed line splitting and the implied extreme broadening of the spectra illustrated by Fig.~\ref{spectra-width} strongly suggest that the molecular gas in the CND has also been forced to leave the plane of the galaxy and has therefore adopted a  3D shell-like geometry.  We defer to Sects.~\ref{torus-conn} and ~\ref{BBarolo} for a more detailed discussion of  the kinematics of the large-scale outflow and its relation to the dynamics of the molecular torus and its immediate surroundings.

  %

  \begin{table*}[t!]
\caption{\label{Tab2} AGN position and fitted torus parameters.}
\centering
\resizebox{0.99\textwidth}{!}{
\begin{tabular}{lcccll}
\noalign{\smallskip} 
\hline
\hline
\noalign{\smallskip} 
Tracer & $\alpha_{\rm 2000}$ & $\delta_{\rm 2000}$ &  Offset-AGN &  FWHM-Size & $PA$	 \\
\noalign{\smallskip}
	    & $^h$ ~~~~~   $^m$ ~~~~ $^s$ &  $^o$~~~~  $'$ ~~~~  $\arcsec$ &  [$\arcsec$,~$\arcsec$]~=~[pc, pc]	&	
	    $\arcsec$~$\times$~$\arcsec$~=~pc~$\times$~pc & degrees \\
\noalign{\smallskip} 	      	      
\hline
\noalign{\smallskip} 	
continuum@229.7GHz & 02$^h$:42$^m$:40.709$^s$ & $-00^{\circ}00^{\prime}47.94\arcsec$ & ---  & --- & --- \\
continuum@344.5GHz & 02$^h$:42$^m$:40.709$^s$ & $-00^{\circ}00^{\prime}47.94\arcsec$ & ---  & --- & --- \\
CO(2--1) & 02$^h$:42$^m$:40.710$^s$ & $-00^{\circ}00^{\prime}47.94\arcsec$ & [+0.02$\arcsec$, +0.00$\arcsec$]~=~[+1.4~pc,~+0.1~pc] & $0.24\arcsec \times 0.12\arcsec=(16.8\pm0.3)$~pc~$\times~(8.5\pm0.2)$~pc& 107$\pm$5$^{\circ}$ \\
CO(2--1)-deconvolved & --- & --- & --- & $0.22\arcsec \times 0.09\arcsec=(15.4\pm0.3)$~pc~$\times~(6.3\pm0.2)$~pc& 112$\pm$5$^{\circ}$ \\
CO(3--2) & 02$^h$:42$^m$:40.711$^s$ & $-00^{\circ}00^{\prime}47.93\arcsec$ & [+0.03$\arcsec$, --0.01$\arcsec$]~=~[+2.0~pc, --0.5~pc] & $0.22\arcsec \times 0.09\arcsec=(15.4\pm0.3)$~pc~$\times~(6.3\pm0.2)$~pc& 115$\pm$5$^{\circ}$ \\
CO(3--2)-deconvolved & --- & --- &  --- & $0.19\arcsec \times 0.06\arcsec=(13.3\pm0.3)$~pc~$\times~(4.2\pm0.2)$~pc& 119$\pm$5$^{\circ}$ \\
HCO$^+$(4--3) &  02$^h$:42$^m$:40.710$^s$ & $-00^{\circ}00^{\prime}47.93\arcsec$ & [+0.01$\arcsec$,  --0.01$\arcsec$]~=~[+0.7~pc, --0.5~pc] & $0.11\arcsec \times 0.08\arcsec=(7.7\pm0.3)$~pc~$\times~(5.6\pm0.2)$~pc& 115$\pm$5$^{\circ}$ \\  
HCO$^+$(4--3)-deconvolved & --- & --- & --- & $0.09\arcsec \times 0.05\arcsec=(6.3\pm0.3)$~pc~$\times~(3.5\pm0.2)$~pc& 138$\pm$5$^{\circ}$ \\
 \noalign{\smallskip}    
\hline
\hline 
\end{tabular}} 
\tablefoot{Columns (2) and (3) list the AGN position in absolute coordinates derived from the continuum observations at 229.7GHz and 344.5GHz. We also list the parameters of the molecular tori derived from the elliptical Gaussian fits of the images obtained in the different lines: center (columns [2] and [3]), offset relative to the AGN position (column [4]), FWHM-sizes (column [5]) , and orientation ($PA$; column [6]). We also list the sizes and orientation of the molecular tori after deconvolution by the beams listed in Table~\ref{Tab1}.}\\
\end{table*}


\section{Molecular line emission: the torus}\label{torus}

\subsection{Morphology, size and orientation}\label{torus-parameters}

Figure~\ref{torus-maps-lr} shows a zoom on the inner $r\leq20$~pc-region of the CO and HCO$^+$ maps around the position of the AGN. The maps  were obtained using the MSR data set, as defined in Table~\ref{Tab1}.  
We also show in Fig.~\ref{torus-maps-hr} the maps obtained using the HSR data set (see Table~\ref{Tab1}).  

The images shown in Figs.~\ref{torus-maps-lr} and \ref{torus-maps-hr} prove the existence of a spatially-resolved molecular disk/torus that extends over a significantly large range of spatial scales  $\simeq10-30$~pc around the central engine. This result adds supporting evidence of the existence of  a spatially-resolved molecular disk/torus in NGC~1068, which has been claimed based on previous high-resolution ALMA images of the galaxy obtained in a different set of molecular lines and dust continuum emission: CO(6--5) and dust emission \citep{GB16, Gal16}; HCN(3--2) and HCO$^+$(3--2) \citep{Ima16, Ima18, Imp19}. 

Table~\ref{Tab2} lists the size, position and orientation of the molecular disks identified in Fig.~\ref{torus-maps-lr}, derived using elliptical Gaussians to fit the images in the plane of the sky\footnote{We opted to use the images obtained  from the MSR data set to derive the physical parameters of the torus so as to take advantage of the higher sensitivity of these data.}. We also list the sizes and orientations derived after deconvolution by the corresponding beams at each frequency.

   \begin{figure*}[bth!]
   \centering
      \includegraphics[width=0.7\textwidth]{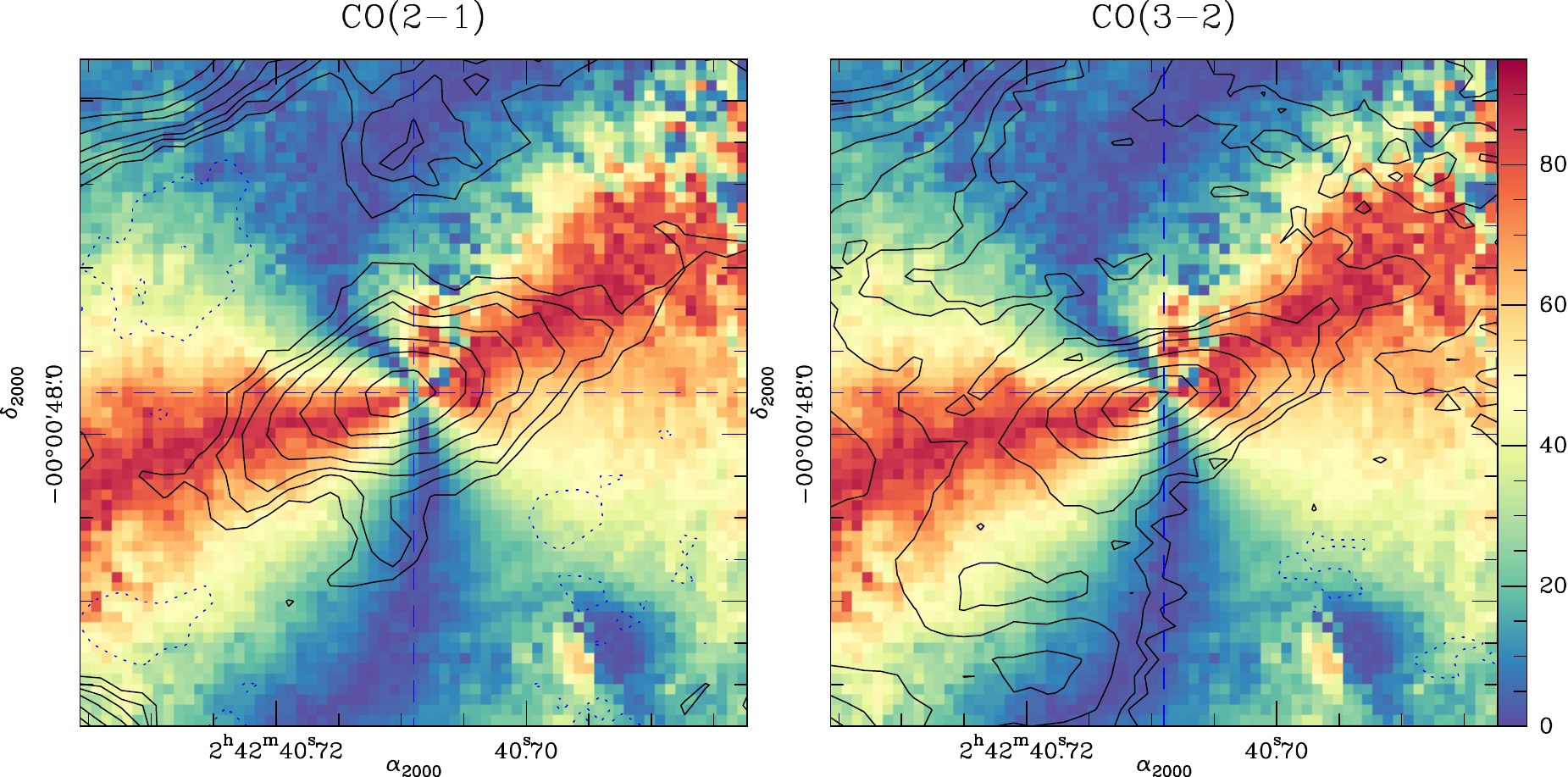} 
       \caption{{\it Left panel:}~We overlay  the CO(2--1) contours of Fig.~\ref{torus-maps-lr} on the difference between the polarization angle map and a purely centro-symmetric pattern derived from the H-band map of \citet{Gra15} (in color scale in units of degrees) in the central $r\leq0\farcs4$ ($\simeq$30~pc) region around the AGN of NGC~1068. {\it Right panel:}~Same as {\it left panel} but showing the comparison between  the CO(3--2) contours and the angle map.}       
   \label{torus-angle}
    \end{figure*}


The physical parameters listed in Table~\ref{Tab2} differ notably depending on the line transition used to image the torus, a result that brings into light the many faces of the molecular torus in NGC~1068. As a common pattern, the torus is shifted overall $\simeq0.7-2$~pc east relative to the AGN. However, the off-centering  is a factor of $\simeq2-3$ larger in the CO lines  compared to  that derived in HCO$^+$(4--3). Furthermore, there is a clear trend showing comparatively  larger sizes for the torus ($D_{\rm major}$) imaged in the lower density tracers. In particular, the deconvolved  FWHM-sizes of the torus go from $\simeq15\pm0.3$~pc in CO(2--1) to $\simeq13\pm0.3$~pc  in CO(3--2), and down to $\simeq6\pm0.3$~pc in HCO$^+$(4--3). The equivalent full-sizes of the tori, defined as the sizes of the disks measured at a  $\simeq3\sigma$ intensity level of the Gaussian used in the fit, range from $\simeq28\pm0.6$~pc in CO(2--1),  to  $\simeq26\pm0.6$~pc in CO(3--2), and $\simeq11\pm0.6$~pc in HCO$^+$(4--3). 
 
 The orientations of the molecular tori, measured anticlockwise from north, are found within the range $PA\subset [112^{\circ}-138^{\circ}]\pm5^{\circ}$. This range encompasses within the errors the values previously reported for the CO(6--5) and dust/continuum tori \citep[$PA_{\rm CO(6-5)}=112^{\circ}\pm20^{\circ}$; $PA_{\rm dust}=143^{\circ}\pm23^{\circ}$;][]{GB16,Gal16}, as well as 
 for the maser disk  \citep [$PA_{\rm maser}=140^{\circ}\pm5^{\circ}$;][]{Gre96,Gal01}.
 
 The orientation of the molecular torus, $PA=113^{\circ}$, derived from the average of the values listed in Table~\ref{Tab2}, is roughly perpendicular to the orientation of the jet and the ionized wind axes  ($PA\simeq10-30^{\circ}\pm5^{\circ}$). Furthermore, the molecular torus is closely aligned with the $\simeq$1-1.5~pc-scale disk imaged in radio continuum by \citet{Gal04}. The described geometry adds supporting evidence for this to be the  orientation of the accretion disk in NGC~1068.  In this context it is noteworthy that a recent high-resolution radio continuum imaging of four low-luminosity AGNs published by \citet{Kam19} has found that  the orientations of the radio jets are misaligned with the normal to the known maser disks in these sources  by less than 30$^{\circ}$.

 We measured aspect ratios for the tori within the range  $D_{\rm major}/D_{\rm minor}\simeq2-3$. The implied lower limit to the inclination of the disks is $i\geq60-70^{\circ}$.  This indicates that molecular gas is settling around the central engine of NGC~1068 on the observed spatial scales $\simeq10-30$~pc on a highly-inclined disk, which is significantly tilted relative to the large-scale disk of the galaxy \citep[$i\simeq41^{\circ}\pm2^{\circ}$;][]{Bla97, GB14}. The high inclination of the molecular torus of NGC~1068 is expected given the Type 2 classification for this Seyfert.
Recent ALMA observations of a sample of seven nearby Seyferts, published by \citet{Com19} \citep[see also][]{Alo18, Alo19}, have  confirmed that NGC~1068 is not an exception: the molecular  tori identified in the targets observed by ALMA are  generally not aligned with the main disks of the AGN hosts, but their measured orientation tend to be in agreement with their optical Type 1--2 classification. Most remarkably, this decoupling is seen to be taking place on spatial scales $\geq10-30$~pc, similar to those identified in NGC~1068.

   \begin{figure*}[bth!]
   \centering
    \includegraphics[width=0.85\textwidth]{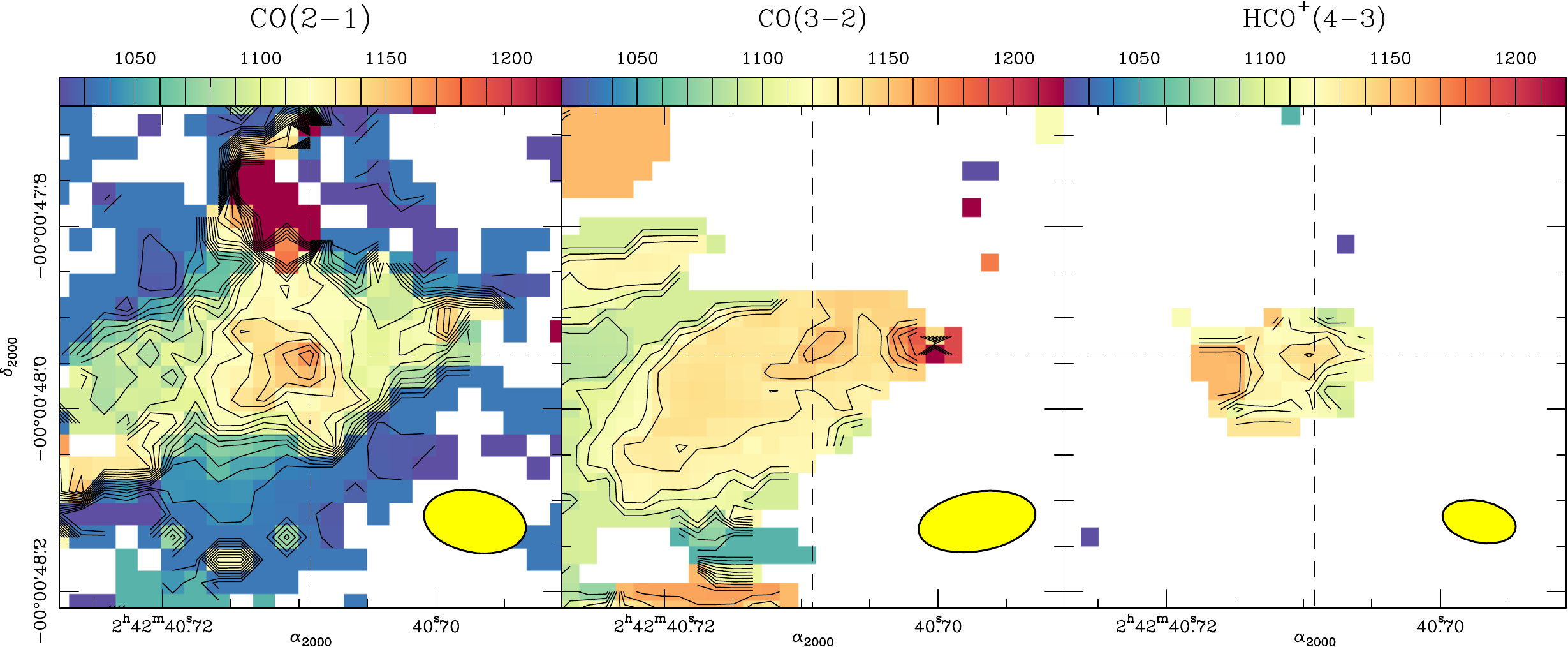} 
    \caption{Mean-velocity fields derived from the CO(2--1) ({\it left panel}), CO(3--2) ({\it middle panel}) and HCO$^+$(4--3) ({\it right panel}) lines in the central $r\leq0\farcs28$ ($\simeq$20~pc) region around the central engine of NGC~1068. Isovelocity contours and color scale span the range [-100, 100]~km~s$^{-1}$ relative to $v_{\rm sys}^{\rm LSR}=1120$km~s$^{-1}$ in steps of 10~km~s$^{-1}$.  The (yellow) filled
ellipses at the bottom right corners of the panels represent the beam sizes of ALMA.}
   \label{torus-velo}
    \end{figure*}

   \begin{figure*}
   \centering
    \includegraphics[width=0.85\textwidth]{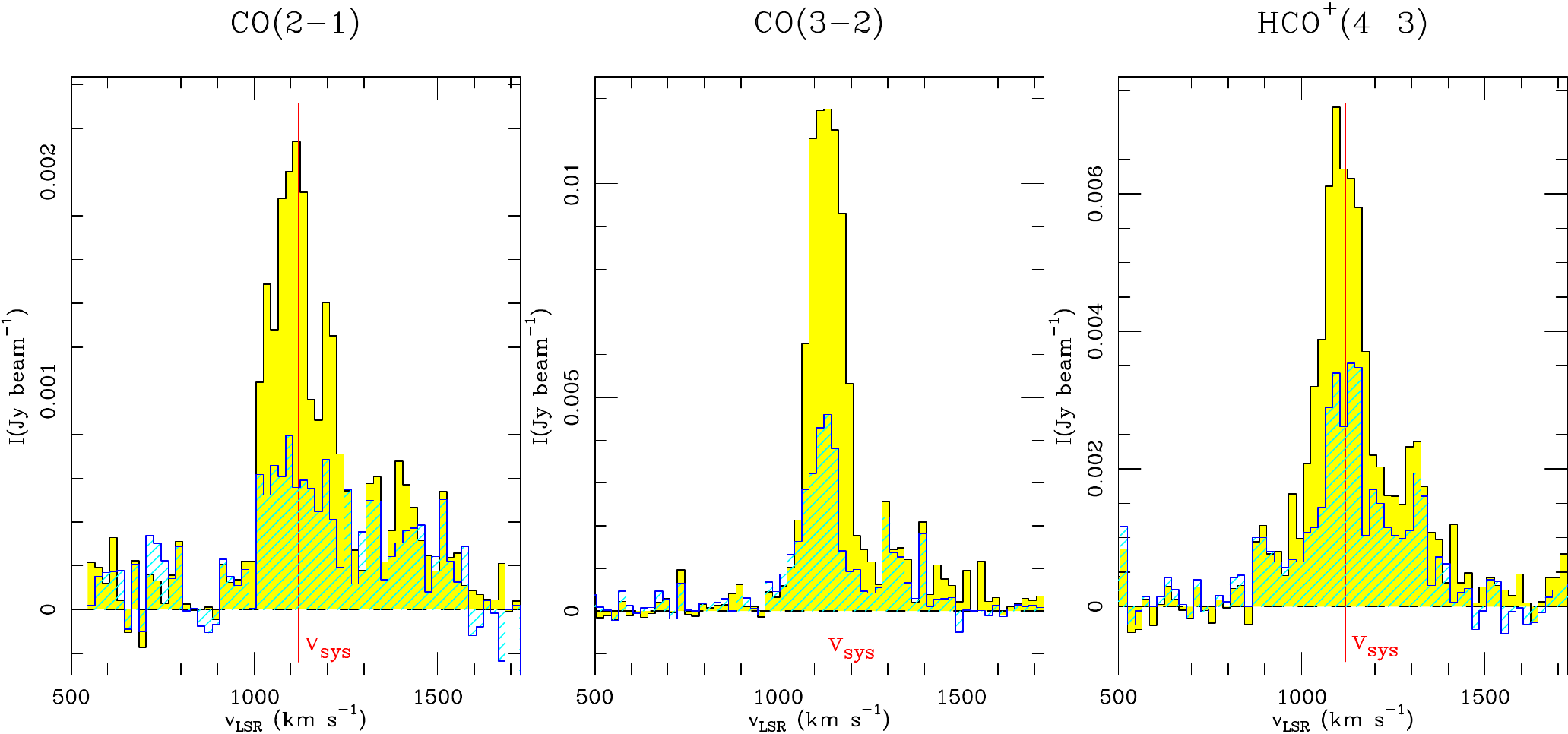} 
    \caption{We show the emission spectra extracted at the position of the AGN from the MSR data set for the $J=2-1$ line of CO  ({\it left panel}), the $J=3-2$ line of CO ({\it middle panel}), and  the $J=4-3$ line of HCO$^+$  ({\it right panel}) using a common aperture of $\simeq$6--7~pc-size (yellow-filled histograms). We also show in blue-hatched histograms the corresponding spectra obtained from  the HSR data set using a common aperture of $\simeq$2--3~pc-size. The red line highlights the value of $v_{\rm sys}^{\rm LSR}=1120$~km~s$^{-1}$ in each panel.}
   \label{agn}
    \end{figure*}

\subsection{Masses and column densities}\label{torus-mass}

We derived the mass of the molecular torus by integrating the CO(2--1) line emission inside the area defined by the full-size of the best-fit Gaussian disk derived in Sect.~\ref{torus-parameters} ($\simeq1.43$~Jy~km~s$^{-1}$).  Based on Equation~(3) of \citet{Bol13}, we estimated the molecular gas mass of the 2--1 torus to be $M^{\rm torus}_{\rm gas}\simeq2.9\times10^5~M_{\sun}$, including the mass of Helium. In our estimate we assumed a 2--1/1--0 brightness temperature ratio $\simeq2.5$, as measured by \citet{Vit14} at the AGN knot, and a galactic CO--to--H$_2$ conversion factor  ($X_{\rm CO}=2\times10^{20}$mol~cm$^{-2}$~(K~km~s$^{-1}$)$^{-1}$). Similarly, the torus mass derived using the  CO(3--2) flux integrated inside the region delimited by the red ellipse of the middle panel of Fig.~\ref{torus-maps-lr} ($\simeq3.54$~Jy~km~s$^{-1}$) is $M^{\rm torus}_{\rm gas}\simeq2.8\times10^5~M_{\sun}$, assuming a 3--2/1--0 brightness temperature ratio $\simeq2.9$ at the AGN knot \citep{Vit14}. Overall, the 3--2 and 2--1 line-based estimates for $M^{\rm torus}_{\rm gas}$ are both a factor of $\simeq3$ higher than the value obtained by \citet{GB16} from the dust continuum emission measured at 694~GHz. This is not unexpected in view of the 
comparatively smaller size measured by \citet{GB16} for the dusty torus/disk (diameter$\simeq7$~pc). 

The H$_2$ column densities measured towards the position of the AGN are $N$(H$_2$)~=~(1.1--3.2)~$\times$~10$^{23}$~mol~cm$^{-2}$ and 
$N$(H$_2$)~=~(1.5--4.4)~$\times$~10$^{23}$~mol~cm$^{-2}$ from the 2--1 and 3--2 intensities, respectively, measured at the corresponding apertures. In either case, the range of column densities encompasses the values derived from the moderate  ($\simeq$6--7~pc) and the high ($\simeq$2--3~pc)  spatial resolution data sets. As expected, the higher column densities are obtained at the highest spatial resolution due to the lower beam dilution of the obscuring material. These values are still a factor of $\simeq$2--3 below the Compton-thick limit ($N$(H$_2$)~$\geq$10$^{24}$~mol~cm$^{-2}$) required to explain the 
 nature of the Type~2 nucleus of NGC~1068.   This mismatch may be an indication that X-ray absorption by clouds which lie in the dust-free inner region of the  torus would be responsible for most of the obscuration \citep[see, e.g., discussion in][]{Eli08}. However, within the uncertainties associated to the values of $X_{\rm CO}$ in a harsh AGN environment \citep[of $\simeq$ up to an oder of magnitude; see, e.g.,][]{Wad18}, the molecular torus detected by ALMA could nevertheless contribute significantly to the properties of obscuration of the central engine of NGC~1068 on the spatial scales of $\simeq$2--3~pc.  In this context it is noteworthy that the recent NuSTAR observations of NGC~1068, published by \citet{Mar16}, found evidence of a $\sim40\%$ decrease of the absorbing column density from $N$(H$_2$)~$\sim$~5~$\times$~10$^{24}$~mol~cm$^{-2}$ to $\sim$~3.3~$\times$10$^{24}$~mol~cm$^{-2}$ on a timescale of 2 years (December 2012-February 2015). The change of column densities required to explain the X-ray variability of the AGN in NGC~1068 could be associated with the small-scale structure of the molecular torus imaged by ALMA.   

The range of $N$(H$_2$) reported above for the 2--1 line of CO at the AGN locus translates into a range of $A_{\rm v}\simeq$~120--350, derived using the ratio $N$(H$_{\rm 2})/ A_{\rm v} = 0.9\times10^{21}$~mol~cm$^{-2}$mag$^{-1}$ of \citet{Boh78}. This range of $A_{\rm v}$  is compatible within the errors with the value inferred by \citet{Gro18} from the 20 magnitude extinction in the Ks band that these authors used to explain their NIR polarization data, which would imply  $A_{\rm v}\simeq$~180. Furthermore, our estimate is only a factor of 2 higher than the value used by \citet{Rou18} to explain the measured radial profile in the Ks band.  

\subsection{Previous evidence of a molecular torus}\label{torus-previous}

 The use of different transitions, which purposely span a wide range of physical conditions of molecular gas  ($n$(H$_2$)~$\subset10^3-10^7$cm$^{-3}$), is instrumental to help reveal the density radial stratification of the gas in the torus. As expected for the contribution of a comparatively cooler gas component, the sizes reported for the molecular torus traced by the low-J CO lines are more than an order of magnitude larger compared with the sizes of the NIR and MIR sources of the host dust \citep{Jaf04, Wei04, Rab09, Bur13, Lop14}
  
 The parameters of the HCO$^+$ torus, listed in Table~\ref{Tab2}, are similar within the errors to those derived by \citet{GB16}, \citet{Gal16}, and \citet{Ima18} for the molecular torus imaged in the CO(6--5), HCO$^+$(3--2) and HCN(3--2) lines. 
Such an agreement is expected among these transitions as they probe a similar range of high densities for molecular gas ($n$(H$_2$)~$\subset10^5-10^7$cm$^{-3}$). In contrast, the lower densities to which the 2--1 and 3--2 transitions are sensitive ($n$(H$_2$)~$\subset10^3-10^4$cm$^{-3}$) seem to trace preferentially the outer region of the molecular torus of NGC~1068. 

Previous evidence supporting the existence of a  dusty disk that extends well beyond the canonical parsec-scale in NGC~1068 was discussed by \citet{Gra15}, based on NIR polarimetric imaging data. In particular,  \citet{Gra15} found that the inferred linear polarization vectors in the CND change their orientation from a centro-symmetric pattern, associated with the hourglass structure shown in Fig.~\ref{torus-polar}, to an orientation  perpendicular to the bicone axis, which is oriented along  $PA\simeq~30^{\circ}$.  These observations show the existence of an extended patch of linear polarization arising from a spatially-resolved  elongated nuclear disk of dust of $\simeq50-60$~pc-diameter and oriented along $PA\simeq~120^{\circ}$.  The change of orientation of the polarization vectors is interpreted by \citet{Gra15} as the result of a double scattering of the radiation from the central engine by a dusty disk:  photons in the ionization cone are scattered a first time, on electrons or on dust grains, towards the outskirts of the torus where they suffer a second scattering towards the observer \citep[see, e.g.,][for a general description of a similar phenomenology expected in the dusty disks around young stellar objects]{Mur10}. Photons scattered towards the inner regions of the torus would not be able to escape because of extinction. This would explain why polarization traces preferentially the outer and less dense regions of the torus. 

Figure~\ref{torus-angle} shows the overlay  of the CO maps on the image showing the difference between the polarization angle derived from the H-band map of \citet{Gra15} and a purely centro-symmetric pattern. The angle-difference map by itself cannot be used as a direct tracer of the dust column densities in the disk. However,  Fig.~\ref{torus-angle}  shows a tantalizing agreement between the maximum extent and overall orientation of the putative dusty disk, identified by  \citet{Gra15}, and those of the CO torus imaged by ALMA.

   \begin{figure*}
   \centering
    \includegraphics[width=0.85\textwidth]{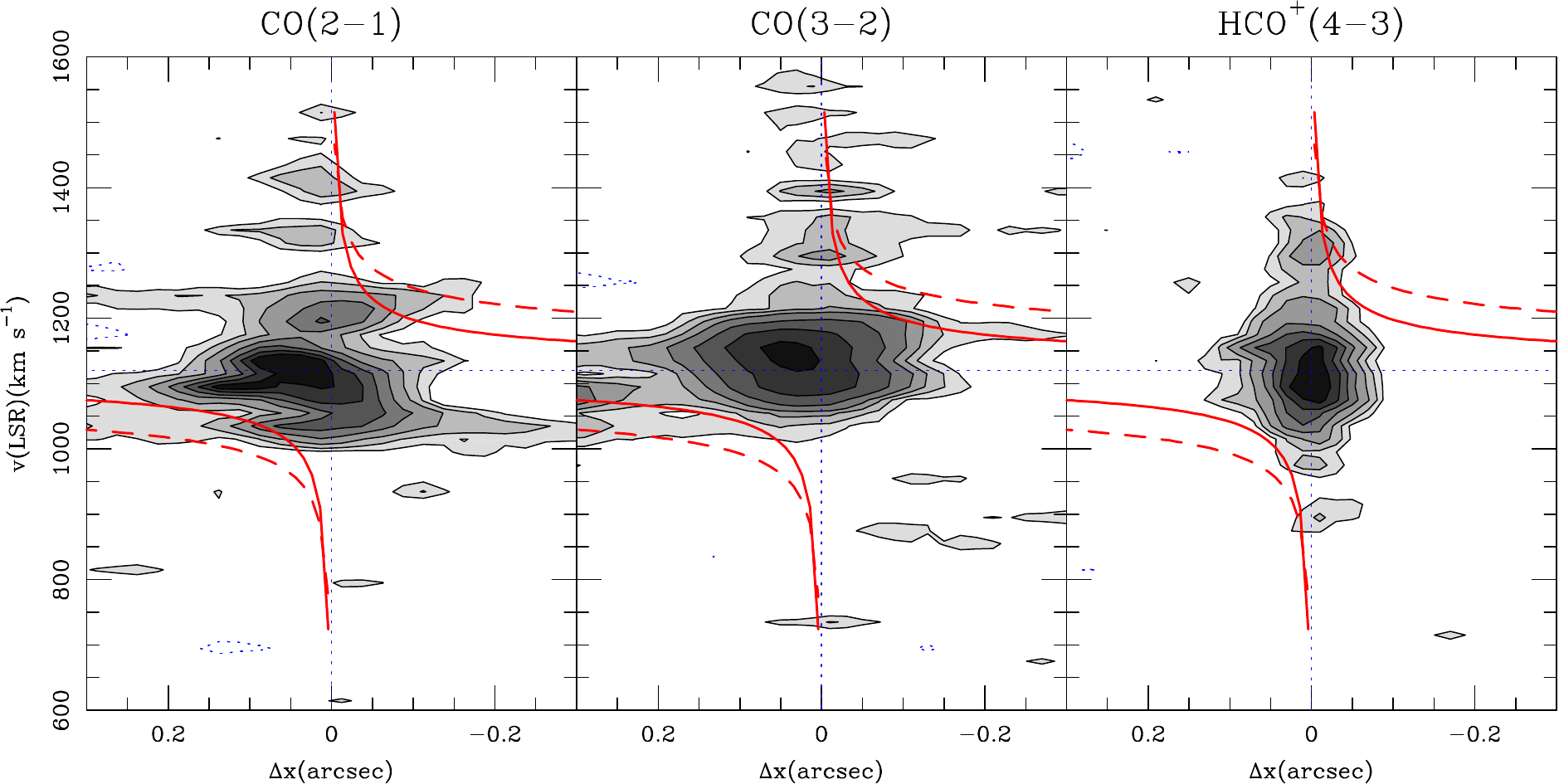}
    \caption {Position-velocity (p-v) diagrams obtained with the MSR data set along the major axis of the torus of NGC~1068 along $PA=113^{\circ}$ in the $J=2-1$ line of CO  ({\it left panel}), the $J=3-2$ of CO ({\it middle panel}), and  the $J=4-3$ line of HCO$^+$  ({\it right panel}). Grey-scale and contours are -2.5$\sigma$,  2.5$\sigma$,    
    4$\sigma$, 6$\sigma$ to 18$\sigma$ in steps of 3$\sigma$, with 1$\sigma$=0.11~mJy~beam$^{-1}$ ({\it left panel}), -2.5$\sigma$,  2.5$\sigma$,    
    5$\sigma$, 8$\sigma$,12$\sigma$, 20$\sigma$, 35$\sigma$ and 50$\sigma$, with 1$\sigma$=0.23~mJy~beam$^{-1}$ ({\it middle panel}), and    
    -2.5$\sigma$,  2.5$\sigma$, 4$\sigma$, 6$\sigma$, 9$\sigma$, 12$\sigma$, 15$\sigma$, and 19$\sigma$,  with 1$\sigma$=0.28~mJy~beam$^{-1}$ ({\it right panel}).   
Offsets along the x axis ($\Delta x$) are measured in arc seconds relative to the AGN locus (vertical dashed lines) with positive (negative) values corresponding to the SE (NW) side of the disk. Velocities are measured in LSR scale with $v_{\rm sys}^{\rm LSR}$=1120~km~s$^{-1}$ (horizontal dashed lines). The dashed (solid) red curves show the best-fit sub-Keplerian (Keplerian) rotation curve  $v_{\rm rot}\propto r^{-\alpha}$  of \cite{Gre96} with $\alpha=0.31$ (0.50); the curves were fitted to the H$_2$O megamaser spots detected along  $PA^{\rm maser}=140^{\circ}\pm5^{\circ}$ \citep{Gre96, Gal01} and subsequently projected along $PA=113^{\circ}$. The curves account for a black hole mass $M_{\rm BH}\simeq1\times10^7M_{\sun}$.}
   \label{torus-major}
    \end{figure*}
   

   \begin{figure*}
   \centering
    \includegraphics[width=0.85\textwidth]{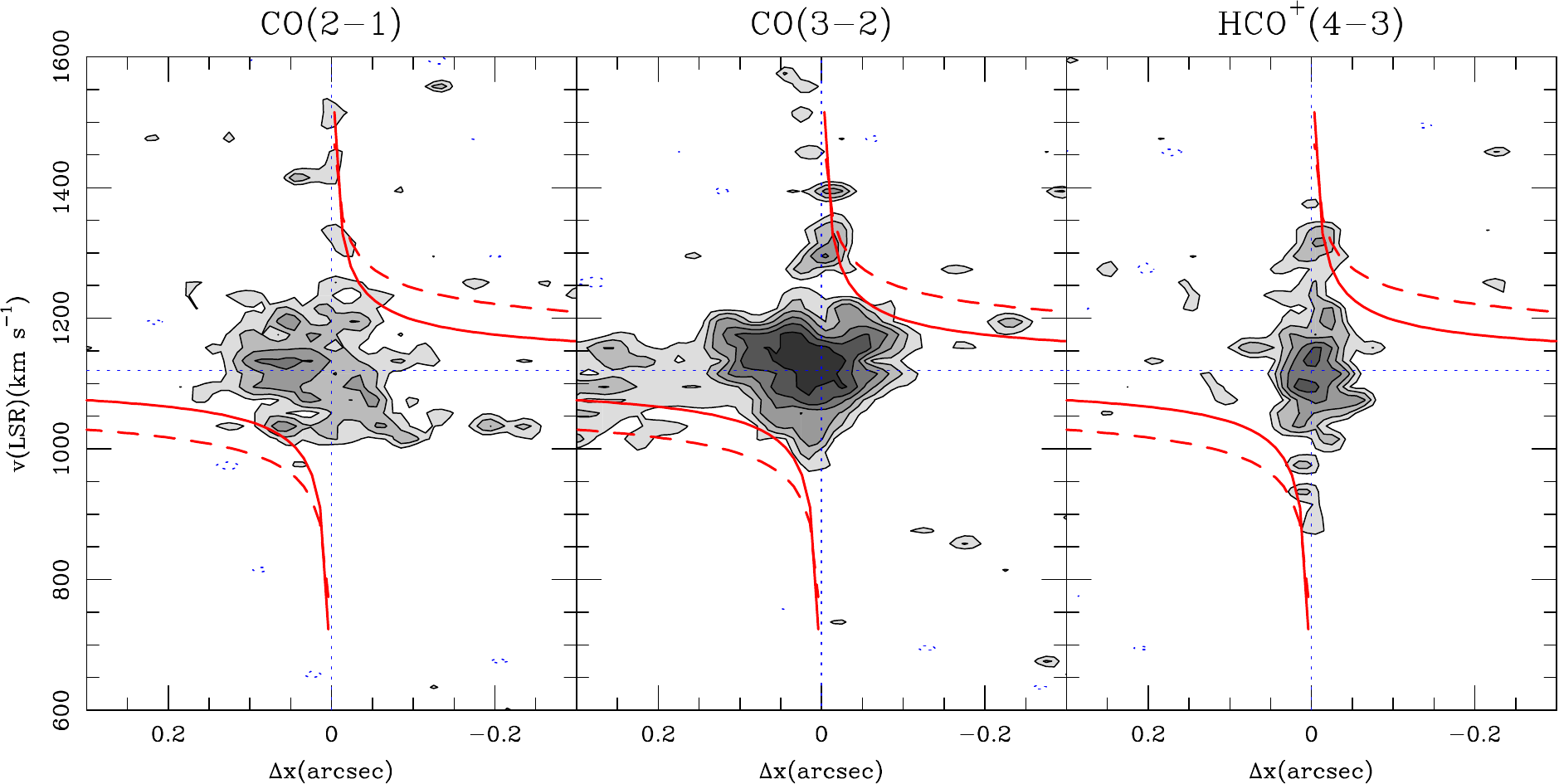}
    \caption {Same as Fig.~\ref{torus-major} but obtained with the HSR data set. Contours are -2.5$\sigma$,  2.5$\sigma$,    
    4$\sigma$, 6$\sigma$, and  to 9$\sigma$, with 1$\sigma$=0.14~mJy~beam$^{-1}$ ({\it left panel}), -2.5$\sigma$,  2.5$\sigma$,    
    4$\sigma$, 6$\sigma$, 9$\sigma$, 12$\sigma$, and 16$\sigma$, with 1$\sigma$=0.23~mJy~beam$^{-1}$ ({\it middle panel}), and    
    -2.5$\sigma$,  2.5$\sigma$, 4$\sigma$, 6$\sigma$, 9$\sigma$, and 12$\sigma$,  with 1$\sigma$=0.28~mJy~beam$^{-1}$ ({\it right panel}).}
   \label{torus-major-HR}
    \end{figure*}
   

 \begin{figure*}
   \centering
    \includegraphics[width=12cm]{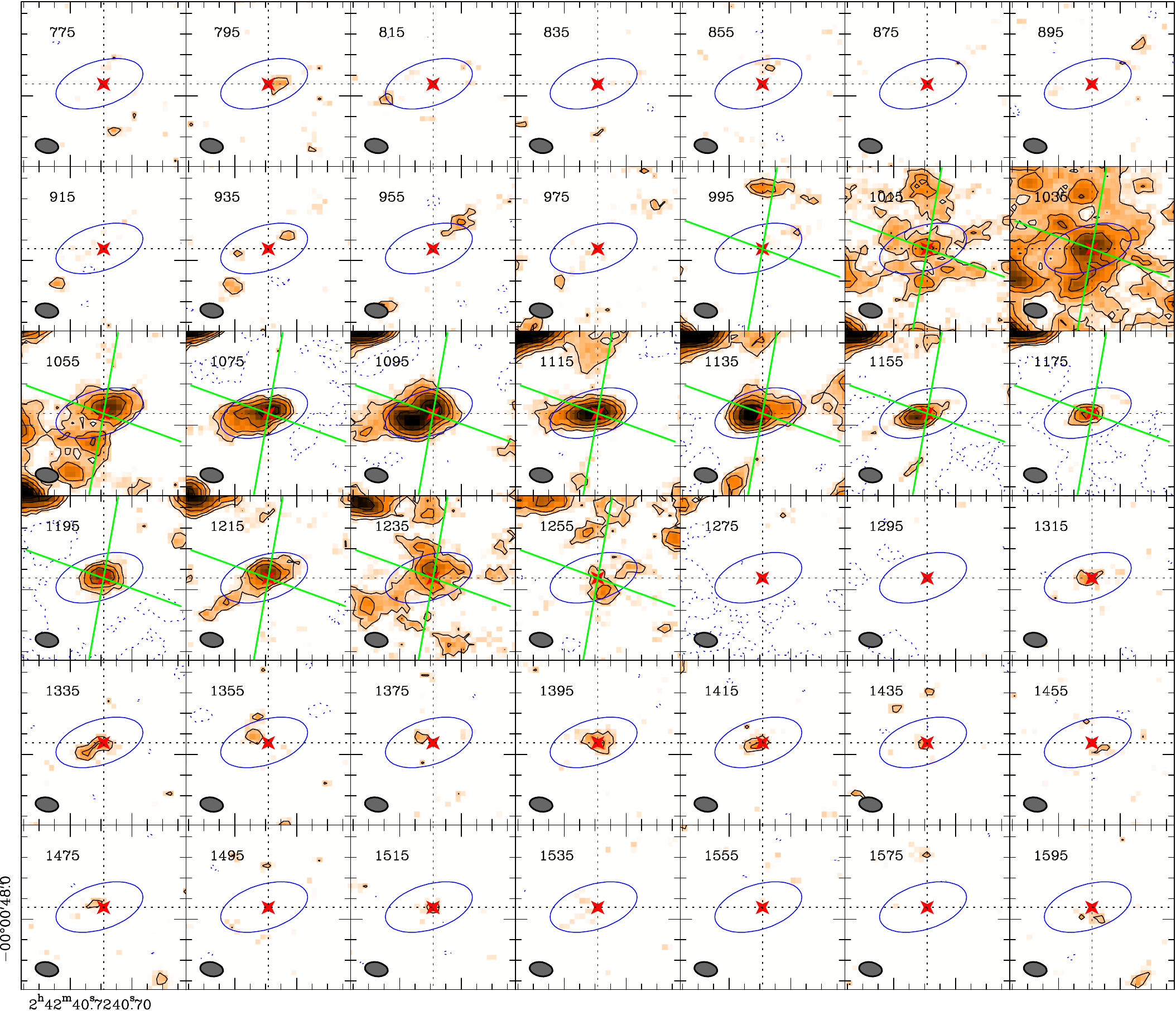}
    \caption{The CO(2--1) channel maps obtained in the central $0\farcs8\times0\farcs8\simeq60$~pc~$\times$~60~pc region of NGC~1068. Intensity contours 
    are  -3$\sigma$,  3$\sigma$,  5$\sigma$, 8$\sigma$ to  20$\sigma$ in steps of 6$\sigma$, with 1$\sigma=0.11$~mJy~beam$^{-1}$. 
    The central velocity in the LSR reference frame is displayed at the upper left corner of each panel. The position of the AGN is 
    identified by the (red) star markers. The (grey) filled ellipses in the lower left corners of each panel represent the beam size of ALMA. The blue empty ellipses highlight the
    position and full size of the CO(2--1) torus as determined  in Sect.~\ref{torus-parameters}. The green lines in the selected velocity channels, i.e., $\simeq v_{\rm sys}\pm140$~km~s$^{-1}$, identify the region occupied by the AGN ionized wind. The tick sizes on the x and y axes are $0\farcs075$ and $0\farcs1$, respectively.}       
   \label{channels}
\end{figure*}

   \begin{figure*}
   \centering
    \includegraphics[width=1\textwidth]{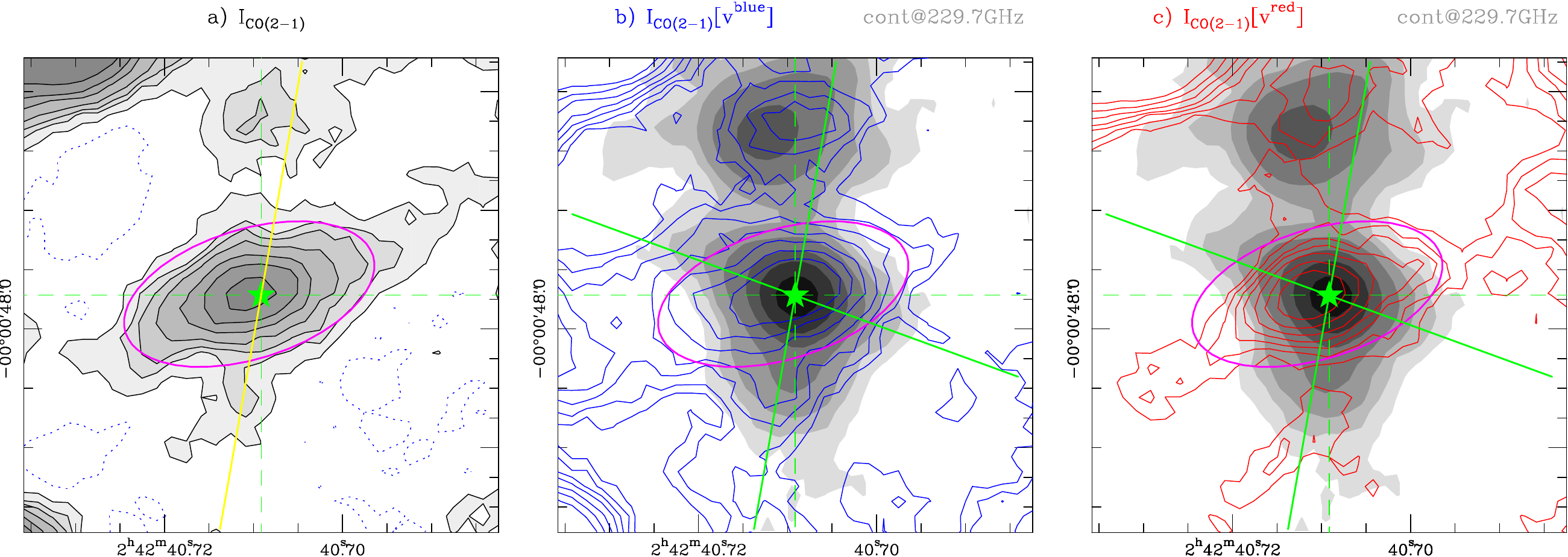}  
    \caption{{\it Left panel:}~The central $r\leq0\farcs4$ ($\simeq$30~pc) region of the CO(2--1) map of NGC~1068 showing the torus and its connections with the CND. The yellow line identifies the orientation of the axis chosen to derive the position-velocity diagram shown in Fig.~\ref{torus-connect-pv} ($PA$~=--10$^{\circ}$). {\it Middle} and {\it right panels:}  Overlay of the CO(2--1) intensity contours integrated inside $v^{\rm blue}$~=~[v$_{\rm sys}$-140, v$_{\rm sys}$]~km~s$^{-1}$ (blue contours in {\it middle panel}) and 
$v^{\rm red}$~=~[v$_{\rm sys}$, v$_{\rm sys}$+140]~km~s$^{-1}$ (red contours in {\it right panel}) on the continuum emission image of Fig.~\ref{cont-maps} obtained at 229.7~GHz (grey scale).    Blue and red contours are  2.5$\sigma$, 4$\sigma$, 6$\sigma$, 8$\sigma$, 11$\sigma$, 15$\sigma$ and 20$\sigma$  where 1$\sigma=5.8$~mJy~km~s$^{-1}$beam$^{-1}$. Grey contours as in Fig.~\ref{cont-maps}({\it left panel}). The (solid) green lines identify the region occupied by the AGN wind. In all panels, the AGN locus is identified by the green marker and the magenta ellipse highlights the position and full size of the CO(2--1) torus as determined  in Sect.~\ref{torus-parameters}.}       
   \label{torus-connect}
    \end{figure*}
    

\subsection{Kinematics of the molecular torus}\label{torus-kin}

Figure~\ref{torus-velo} shows the mean-velocity fields derived from the CO(2--1), CO(3--2)  and HCO$^+$(4--3)  lines in the central $r\leq20$~pc region around the central engine of NGC~1068. By contrast with the CND, the three lines show  a surprising lack of a regular pattern and/or a defined velocity gradient in the torus. As argued below, the existence of different velocity components inside the torus, but also in the surrounding region that connects the torus to the host (see Sect.~\ref{torus-conn}), adds to the extreme complexity of the velocity field, most particularly in the case of the CO(2--1) line.

Figure~\ref{agn} shows the CO and HCO$^+$ emission spectra extracted at the position of the AGN using the MSR and HSR data sets for a common aperture of $\simeq$6--7~pc-size and $\simeq$2--3~pc-size, respectively. Most of the emission for the three lines stems from a core component, which corresponds to gas at low velocities:  $| v-v_{\rm sys} | \leq100$~km~s$^{-1}$.  Furthermore we detect significant emission at highly redshifted velocities in the 2--1 and 3--2 lines of CO up to  $v-v_{\rm sys}  \simeq +450$~km~s$^{-1}$. There is a remarkable absence of any significant blue-shifted counterpart of the high-velocity gas in the CO lines.  A similar yet less pronounced asymmetry in the emission of high-velocity gas was present in the AGN profile of the CO(6--5) line of \citet{Gal16}.  By contrast, we detect both red-shifted and blue-shifted high-velocity gas emission in the HCO$^+$(4--3) spectrum of the AGN, although with a maximum velocity span that is smaller compared to that reported for the CO lines:  $| v-v_{\rm sys} | \leq 300-350$~km~s$^{-1}$ . As argued below, this high-velocity regime corresponds to the range of velocities displayed by the H$_2$O megamaser spots identified in the inner $r\leq1$~pc-region \citep{Gre96, Gal01}. Unlike the emission at low velocities, the high-velocity gas component is mostly spatially unresolved, as expected if  this emission comes from a region localized very close to the inner edge of the torus ($r\leq1$~pc) (see Fig.~\ref{agn}). 

Assuming a geometry similar to that of the H$_2$O megamaser spots, where redshifted velocities are located on the west side of the disk,  the reported asymmetry of high-velocity CO emission  would imply that the gas at the inner edge of the torus is lopsided to the west. 
However, the  global off-centering described in Sect.~\ref{torus-parameters} would imply that the overall gas lopsidedness of the torus changes side to the east.  As the bulk of the gas in the torus emits at low velocities and at correspondingly large radii, this behavior is suggestive of the presence of a $m=1$--spiral in the gas. This asymmetry is expected if the molecular torus reflects the propagation of the Papaloizou-Pringle instability (PPI) \citep{Pap84, Kiu11, Kor13, Don14, Don17}.

The kinematics of the molecular torus is further illustrated in Figs.~\ref{torus-major} and ~\ref{torus-major-HR}, which show the position-velocity (p-v) diagrams obtained along the average fitted major axis of the nuclear disk/torus ($PA=113^{\circ}$: taken from the average of the values listed in Table~\ref{Tab2}).  Figs.~\ref{torus-major} and ~\ref{torus-major-HR} show that a significant fraction of the emission spreads over the bottom left and top right quadrants of the diagrams. Furthermore, 
the observed terminal velocities are fairly compatible with the predicted extrapolation of the  sub-Keplerian (Keplerian) rotation curve  $v_{\rm rot}\propto r^{-\alpha}$  of \cite{Gre96} with $\alpha=0.31$ (0.50). 
\citet{Gal16} also found evidence for Keplerian rotation fairly consistent with the compact maser disk based on their CO(6--5) ALMA data. However, we also find significant emission arising  from the
upper left and lower right quadrants of the p-v diagrams, especially in the 2--1 line of CO. Taken at face value, this anomalous component would indicate the existence of gas in {\em apparent} counter-rotation. The emission at velocities {\em forbidden} in the frame of a circular velocity model is also identified in the CO(3--2) line although this component appears at comparatively lower velocities in the p-v diagram.

The configuration of two  {\em apparently} counter-rotating disks described above would be nevertheless dynamically unstable: the dissipative nature of the gas would make the two disks collapse on the time scale of the orbital time. The implied fast evolution of the system would make very unlikely to observe this configuration. As we shall argue in Sect.~\ref{BBarolo}, we favor an alternative scenario where the {\em apparent} counter-rotation signature in the major axis p-v diagram would rather reflect the presence of non-circular motions in the gas.

   \begin{figure*}
   \centering
    \includegraphics[width=1\textwidth]{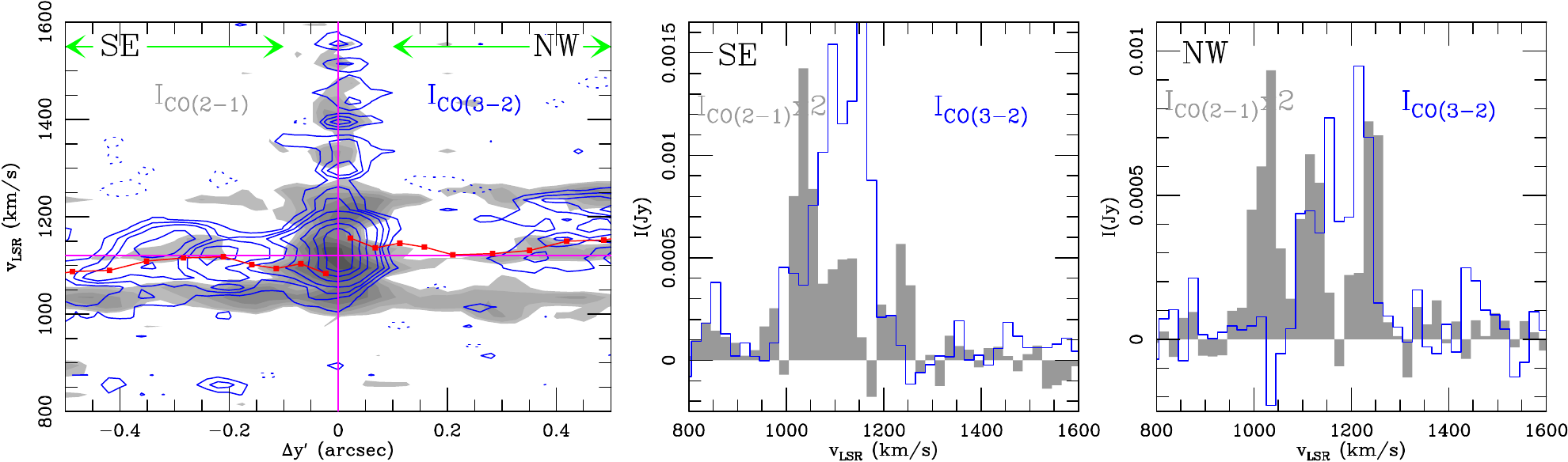}
    \caption{{\it Left panel:}~The CO position-velocity diagrams taken along the gas lanes that connect the torus with the CND  at $PA$~=~--10$^{\circ}$.  The CO(2--1) line emission is shown in grey linear scale  spanning the range [2$\sigma$, 18$\sigma$] with 
    1$\sigma=0.11$~mJy~beam$^{-1}$.  The CO(3--2) position-velocity diagram is shown in (blue) contours: -2$\sigma$ (dashed), 2$\sigma$, 4$\sigma$, 6$\sigma$, 10$\sigma$ to 40$\sigma$ in steps of 10$\sigma$, with 1$\sigma=0.23$~mJy~beam$^{-1}$. The (red) line and square markers show the mean velocities that can be attributed to circular rotation along $PA$~=~--10$^{\circ}$
    (according to the fit of Sect~\ref{BBarolo}).
      The spatially-integrated CO  2--1 and 3--2 line profiles averaged over the $SE$ and $NW$ segments of the connecting gas lanes (as defined in {\it left panel}) are shown, respectively,  in the  {\it middle panel}  and in the {\it right panel}.}       
   \label{torus-connect-pv}
    \end{figure*}


   \begin{figure}
   \centering
     \includegraphics[width=0.5\textwidth]{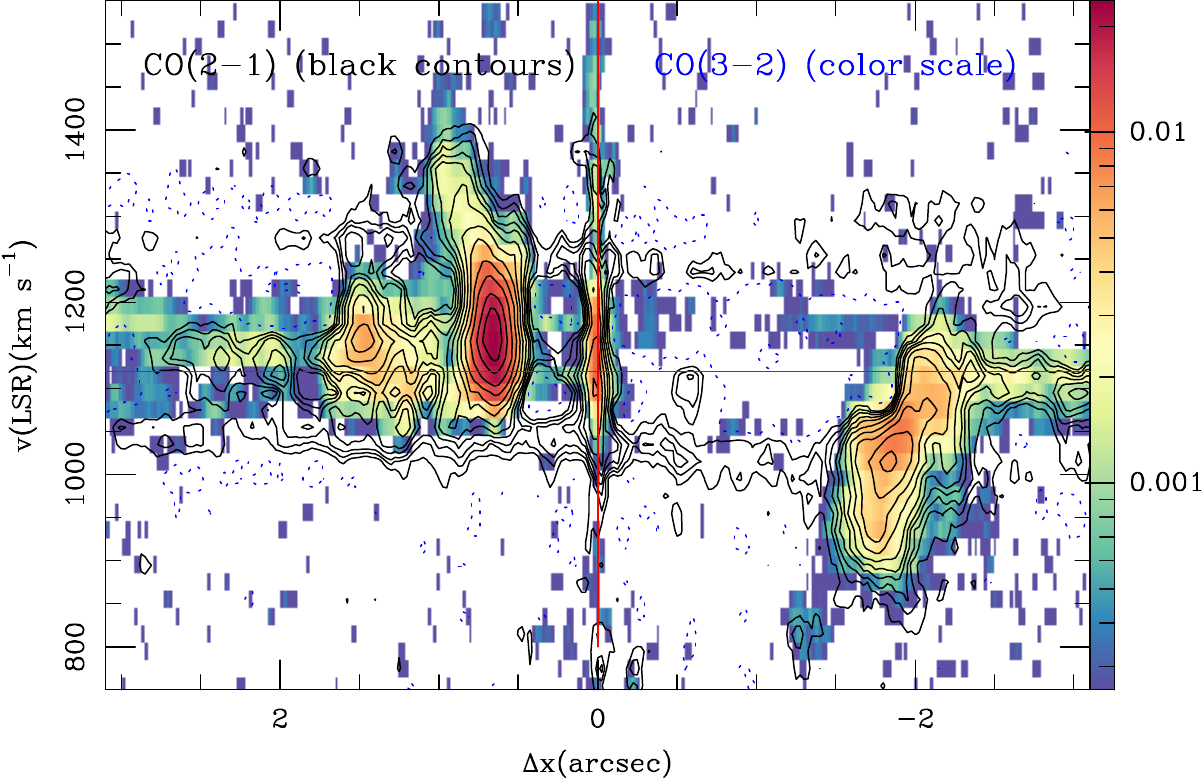}  
    \caption{Comparison of
    the average position-velocity (p-v) plot obtained for the outflow region out to $r=3\arcsec$~($\simeq210$~pc) in CO(2-1) (contours: -2.5$\sigma$ [dashed], 2.5$\sigma$ 
    4$\sigma$, 6$\sigma$, 9$\sigma$, 12$\sigma$, 17$\sigma$, 25$\sigma$, 35$\sigma$, 50$\sigma$, 70$\sigma$ to 160$\sigma$ in steps of  30$\sigma$, with 
    1$\sigma=0.04$~mJy~beam$^{ -1}$ and CO(3-2) (color logarithmic scale: [2.5$\sigma$, 230$\sigma$]  with 1$\sigma=0.08$~mJy~beam$^{-1}$). The emission has 
    been averaged using data cubes where the rotation curve model of Sect.~\ref{kinemetry} has been subtracted over a range of $PA$:  [$PA_{\rm outflow}-20^{\circ}$,  $PA_{\rm outflow}
    +20^{\circ}$] , where $PA_{\rm outflow}=30^{\circ}$. 
    Offsets along the x-axis are measured in arc seconds relative to the AGN locus.}       
   \label{outflow-pv-CO}
    \end{figure}

   \begin{SCfigure*}
   \centering
    \includegraphics[width=13cm]{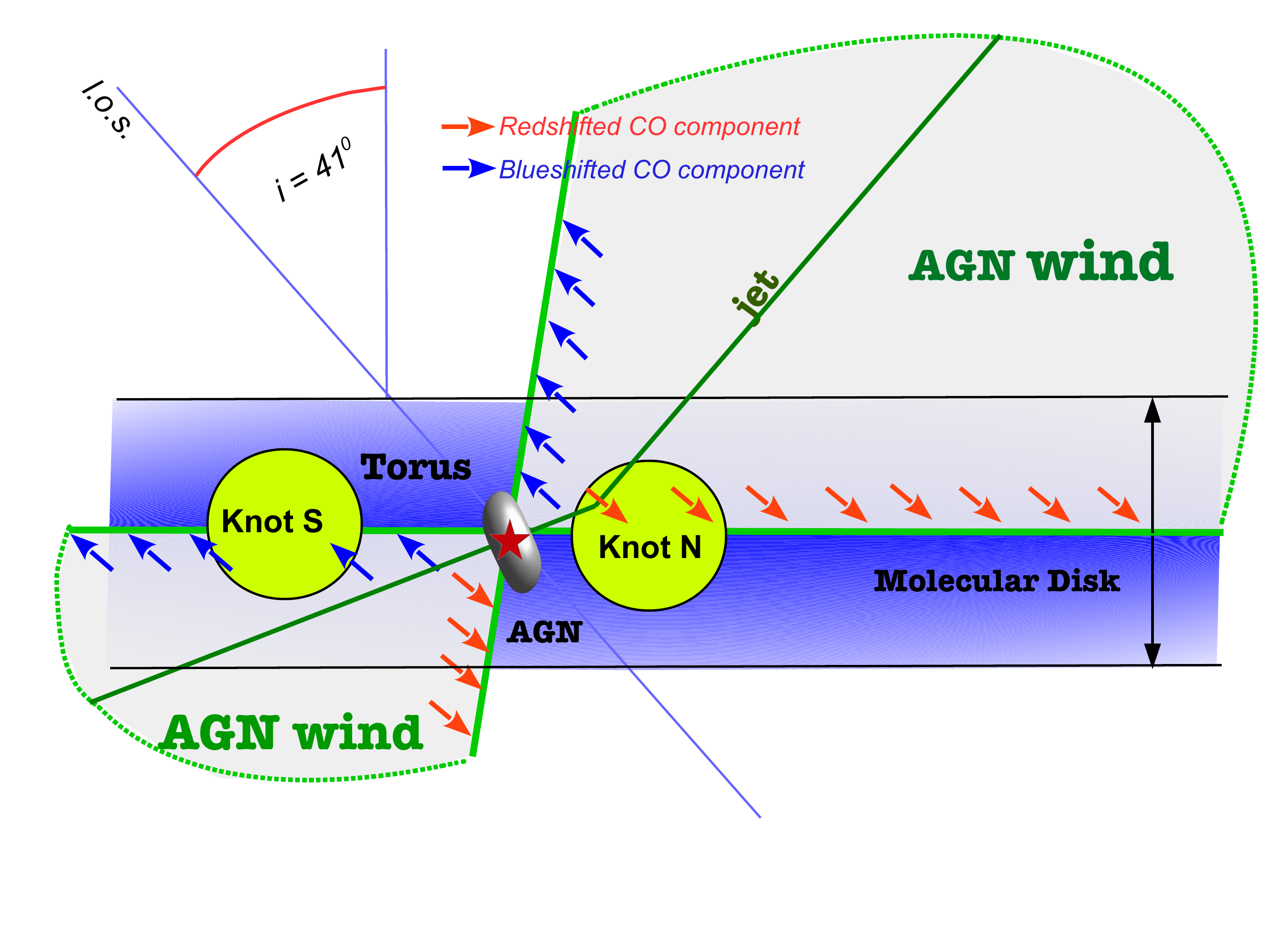}
    \caption{A revised scheme presenting a cross-cut view of the NLR of NGC~1068 along the projected direction of the ionized outflow axis  ($PA\sim30^{\circ}$; shown by the green line).  The new ALMA data shows that CO line emission is split into blueshifted and redshifted components (relative to $v_{\rm sys}$)  along the outflow axis, from the torus out to the edge of the CND, here represented by Knot N and Knot S. 
We highlight the extent of the AGN ionized wind (grey shade) and the assumed geometry defined by the inclination of the galaxy disk, $i=41^{\circ}$, the inclination of the outflow axis, $i_{\rm outflow}\geq80^{\circ}$, and the full opening angle of the outflow, FWHM$_{\rm outer}\simeq80^{\circ}$.}        
   \label{outflow-scheme}
    \end{SCfigure*}


\section{The torus-CND connections}\label{torus-conn}

\subsection{The central $r\leq30$~pc region: the gas lanes /streamers}\label{torus-streamers}

Figure~\ref{channels} shows a zoom view of the CO(2--1) channel map obtained in the central $0\farcs8\times0\farcs8\simeq60$~pc~$\times$~60~pc region located around the AGN torus. The maps show that beyond the 
region identified in Sect.~\ref{torus-parameters} as  the torus  (the blue empty ellipse in Fig.~\ref{channels}) there is  significant emission that bridges the torus with the inner edge of the CND ring out to $r=0.4\arcsec$ ($\simeq30$~pc).  A sizable fraction of this component is detected  inside or at the boundaries of the region in the disk that is suspected to be swept by the  AGN 
wind  \citep[$PA_{\rm outflow}\pm40^{\circ}$, with $PA_{\rm outflow}\simeq30^{\circ}$; e.g.,][]{Das06,Bar14} and the radio jet \citep[$PA_{\rm jet}\simeq10^{\circ}$; e.g.][]{Gal96,Gal04}. The strongest emission of these features  arises at velocities that are shifted by up to  $\simeq\pm140$~km~s$^{-1}$ relative to   $v_{\rm sys}$.

Figure~\ref{torus-connect} shows  the CO(2--1) intensities integrated inside the two velocity ranges that encompass the bulk of the emission in the CND-torus connections ($v^{\rm blue}$~=~[v$_{\rm sys}$-140, v$_{\rm sys}$]~km~s$^{-1}$  and $v^{\rm red}$~=~[v$_{\rm sys}$, v$_{\rm sys}$+140]~km~s$^{-1}$) overlaid on the millimeter continuum image of Fig.~\ref{cont-maps}.  An inspection of Fig.~
\ref{torus-connect} confirms that the brightest emission features come from two gas lanes/streamers that spring northward and southward from the torus. The emission in the blueshifted channels is stronger than in its redshifted counterpart in the southern streamer, while blueshifted and redshifted emissions are present with a comparable strength in the northern streamer.  The uneven balance between blueshifted and redshifted emission explains the appearance of the mean-velocity field of Fig.~\ref{torus-velo}, which shows blueshifted velocities in the southern streamer and a more chaotic pattern in the northern streamer.  Further emission with a similar split into redshifted and blueshifted channels is detected in a gas lane connecting the torus with the CND eastward, as well as at a fairly large number of positions located inside the region occupied by the AGN bicone.

 The northern and southern gas streamers are located close to the trajectory of the inner radio jet and the AGN wind. Overall, this association suggests that the interaction of the radio jet and the AGN wind with the molecular gas in the disk creates the observed line splitting reported above.  In this scenario, besides a radial expansion in the disk (the {\em 
coplanar} outflow discussed in Sect.~\ref{kinemetry}), 
 the gas may have also been forced to leave the plane of the galaxy and adopt a  3D shell geometry (the {\em vertical} outflow) in the region connecting the torus with the CND. 

Figure~\ref{torus-connect-pv} shows the CO(2--1) and CO(3--2) position-velocity diagrams taken along $PA$~=--10$^{\circ}$  (see the yellow line in the left panel of Fig.~\ref{torus-connect}) as well as the corresponding profiles representative of the NW and SE sections of the strip. The chosen strip goes across the NW and SE boundaries of the AGN bicone and, also, across the peaks of the connecting structures detected in the 2--1 line.  The CO(3--2) profiles along  
$PA$~=--10$^{\circ}$ indicate that the emission of this line is spread over a narrower range of velocities compared to the 2--1 line. The 2--1 line profile shows several components that encompass those of the  3--2 line.
By contrast with the CO(3--2) line, the CO(2--1) emission peaks at the most extreme velocities relative to the centroid dictated by circular rotation. Although the gas streamers are detected in both molecular lines,  the described nested structure suggests that the lower density gas traced by the CO(2--1) line in the {\em vertical} outflow has been pushed to higher velocities compared to the denser gas traced by the CO(3--2) line.
This may be the scenario resulting from the passage of C- and J-type shocks impacting on a stratified medium, whereby the CO (2--1) transition traces a lower density gas compared to the CO (3--2) transition: the piston-effect from the shock would then lead to the lower density gas moving at higher velocities than the denser gas. We can not exclude, however, also an evolutionary effect whereby depending on the flow time the transitions peak at different velocities with higher flow times corresponding to higher velocities, which means a more evolved shock. Time-dependent simulations of oblique C-type shocks in inhomogeneous media by \citet{Ash10} show very complex density structures that vary with time. A similar nested structure is found on the larger scales of the outflow when we compare the CO lines with the ionized gas lines of the AGN wind, as shall be discussed in Sect.~\ref{torus-outflow}.

\subsubsection{The gas streamers seen at other wavelengths: evidence of outflow}

As mentioned in Sect.~\ref{CND-ring}, there is independent evidence of a significant pile-up of dusty material along the location of the northern and southern molecular streamers detected by ALMA. As illustrated in Fig.~\ref{torus-polar}, both gas streamers are characterized by an increased degree of polarization of their dust-scattered NIR emission. There is observational evidence that the C knot of the northern streamer, located at $0\farcs3$ ($\simeq$20~pc) from the AGN core, is spatially co-incident with the interaction of the jet and the molecular ISM. \citet{LopR16} argue that this interaction can explain the increase of the polarization degree of the MIR emission in this region. Furthermore, gas streamers connecting the CND and the torus with an extent and morphology similar to the ones imaged by ALMA were previously detected in the 2.12~$\mu m$ H$_2$ line, a tracer of hot molecular gas ($T_{\rm k}\geq10^3$~K) \citep{Mue09,Bar14, May17}.

Our interpretation of the kinematics of  the gas detected by ALMA in these structures favors the outflow scenario, also invoked by  \citet{Bar14} and \citet{May17}, rather than the inflow scenario, first proposed by \citet{Mue09}. We note that the inflow interpretation of \citet{Mue09} was based on an analysis of the mean-velocity field of the gas streamers. In our case, and due to the extreme complexity of the CO(2--1) line profiles in the streamers, characterized by the existence of different velocity components of comparable strengths, the use of mean-velocity fields to diagnose the prevalence of inflow or outflow motions in these structures would be
misleading. Similarly, the detection of multiple velocity components in the gas streamers traced by the  2.12~$\mu m$ H$_2$ line, put forward  after a reanalysis of  \citet{Mue09} data  (Vives-Arias+et al in prep.), has confirmed the 3D geometry of the molecular outflow. 

\subsubsection{Molecular gas mass of the streamers}

The spatially integrated flux of the torus-CND connections out to $r\simeq30$~pc amounts to 820~mJy~km~s$^{-1}$. This is equivalent to a molecular gas mass $\simeq1.7\times10^{5}~M_{\sun}$\footnote{We 
base our estimates on the same hypotheses used in Sect~\ref{torus-mass}.}.  Overall, the mass of the bridging gas lanes ($M_{\rm lanes}$) represents 
a sizable fraction of the mass of the molecular torus ($M_{\rm lanes} \simeq 0.6 \times M^{\rm torus}_{\rm gas}$), while it is only a tiny fraction of  the mass of the outflowing CND \citep[$M^{\rm CND}_{\rm out}\simeq7\times10^7M_{\sun}\simeq 400 \times M_{\rm lanes}$; ][and this work]{GB14}. The remarkably different masses but similar  outflow velocities estimated through modeling in Sect.~\ref{BBarolo} for the torus-CND connections and the CND  imply that the radial profile of the molecular outflow rate  shows a sharp discontinuity beyond $r\simeq50$~pc.

   \begin{figure}
   \centering
    \includegraphics[width=0.495\textwidth]{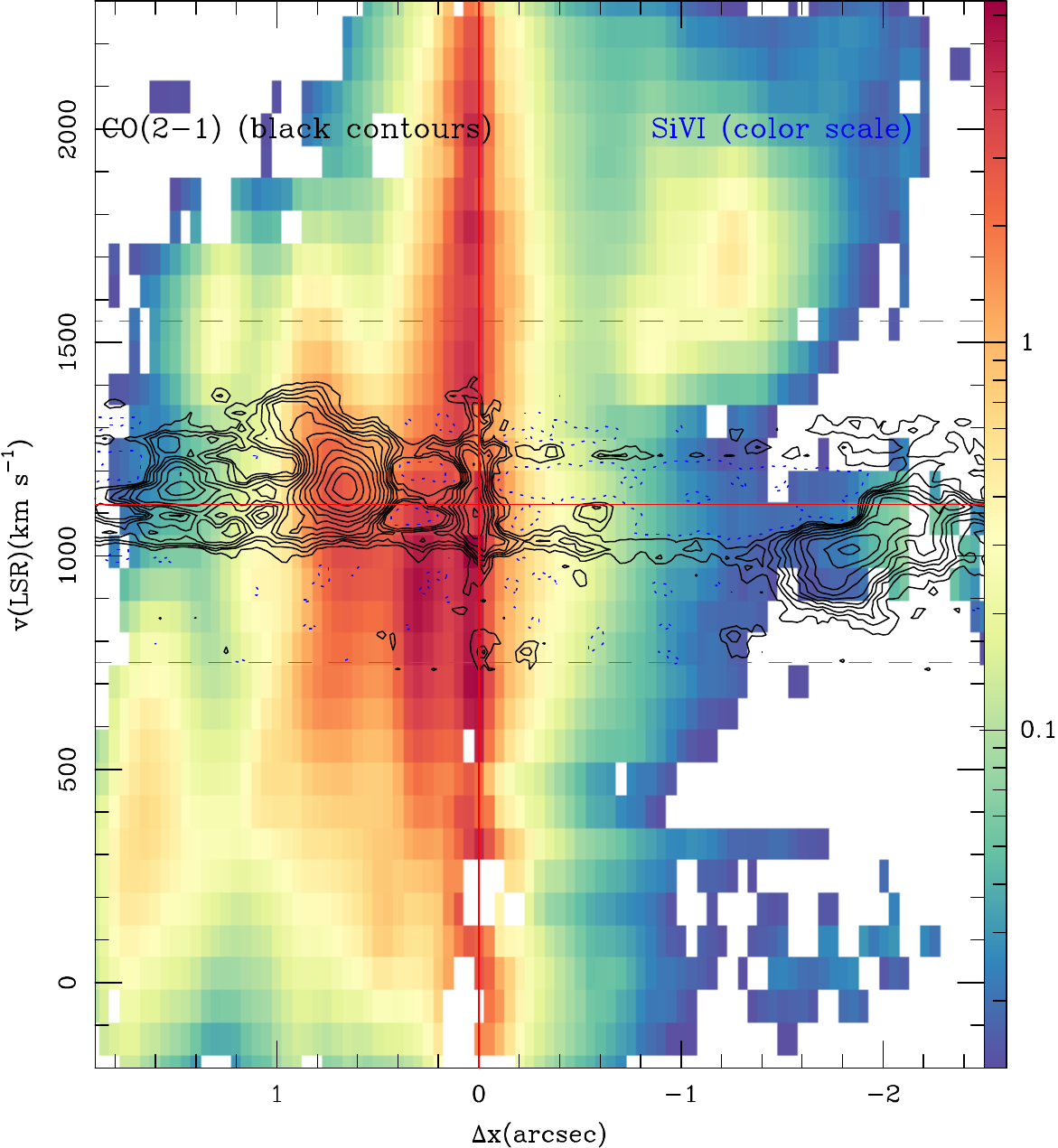} 
    \caption{Same as Fig.~\ref{outflow-pv-CO} but showing the comparison between CO(2--1) (contours) and [SiVI] 
    (color scale spanning the range [3$\sigma$,~1700$\sigma$]). For visualization purposes the v-scale has been highly compressed relative to that of Fig.~\ref{outflow-pv-CO} (identified by the magenta dashed lines). [SiVI]  is a high-ionization line ($\simeq205$~eV) that traces the emission of high-velocity ionized gas in the outflow  up to $v-v_{\rm sys} \simeq \pm 800$~km~s$^{-1}$ inside the CND region.}       
   \label{outflow-pv-SiVI}
    \end{figure}


\subsection{The connection with the large-scale molecular outflow: the 30~pc~$\leq r \leq$200~pc region}\label{torus-outflow}

The splitting of the line emission into two velocity ranges identified in the gas lanes /streamers (Sect.~\ref{torus-streamers}) is similar to the one  observed at larger radial distances over the  northern and southern extremes  of the CND ring. 
As discussed in Sect.~\ref{CND-width}, these regions of the CND lie along the path of the jet and AGN wind trajectories. 
Zooming out on the torus and its surroundings allows us to obtain a wider view of the entire outflow region out to  $r \leq$200~pc. 

In an attempt to characterize the global kinematics of the outflow we have constructed a representative p-v plot of the region as follows. Firstly, we have  subtracted the contribution from circular motions, derived from the rotation curve model of Sect.~\ref{kinemetry}, from the CO data cubes on a pixel-by-pixel basis. Secondly, the re-shuffled version of the data cubes have been used to average  the emission along a range of strips derived for a set of $PA$ and radial distances, which are designed to cover a large fraction of the outflow region under study, here defined by: $r\subset[0,200]$~pc and $PA\subset[PA_{\rm outflow}-20^{\circ},  PA_{\rm outflow}+20^{\circ}]$, i.e., covering the outflow section over its FWHM$_{\rm inner} \simeq$40$^{\circ}$. In this diagram obtained from the rotation-subtracted cube,  the line emission outside a range generously defined as $\simeq~v_{\rm sys}~\pm~(25-50)$km~s$^{-1}$, which can be attributed to turbulence, would be produced by the contribution of (unsubtracted) non-circular motions. The latter could result from streaming motions in the plane of the galaxy ({\em coplanar} radial inflow/outflow or tangential motions) or outside the plane of the galaxy ({\em vertical} inflow or outflow). As the range of $PA$ values used in the averages are close to $PA_{\rm minor}=19^{\circ}$, we can foresee that the contribution from the tangential component of streaming motions will be negligible\footnote{The pattern of non-circular motions displayed in Fig.~\ref{outflow-pv-CO} is virtually the same as the one displayed in the minor-axis p-v plot where the projection of the tangential component of streaming motions is zero by construction; this confirms our working hypothesis.}.  The result is shown in Fig.~\ref{outflow-pv-CO} for the CO(2--1) and CO(3--2) lines.  

An inspection of Fig.~\ref{outflow-pv-CO} shows that for a large range of radii a significant fraction of the CO emission stems from gas in non-circular motions.  For the brightest features, which correspond to the crossings of the CND ring, the emission is on average $\geq50$~km/s-redshifted on the northern side of the CND ($0\farcs5$ < $\Delta x$ < $1\farcs5$), while it is $\geq100$~km/s-blueshifted on the southern side ($-0\farcs5$ < $\Delta x$ < $-2\farcs0$). This reflects the sign and the right order of magnitude of the mean-velocity field deviations seen in the CO map of the CND (Fig.~\ref{residuals}), which have been interpreted in terms of a coplanar outflow. However, the emission at the crossings of the CND ring spreads over a surprisingly large velocity range (up to $\simeq 400-500$ km~s$^{-1}$) around the reported red(blue)shifted values and it shows signs of multiple velocity components. In particular, a smaller yet significant fraction of the emission is blueshifted (redshifted)  on the northern (southern) side of the CND.  In conclusion, although the coplanar outflow scenario can explain the bulk of the velocity distortions identified in the outflow region, the large linewidths and the reported line-splitting of the CO lines along the outflow axis require the inclusion of a vertical component.   

The proposed geometry is illustrated in Fig.~\ref{outflow-scheme}, which shows a scaled cross-cut view of the NLR of NGC~1068 along the projected direction of the outflow axis ($PA_{\rm outflow}\simeq30^{\circ}\pm5^{\circ}$). This is a revised version of the scheme proposed by  \citet{GB14} (see Fig.~17 of their paper), who first 
accounted for the molecular outflow  detected by ALMA in the CND by including only a coplanar outflowing 
component. The existence of a {\em vertical} molecular outflow would explain the simultaneous detection of blueshifts and redshifts on  either  side  of  the  AGN locus  in the torus-CND connections and throughout the entire outflow region, as evidenced by the new ALMA data.

Figure~\ref{outflow-pv-SiVI} compares the  p-v diagrams of the outflow obtained in the CO(2--1) line and in the NIR line of [SiVI], a highly ionized species imaged with the SINFONI instrument of the VLT (Ric Davies, private communication).  Compared to the cold molecular gas traced by the CO lines,  the emission of this tracer of highly-ionized gas in the AGN wind is spread throughout the outflow region over a much wider velocity range: up to $v-v_{\rm sys} \simeq \pm 900-1000$~km~s$^{-1}$ (see Fig.~\ref{outflow-pv-SiVI}). In the scheme depicted by Fig.~\ref{outflow-scheme}, the ionized wind would encounter less molecular gas, and consequently less resistance against expansion, above (below) the midplane of the galaxy on the northern (southern) side of the disk. This would explain why the sign of the velocity shifts for the strongest emission components of the [SiVI] line is noticeably reversed with respect to CO along the outflow axis.

\section{Towards a global model for the outflow}\label{BBarolo}

\subsection{{\tt $^{3D}$Barolo} fit}\label{Barolo}

As argued in Sects.~\ref{CND}, \ref{torus} and \ref{torus-conn}, the kinematics of molecular gas show strong departures from circular motions in  the torus,  the  torus-CND  connections, and the CND ring.  
These velocity field distortions are interconnected and reflect the effects of feedback of the AGN on the
kinematics of molecular gas on a wide range of  spatial scales around the central engine. In the following we attempt to define the  main features of a global model that best accounts for the complex kinematics of the molecular gas revealed by ALMA in \object{NGC~1068}.

To this aim we have used the software {\tt $^{3D}$Barolo} (3D-Based Analysis of Rotating Objects from Line Observations), developed by \citet{Dit15}.  {\tt $^{3D}$Barolo} uses a set of tilted-ring models to fit the emission-line 3D data cubes, which have two spatial dimensions and one spectral dimension. Compared to the {\tt kinemetry} software used in Sect.~\ref{kinemetry}, which fits with tilted-ring models the 2D mean-velocity images, {\tt $^{3D}$Barolo} provides a more complete exploration of the 3D geometrical and kinematic parameter space and also takes into account the effects of beam smearing in finding the best-fit model. Furthermore, when we 
ran the fit routine we adopted the pixel-based normalization option of the program to generate the modeled maps. This  allows us to have a non-axisymmetric model in density, which is excluded from the fit, while the dependance of all the physical parameters found by {\tt $^{3D}$Barolo} is by construction axisymmetric.

Among the set of output parameters that can be obtained for the optimum solution, {\tt $^{3D}$Barolo} provides the profiles of the circular rotation as a function of radius ($v_{\rm rot}$, equivalent to the $c_{1}$/sin$(i)$ term of  {\tt kinemetry}), and the corresponding profiles of the radial motions ($v_{\rm rad}$, equivalent to the $s_{1}$/sin$(i)$ term of  {\tt kinemetry}).  The list of parameters that can be fitted also includes the velocity dispersion
 ($\sigma_{\rm gas}$), the position angle ($PA$), the inclination ($i$), and the Gaussian scale-height for the disk ($H$).
In all the fit runs we used a set of 103 radii sampling the CND from $r=0\farcs03~(\simeq2$~pc) to   $r=3\arcsec~(\simeq210$~pc), i.e., the same region used to fit the velocity field with  {\tt kinemetry} in Sect.~\ref{kinemetry}.

 As a sanity check we first benchmarked the solutions found by {\tt $^{3D}$Barolo} against the output of {\tt kinemetry}, using for this purpose a common set of fixed parameters, as detailed in Appendix~\ref{App1}.
 Notwithstanding the different approach used by {\tt $^{3D}$Barolo} compared to  {\tt kinemetry}, the $v_{\rm rad}$ and $v_{\rm rot}$ profiles found are in fair agreement. In particular, similarly to {\tt kinemetry}, {\tt $^{3D}$Barolo} finds a $v_{\rm rad}$--profile indicative of radial outflow motions of $\simeq50$-100~km~s$^{-1}$ in the plane of the galaxy in a range of radii  $r\simeq50$-210~pc (the {\em coplanar} outflow).  
 
In a second {\tt $^{3D}$Barolo} run, which is described in detail in Appendix~\ref{App1}, we allowed for a more generous exploration of the parameter space by only fixing the AGN position and $v_{\rm sys}$ and letting free $v_{\rm rot}$, $v_{\rm rad}$, $v_{\rm disp}$,  $PA$,  $i$, and $H$. As an additional degree of freedom we also introduced a velocity field perpendicular to the plane of the galaxy at each radius, $v_{\rm vert}$, in an attempt to account for the 3D nature of the outflow suggested by observations, which call for the inclusion of a {\em vertical} outflow component. 
This second run was executed in two steps. We first derived a model by fitting all the parameters simultaneously, while in a second iteration we used values interpolated from the previous iteration for $PA$ and  $i$ while the rest of parameters are let free. The radial profiles of the best-fit parameters in this two-step process are shown in Fig.~\ref{Barolo-parameters}  of Appendix~\ref{App1}.  We summarize below the main features describing the best-fit solution found as follows:

\begin{itemize}

\item

The overall fit reflects that the angular momentum axis of the gas in the molecular torus is tilted by $\alpha \simeq60^{\circ}$ relative to the angular momentum of the gas in the large-scale disk: $\alpha=180^{\circ}-(<i_{\rm disk}> -<i_{\rm torus}>) \simeq 60^{\circ}$, where $<i_{\rm disk}>\simeq41^{\circ}$ and $<i_{\rm torus}>\simeq-80^{\circ}$. 

\item

The $v_{\rm rad}$  profile, which accounts for  the {\em coplanar} outflow component, reflects a change of sign, similar to the one described in Appendix~\ref{App1} for $v_{\rm rot}$, from the torus region, where $v_{\rm rad}<0$, to the outer disk, where $v_{\rm rad}>0$. In either case the sign and order of magnitude of the $v_{\rm rad}$ solution implies the existence a significant outward radial component of $\simeq50-100$~km~s$^{-1}$ throughout the fitted region $r\leq210$~pc. 

\item

The best-fit favors a value of  $v_{\rm vert}\simeq100$~km~s$^{-1}$ for the {\em vertical} outflow component up to at least $r\simeq150$~pc. Combined with the $v_{\rm rad}$ profile, the $v_{\rm vert}$ value implies the existence of 3D outflow velocities  $\simeq$100-140~km~s$^{-1}$ up to $r\simeq150$~pc.
  
\end{itemize}
 
   \begin{figure}
   \centering
    \includegraphics[width=9cm]{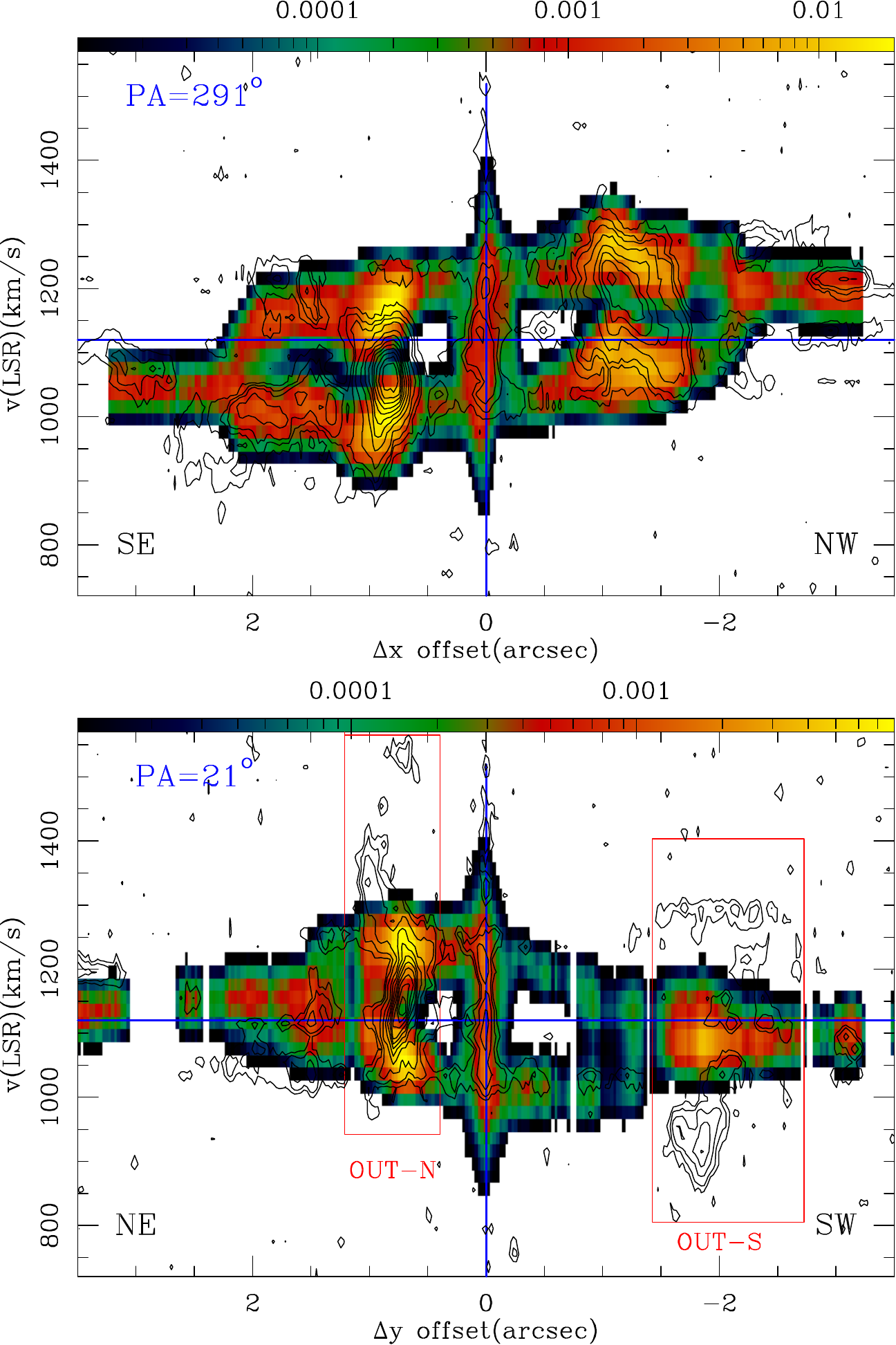}
    \caption{Overlay of CO(2--1) p-v plots (contours in Jy~beam$^{-1}$-units) along the average major axis ({\it upper panel}: $PA=291^{\circ}=180^{\circ}+111^{\circ}$; where we adopt the orientation of the receding side of the major axis) and average minor axis ({\it lower panel}: $PA=21^{\circ}$) on the corresponding synthetic p-v plots (in logarithmic color scale) generated by  {\tt $^{3D}$Barolo} for the best-fit model described in Sect.~\ref{BBarolo}.  In our convention $\Delta x$-offsets are $>0$ to the SE and $\Delta y$-offsets  are $>0$ to the NE. Contour levels in the {\it upper panel} are 1.3$\%$, 3$\%$, 5$\%$, 10$\%$to 90$\%$ in steps of 10$\%$ of the peak value = 23.1~mJy~beam$^{-1}$ . Contour levels in the {\it lower panel} are 2.7$\%$, 5$\%$, 10$\%$to 90$\%$ in steps of 10$\%$ of the peak value = 9.1~mJy~beam$^{-1}$. The regions identified as {\em OUT-N} and  {\em OUT-S} in the {\it lower panel} correspond to the crossings  of the AGN-wind working surface with  the molecular gas at the N and S extremes of the CND.}      
   \label{pv-models}
    \end{figure}

   \begin{figure*}
   \centering
    \includegraphics[width=14cm]{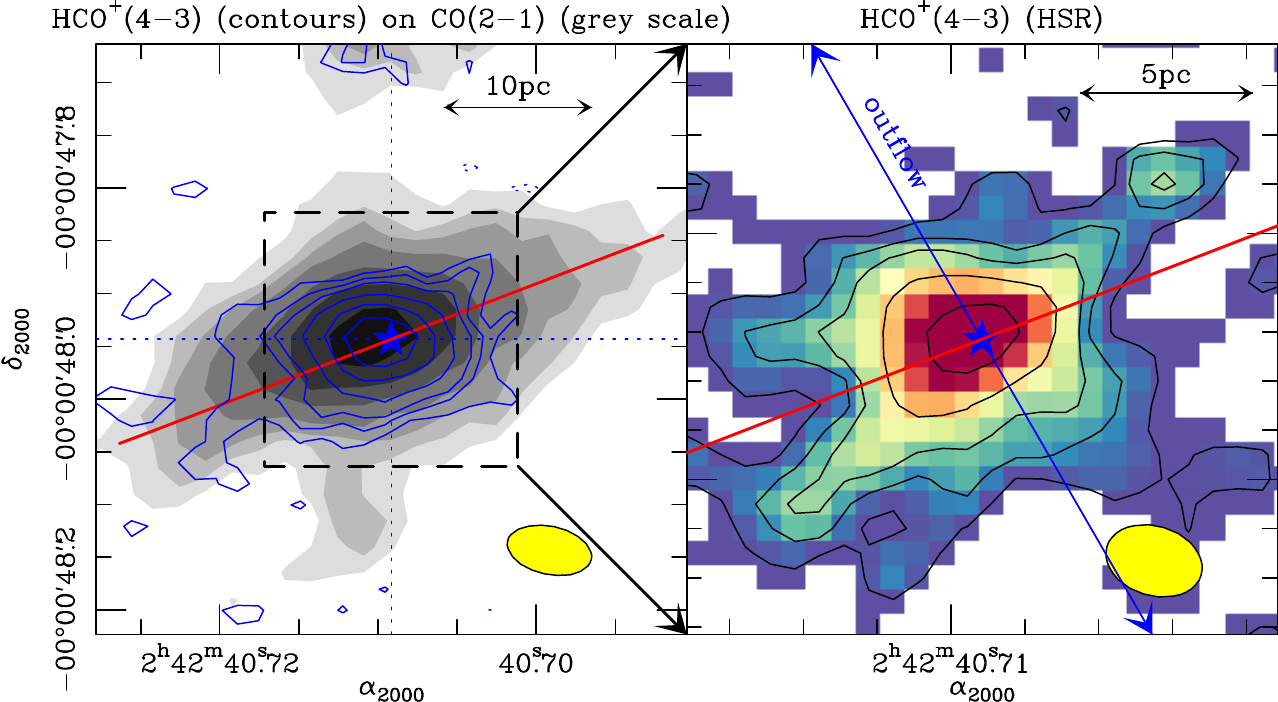}
    \caption{Overlay of the HCO$^+$(4--3) (blue) contours on the CO(2--1) grey-scale image  of the molecular torus of NGC~1068  derived from the MSR  data set ({\it left panel}). We show a zoomed view of the inner $r=0\farcs12\simeq8$~pc of the molecular torus obtained from the HSR data set in the {\it right panel}. Contours and intensity scales are the same as in Fig.~\ref{torus-maps-lr} and Fig.~\ref{torus-maps-hr}.  The continuous (blue) line shows the axis of the large-scale ionized outflow. The (red) line shows the orientation of the kinematic major axis of the CO(2--1) torus derived in Sect.~\ref{Barolo} by {\tt $^{3D}$Barolo} ($PA=291^{\circ}=180^{\circ}+111^{\circ}$). The (yellow) filled  ellipses show the ALMA beam sizes.}      
   \label{torus-overlays}
    \end{figure*}

\begin{figure}
   \centering
    \includegraphics[width=9.0cm, angle=0]{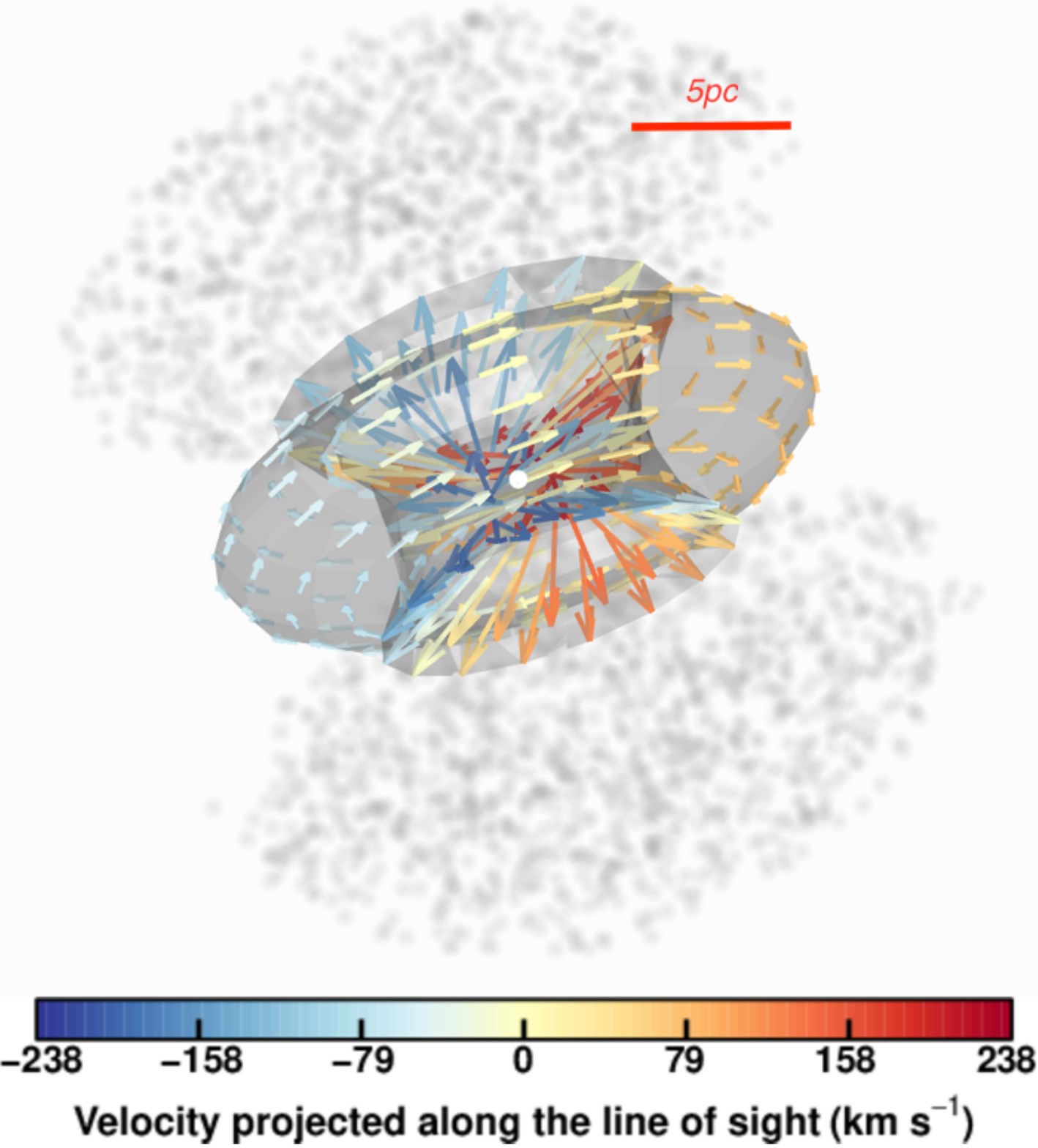}
    \caption{A view of the kinematic model of the molecular torus described in Sect.~\ref{BBarolo} projected along the line of sight. The arrows represent the velocity field of the gas in the torus (with Keplerian rotation) and over the working surface of the AGN wind on the torus (with outflowing motions). The arrows are color-coded using red, yellow and blue colors according to the absolute value and the sign of the radial velocities projected along the line-of-sight measured relative to $v_{\rm sys}$. The major axis of the modeled torus  has the orientation of the observed torus ($PA\simeq113^{\circ}$). The adopted inclination ($i\simeq-70^{\circ}$) is lower than the average value found by {\tt $^{3D}$Barolo} ($i\simeq-80^{\circ}$)  for a better illustration of the different velocity  components of the model.
   }       
   \label{torus-scheme}
\end{figure}


We illustrate the goodness of the best-fit solution in Figure~\ref{pv-models},  where we overlay the CO(2--1) emission contours observed by ALMA on the corresponding emission derived by {\tt $^{3D}$Barolo} (color scale) in two p-v plots oriented along $PA=291^{\circ}=180^{\circ}+111^{\circ}$ and $PA=21^{\circ}$. The latter correspond, respectively,  to the mean values describing the orientation of the major and minor axes, determined inside the region used in the fit (0~pc$<r<210$~pc).

Overall, the parameter space occupied by the observations along the major and minor axis pv-plots is  reasonably accounted for by the {\tt $^{3D}$Barolo} solution. However, there is  a number of noticeable  disagreements between the observations and the model. These discrepancies are explained to some extent by the inability of the model to reproduce any substantial deviation from the axisymmetric dependence of  the kinematic parameters inherent to {\tt $^{3D}$Barolo}. Fig.~\ref{pv-models} shows that the combination of $v_{\rm rad}$ and $v_{\rm vert}$ in the model is able to significantly spread the emission over blueshifted and redshifted velocities relative to $v_{\rm sys}$ through the inner section of the minor axis up to $\Delta y\simeq\pm1\farcs5$ ($\pm$100~pc). However, the observations show a larger velocity spread ($\Delta v_{\rm tot}\simeq400-500$~kms$^{-1}$) of the emission at $\Delta y\simeq+0\farcs7$ (+50~pc) and $\Delta y\simeq-2\arcsec$  (--150~pc). These regions, identified as  ({\em OUT-N}) and  ({\em OUT-S}) in Fig.~\ref{pv-models}, correspond to the suspected contact points of the working surface of  the AGN wind with  the molecular gas at the N and S crossings of the CND, respectively. Observations also show a significant yet comparatively less dramatic spread of the emission ($\Delta v_{\rm tot}\simeq200$~kms$^{-1}$) along the major axis up to $\Delta x\simeq\pm2\arcsec$ ($\pm$150~pc). This observational feature, which is fairly accounted for by the model, suggests that, although the AGN feedback on the kinematics of molecular gas is admittedly more extreme at the regions that lie closer to the outflow axis, the kinematics of the CND seems to be globally affected.

\subsection {An outflowing torus?}\label{torus-out-CR}

The  global {\tt $^{3D}$Barolo} fit of  the kinematics of the CO(2--1) emitting gas discussed in Sect.~\ref{Barolo} favors a 3D-outflow solution that applies to a wide range of spatial scales from the $\simeq400$~pc--CND down to the $\simeq10-30$~pc-molecular torus. This scenario, which explains the distribution and kinematics of the gas in the torus and its connections,  suggests that a significant fraction of the gas traced by the 2--1 line of CO {\em inside} the torus could be entrained by the AGN-driven wind. 

   \begin{figure*}
   \centering
    \includegraphics[width=12cm]{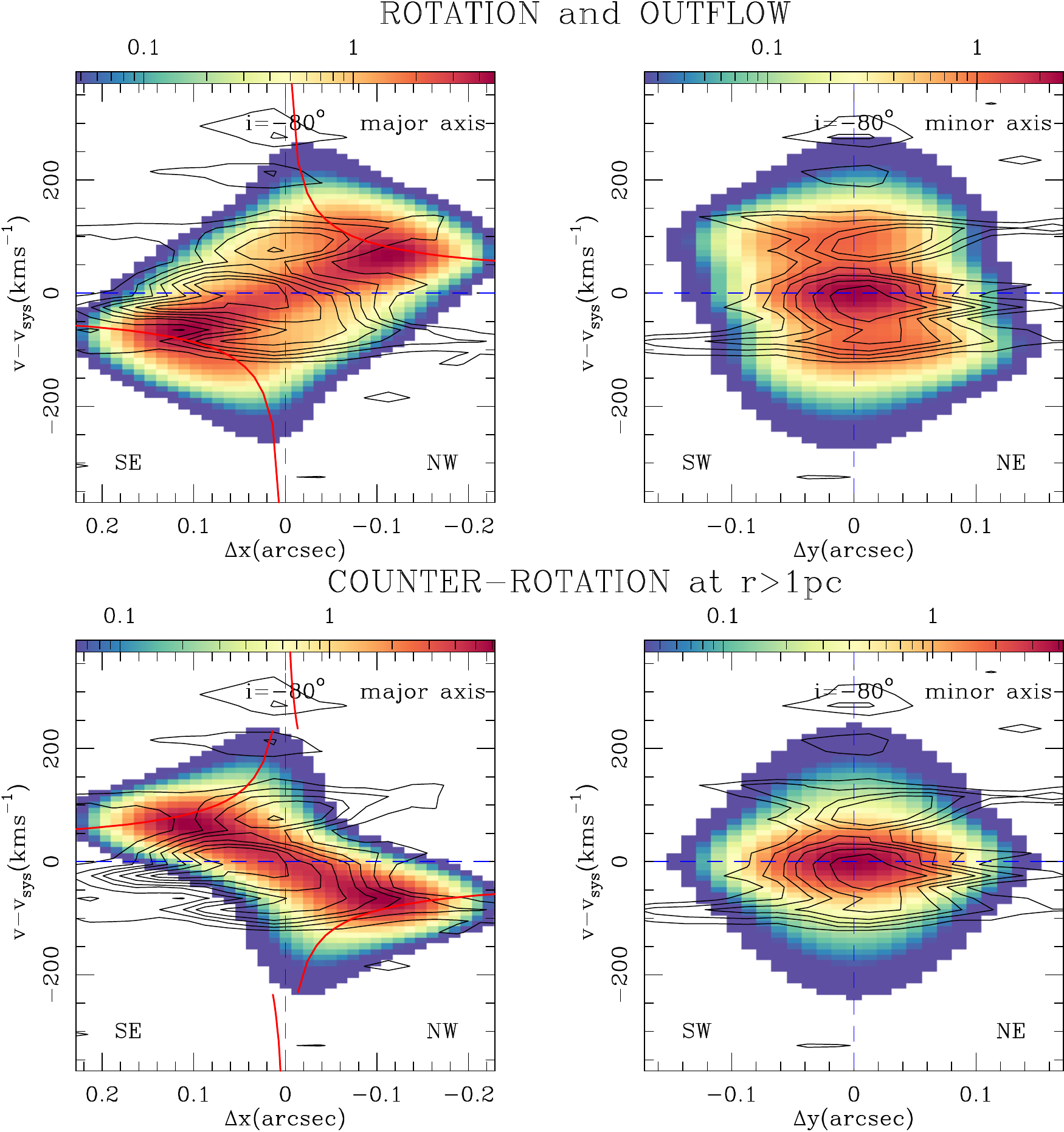}
    \caption{{\it Upper panels:}~Overlay of CO(2--1) p-v plots (in contours) along the  major and minor axis of the torus ({\it upper left} and {\it upper right panels}, respectively) on the corresponding synthetic p-v plots (in logarithmic color scale in K-units) generated by  the best-fit model of the torus described in Sect.~\ref{torus-out-CR} (OUT-model).  The CO(2--1) contour levels are as in Fig.~\ref{torus-major} and \ref{torus-connect-pv}. {\it Lower panels:}~Same as  {\it upper panels} but for the CR-model described in Sect.~\ref{torus-out-CR}.}      
   \label{torus-model-pvs}
    \end{figure*}

The left panel of Fig.~\ref{torus-overlays} compares the distribution of the HCO$^+$(4--3) line emission with that of the CO(2--1) line in the molecular torus, as derived from the MSR data sets used in this work. This overlay underlines that the densest molecular gas probed by HCO$^+$ is located  in the inner region of the torus, as discussed in Sect.~\ref{torus}. Moreover, already at this spatial resolution the CO and the HCO$^+$ images  show a spatially-resolved vertical structure of the torus. In particular, the HCO$^+$ contours of the 
MSR image provide tantalizing evidence of a boxy  structure, which is fully resolved in the image obtained from  the HSR data set, shown in the right panel of Fig.~\ref{torus-overlays}. This higher-resolution HCO$^+$ image shows a network of gas protrusions that stem from the central $r\leq2$~pc-region around the AGN. These gas protrusions extend over a wide range of angles measured from the kinematic major axis of the CO torus  derived in
Sect.~\ref{Barolo}. Taken at face value, these observations 
would lend support to a dynamical model where a substantial fraction of the dense gas inside the torus is being swept along the edges of an ionized wind bicone that emerges from the 
accretion disk, giving rise to a molecular/dusty wind or outflow.  The latter would adopt the geometry of a hollow cone, seen in projection as an X-shaped feature.  The AGN 
wind could peel off the inner layers of the torus and push the gas outward.   An X-shaped structure that extends on spatial scales similar to the ones reported above has been recently identified in the ALMA 256~GHz continuum image of the torus published by \citet{Imp19}.

Besides the direct morphological evidence illustrated by Fig.~\ref{torus-overlays}, the kinematics of molecular gas inside the torus, discussed in Sect.~\ref{torus-kin}, reveal the 
existence of strong non-circular motions:  a sizable fraction of gas emission arises at velocities {\em forbidden} in the frame of a circular velocity model, i.e., gas in {\em 
apparent} counter-rotation. These non-circular motions are to first order accounted for by the {\tt $^{3D}$Barolo} fit of Sect.~\ref{Barolo} in terms of a molecular outflow with a 
significant equatorial component, without invoking directly counter-rotating gas.

In an attempt to further constrain both the geometry and the mass of the molecular gas  that may be entrained by the AGN wind through this type of mechanism, we have 
built up a simple 3D morpho-kinematic model of the NGC1068 torus. In this model the gas is assumed to be in a toroidal ringed disk geometry and to follow circular rotation. A fraction of the gas inside the torus,  as determined by the intersection of the AGN wind bicone with the torus, is perturbed by the AGN wind creating a 3D radial outflow superposed to rotation. The latter is determined by the area of the circular section of the torus that intersects with the AGN wind bicone and it is therefore a function of the half-opening angle of the wind  ($\theta=FWHM/2$) on the scale of the torus. The  parameters assumed for the model are based on the geometry (size, orientation, and aspect ratio) and CO luminosities derived from the observations in Sects.~\ref{torus-parameters} and ~\ref{torus-mass}, as well as on the kinematic parameters provided by the {\tt $^{3D}$Barolo} fit of Sect.~\ref{Barolo}, i.e., rotation ($v_{\rm rot}$) and outflow velocities ($v_{\rm out}=(v_{\rm rad}^2+v_{\rm vert}^2)^{1/2}$). In particular, we assume a purposely fixed toroidal ringed disk geometry where the gas fills with constant density a volume generated by a disk of $\simeq4-5$~pc radius, which revolves about a vertical axis located at a radius of $\simeq5-6$~pc. The full size of the modeled torus  is therefore $\simeq20-22$~pc. We adopt a $PA=113^{\circ}$ and an $i=-80^{\circ}$.  The rotation curve for the gas is taken from the observed maser kinematics.  We scale the data cube of the model (in T$_{\rm mb}$-units) to the observed luminosity of the CO(2--1) line, convolve the gas distribution by a 5~pc-FWHM size Gaussian beam to mimic the effect of beam smearing, and regrid the output to the same pixel scale of the observations.

Figure~\ref{torus-scheme} illustrates the geometry and kinematics of the gas in the torus model  projected along the line-of-sight for $\theta=60^{\circ}$ and  $v_{\rm out}=150$~km~s$^{-1}$. While we explore different values of the outflow radial velocities for the gas swept by the AGN wind  from $v_{\rm out}$~=50 to 150 km~s$^{-1}$,  based on the values of $v_{\rm rad}$ and $v_{\rm vert}$ found by  {\tt $^{3D}$Barolo},  the main free parameter of the model is the amount of mass actually entrained by the outflow, which is a function of $\theta$.  This angle is let to vary from $\theta=0^{\circ}$ (i.e., no outflow) to $\theta=90^{\circ}$ (i.e., the outflow affects the bulk of the gas in the torus). To choose among the range of values assumed for $\theta$ and $v_{\rm out}$, we minimize the standard goodness-of-fit parameter  $\chi^2=\sum\left[D(i,j)-M(i,j)/\sigma\right]^2$ derived for the major and minor axes of the torus and subsequently normalized those by the number of pixels used in the fit. The sum extends over the common set of pixels of the model and the observations ($M(i,j)$ and $D(i,j)$, respectively) and $\sigma$ stands for the observed rms derived from the 2--1 line data cube.

Figure~\ref{torus-model-pvs} shows the p-v plots  along the major and minor axes of the torus for the best solution found for $\theta\simeq80^{\circ}$ and  $v_{\rm out}\simeq100$~km~s$^{-1}$, hereafter referred to as the OUT-model. We also show in Fig.~\ref{torus-model-pvs} the corresponding p-v plots generated by a reference model of the same toroidal disk, hereafter referred to as the CR-model, where we eliminate  outflow motions but include counter-rotation of the gas at $r>1$~pc. A counter-rotating torus scenario, which is the basis of the CR-model explored here, has been previously  invoked by \citet{Ima18} and \citet{Imp19} to explain the complex kinematics of molecular gas in the torus. Not surprisingly, the goodness-of-fit parameters are admittedly large in either case ($\geq$4), due to the number of inherent assumptions in both models, which include the axisymmetry and the smoothness of the gas distribution in the torus. However, the OUT-model qualifies as a more likely scenario compared to the CR-model, based not only on its factor $\simeq1.2-1.3$ lower $\chi^2$ \footnote{Relative to a reference model without counter-rotation and with no outflow, the OUT-model has a factor $\simeq$1.4 lower $\chi^2$}, but mainly because it better explains the following key observational signatures:

   \begin{figure*}
   \centering
    \includegraphics[width=12cm]{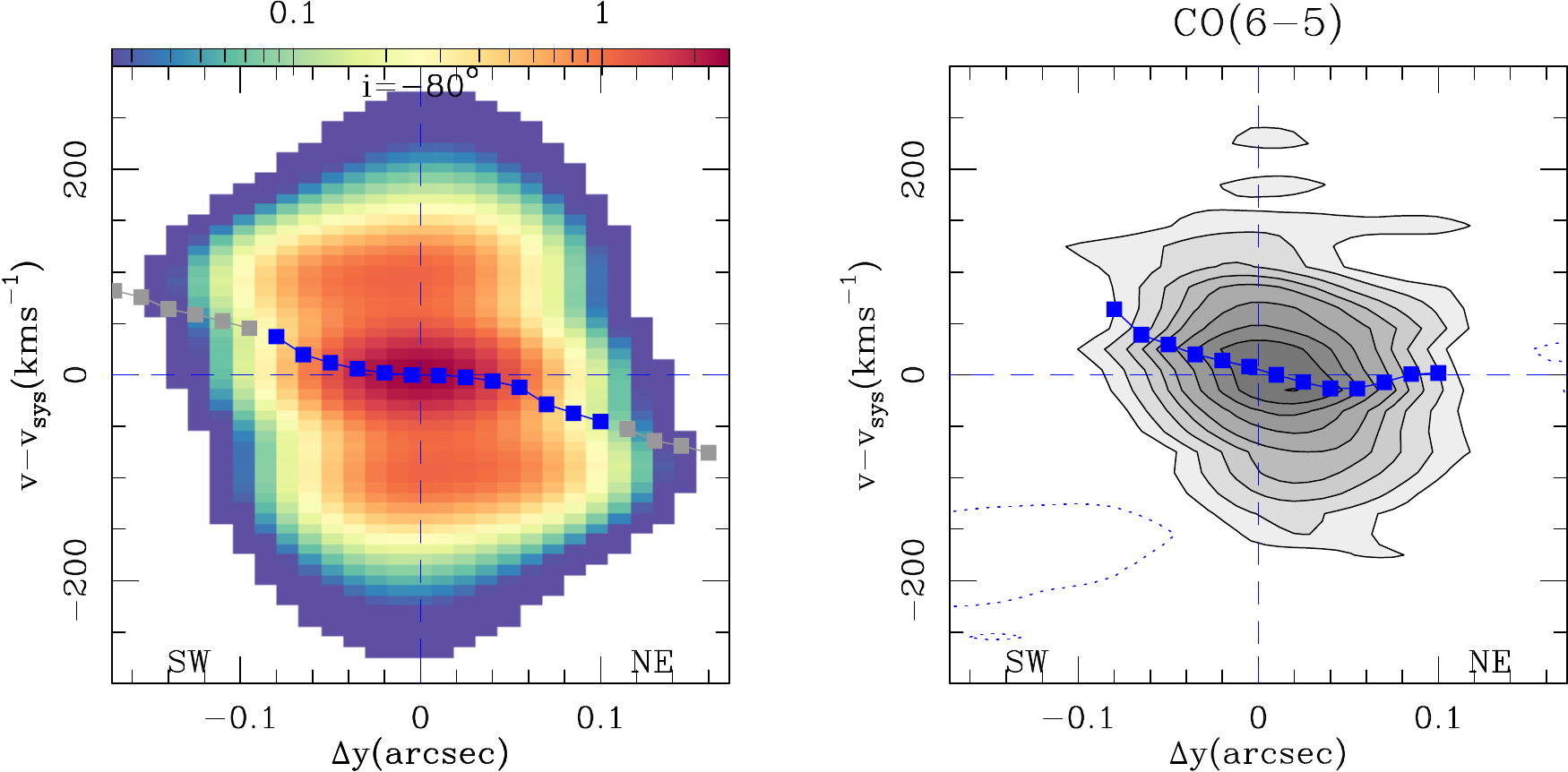}
    \caption{A comparison of the minor axis p-v plot predicted by the OUT-model of the torus, described in Sect.~\ref{torus-out-CR}, with the corresponding CO(6--5) minor axis p-v plot obtained from the combined data set of  \citet{GB16} and \citet{Gal16}. The square markers highlight in both panels the velocity centroid of the emission as function of position along the minor axis.}      
   \label{torus-co65}
    \end{figure*}

\begin{enumerate}

\item

The  gas emission  detected in {\em apparent} counter-rotation in the major-axis p-v plot is to a large extent explained by the OUT-model as due to the projection of outflowing motions that are present at the torus AGN-wind working surface encapsulated in the beam, without having to resort to the addition of an unphysical {\em true} counter-rotating component. 

\item

The contribution of the outflow in the OUT-model also easily explains the presence of three distinct velocity components detected along the minor axis p-v plot. The central component around $v_{\rm sys}$ would be explained by gas unaffected by the outflow close to the equatorial plane of the torus, while the $\simeq$100~km~s$^{-1}$ redshifted (blueshifted) component would reflect the contribution from the outflowing gas below (above) the equatorial plane captured by the beam. This is not reproduced by the CR-model.

\item

Overall, at this spatial  resolution the superposition of circular and non-circular motions in the OUT-model significantly widens the spectra to the level required by observations, in contrast to the CR-model. 

\item

Figure~\ref{torus-co65} shows that an outflow model predicts that there should be a measurable yet shallow gradient of the mean-velocity centroid along the minor axis: from redshifted velocities S of the AGN to  blueshifted velocities N of the AGN. This gradient is not detected in the CO(2--1) data, probably because at this spatial resolution the contribution from gas components located outside the torus in the gas streamers, which are not included in the model,  wash out any potential trend inside the beam. However, the gradient is tentatively detected in the CO(6--5) minor axis p-v plot \footnote{Obtained from the combined data set of  \citet{GB16} and \citet{Gal16}.}, as illustrated in Fig.~\ref{torus-co65}. By construction, the CR-model is unable to reproduce this gradient. The OUT-model easily explains why the velocity shift measured along the minor axis of the CO(6--5) torus, first reported by \citet{GB16} and \citet{Gal16}, is reversed out at larger radii on the scales of the CND.  

\item

The marked asymmetry shown by the high-velocity emission in the CO(2--1) and CO(3--2) AGN spectra, which is not reproduced by HCO$^+$(4--3), could be explained by an outflow model that includes radiative transfer effects in the torus. The different behavior displayed by CO and HCO$^+$  may be due to a larger contribution of absorption of the pc--scale central continuum source by the comparatively lower density  gas traced by CO in the outflow. This asymmetry cannot be easily reproduced by a (counter)rotating disk without imposing a strong asymmetry in the gas distribution (see Sect.~\ref{torus-kin}).

\end{enumerate}

Numerical simulations have long predicted the existence of dusty winds coming out from AGN tori with a large variance of geometries, which depend  on the different treatment of radiation pressure driven by photons at different wavelengths (IR, X-ray and UV), on the assumed degree of anisotropy of the AGN radiation, and on the details of the particular hydrodynamic code used in these simulations \citep[e.g.,][]{Wad15, Wad16, Cha16, Cha17, Nam16, Will19}.  Furthermore, as recently argued by \citet{Giu19}, theoretical analyses predict that radiation-driven accretion disk winds in AGN can have a fairly large equatorial component in Seyfert galaxies that are characterized  by  Eddington ratios ($Edd_{\rm ratio}$) and black hole masses ($M_{\rm BH}$) similar to those of NGC~1068  \citep[$Edd_{\rm ratio}\simeq0.5$, $log_{10}(M_{\rm BH}/M_{\sun})\simeq7.2$;][]{Gre96}. The interaction of this class of AGN winds with the NGC~1068 torus could launch locally a molecular/dusty outflow with a significant equatorial component.   

The value of $\theta\simeq80^{\circ}$ favored by the OUT-model described above implies that this class of wide angle AGN wind is currently impacting a sizable fraction of the gas in the NGC~1068 torus. Based on the model, we estimate this fraction to be $\simeq(0.4-0.6)\times M^{\rm torus}_{\rm gas}$. The gas mass experiencing outflowing motions inside the torus is comparable to the molecular  gas mass derived in Sect.~\ref{torus-streamers} for the streamers, which connect the torus with the CND. Taken at face value, this coincidence highlights the continuity in the outflow mass radial profile out to $r\simeq30$~pc, an indication that the two components are part of the same outflowing feature. However, the bulk of the mass, momentum and energy of the molecular outflow of NGC~1068 is contained in the CND region, where the AGN wind and the radio jet are currently pushing $\simeq7\times10^7M_{\sun}$ of the gas that has been assembled at the ILR ring of the bar at $r\simeq50-200$~pc.

\section{Summary and conclusions}\label{summary}

We have used ALMA to image the emission of molecular gas and dust in the CND and the torus of NGC~1068 using the CO(2--1), CO(3--2) and HCO$^+$(4--3) lines and their underlying continuum emission with a set of spatial resolutions ($0.03\arcsec-0.09\arcsec\simeq2-6$~pc). The use of these transitions, which span a wide range of physical conditions of molecular gas  ($n$(H$_2$)~$\subset10^3-10^7$cm$^{-3}$), is instrumental in revealing the density radial stratification and the complex kinematics of the gas in the torus and its connections to the host. 

We summarize the main results of our study as follows:

\begin{itemize}

\item[$\bullet$]

We detect continuum emission at 229.7\,GHz  and 344.5\,GHz 
stemming from a highly-clumpy $\simeq70$~pc-size  jet-like structure characterized by synchrotron-like emission. The spectral index of the continuum shows the prevalence of dust emission in the regions of the CND located far from the radio jet trajectory itself. In contrast, spectral indexes are flat or marginally positive at the location of the AGN core.

\item[$\bullet$]

The ALMA images resolve the CND, which displays the shape of an asymmetric ringed disk with a de-projected diameter of $\simeq400$~pc and mass of $\simeq1.4\times10^8 M_{\rm sun}$. .  
The CND shows a marked deficit of molecular gas in its central $\simeq130$~pc-region, which is to a large extent filled by the emission stemming from the AGN wind. This deficit leaves the imprint of a contrasted ring morphology on the CND. The inner edge of the ring is associated with the presence of edge-brightened arcs of NIR polarized emission, which are identified with the current working surface of the AGN wind.

\item[$\bullet$]

The kinematics of the gas in the CND  is shaped by an outflow that is captured by the fit of {\tt kinemetry}. The gas is being  pushed by the AGN wind inside the disk creating a radial expansion of the CND with average velocities  $\simeq85$~km~s$^{-1}$ from  $r\simeq50$~pc out to $r\simeq200$~pc.  

 \item[$\bullet$]

Furthermore, the CO emission profiles show extreme line broadening and line splitting into two velocity components, which can be shifted by up to $\simeq250$~km~s$^{-1}$ relative to each other, at a number of positions of the CND that are located in the way of the AGN wind trajectory. This result indicates that the molecular gas in the CND has been forced to leave the plane of the galaxy and has adopted a 3D outflow geometry.  This scenario is confirmed by a global model of the outflow obtained with {\tt $^{3D}$Barolo}.

 \item[$\bullet$]
 
 The new data prove the existence of an elongated molecular disk/torus in NGC~1068 of $3\times10^{5}~M_{\sun}$, which extends over a significantly large range of spatial scales  $D\simeq10-30$~pc around the central engine.  These observations evidence the density radial stratification of the torus: compared to the picture drawn from the HCO$^+$(4--3) line, the molecular disk is bigger and more lopsided in the lower-J CO transitions, a result that brings into light the many faces of the molecular torus. 
 
 \item[$\bullet$] 
 
 The H$_2$ column densities measured towards the position of the AGN are up to $N$(H$_2$)~=~3-4~$\times$~10$^{23}$~mol~cm$^{-2}$. These values are a factor of $\simeq$2--3 below the Compton-thick limit  required to explain the 
nature of the Type~2 nucleus of NGC~1068. Furthermore, the change of column densities required to account for X-ray variability of the AGN could be explained by the small-scale structure of the molecular torus. This is an indication that the molecular torus detected by ALMA may contribute to the obscuration of the central engine of the galaxy on spatial scales of $\simeq$2--3~pc.
 
 \item[$\bullet$] 
 
 The torus is connected to the CND through a network of molecular gas streamers detected in the gas-deficient region. These connecting structures are associated with molecular gas that is being swept along the edges of the bicone structure of the AGN wind detected in ionized gas.  

 \item[$\bullet$] 
 
The kinematics of molecular gas show strong departures from circular motions in  the torus,  the  gas streamers, and the CND ring.  
These velocity field distortions are interconnected and are part of a 3D outflow that reflects the effects of feedback of the AGN on the
kinematics of molecular gas on a wide range of  spatial scales around the central engine. In particular, we estimate through modeling that a significant fraction of the gas  inside the torus ($\simeq0.4-0.6 \times M_{\rm torus}^{\rm gas}$) is outflowing.

\end{itemize}

NGC~1068 is not the only outflowing torus case reported in the literature. The presence of an equatorial outflow  has also been  proposed to account for the complex kinematics of molecular gas in the torus/disk of the Seyfert~2 galaxy NGC~5643 imaged by ALMA \citep{Alo18}.  These equatorial outflows detected close to the central engines of  Seyfert galaxies  would reflect the AGN feedback driven by the ionized wind on the molecular reservoir of the torus, playing the role of a self-regulation mechanism.

Although the total accretion timescales of SMBH  likely exceed 10$^{7-8}$ yr \citep[e.g.,][]{Mar04}, there is both observational and theoretical evidence that AGN can have much shorter {\it flickering} timescales ranging from a few years up to $\simeq$10$^5$ yr \citep{Sch15, Kin15}.
Assuming that the AGN luminosity of NGC\,1068 is $L_{\rm AGN}\simeq4.2^{+1.4}_{-1.1}\times10^{44}$~erg~s$^{-1}$ \citep{GB14}, and for a canonical mass-to-luminosity accretion efficiency of $\epsilon\simeq0.1$, we can estimate an accretion rate for its SMBH of $\dot M$~=~$L_{\rm AGN}/(\epsilon \times c^2)\simeq0.05-0.1 M_{\sun} ~yr^{-1}$. Although the outflowing  gas in the torus is a significant fraction of its total mass ($\simeq0.4-0.6 \times M^{\rm torus}_{\rm gas}$)  a large gas reservoir ($\simeq 1.2-1.8 \times10^5M_{\sun}$) close to its equatorial plane remains mostly unaffected by the feedback of the AGN wind. The latter implies that the molecular torus can continue fueling the AGN for at least $\simeq1-4$~Myr, i.e., for a  time span $\simeq10-40$ longer than the expected {\it flickering} timescale.

We estimate that about half of the molecular gas in the CND is outflowing  \citep[$M_{\rm out}^{\rm CND}\simeq7\times10^7M_{\sun}$; see also][]{GB14}. This molecular outflow at $r\simeq50-200$~pc is likely launched  when the AGN wind and the radio jet impact the gas accumulated at the ILR ring. After accounting for this outflowing component, a large amount of molecular gas in the CND may still serve as a reservoir for future fueling episodes. However AGN fueling seems to be currently thwarted on the intermediate scales corresponding to the torus-CND connections   ($15$~pc~$\leq r \leq50$~pc). We estimate that the outflow rate measured for the gas streamers at $r_{\rm lanes}\simeq30$~pc is $\dot M_{\rm lanes}\simeq M_{\rm lanes}\times v_{\rm out} /  r_{\rm lanes} \simeq 0.6 M_{\sun}~yr^{-1}$, i.e., this is still a factor $\geq100$ lower than the mass outflow rate measured at the CND scales. 

Forthcoming ALMA observations of NGC~1068 will use a set of molecular tracers similar to the ones presented in this work but with a factor of $\simeq$6 smaller beam ($0\farcs013$), corresponding to an unprecedented spatial resolution $\leq1$~pc. These new ALMA images of the torus will be used to better characterize the non-circular motions inside the torus. In particular, we aim to better resolve and disentangle the outflowing component from rotation and identify and model any inflowing component in the gas velocity field. Furthermore, the future ALMA images of the torus and its surroundings will be combined with planned interferometric observations with the MATISSE instrument at the VLT. The latter will probe the warm dust in the polar outflowing component and the hot dust in the inner walls of the torus with spatial resolutions down to $0\farcs01$ (0.07~pc) at 10$\mu$m and $0\farcs003$ (0.02~pc) at 3.3$\mu$m.

\begin{acknowledgements}
         We thank our referee, Prof. Masatoshi Imanishi, for his detailed referee report. We acknowledge the staff of ALMA in Chile and Edwige Chapillon from the ARC at IRAM-Grenoble in France 
for their invaluable help during the data reduction process. This paper
makes use of the following ALMA data: ADS/JAO.ALMA$\#$2016.1.0232.S and $\#$2011.0.00083.S.
ALMA is a partnership of ESO (representing its member states), NSF (USA)
and NINS (Japan), together with NRC (Canada) and NSC and ASIAA (Taiwan),
in cooperation with the Republic of Chile. The Joint ALMA Observatory is
operated by ESO, AUI/NRAO and NAOJ. The National Radio Astronomy
Observatory is a facility of the National Science Foundation operated under cooperative
agreement by Associated Universities, Inc. 
We used observations made
with the NASA/ESA Hubble Space Telescope, and obtained from the Hubble
Legacy Archive, which is a collaboration between the Space Telescope Science
Institute (STScI/NASA), the Space Telescope European Coordinating Facility
(ST-ECF/ESA), and the Canadian Astronomy Data center (CADC/NRC/CSA).     
 SGB and AU acknowledge support from the Spanish MINECO and FEDER funding grants AYA2016-76682-C3-2-P, AYA2016-79006-P and 
 ESP2015-68964-P. CRA acknowledges the Ram\'on y Cajal Program of the Spanish Ministry of Economy and Competitiveness through project 
RYC-2014-15779 and the Spanish Plan Nacional de Astronom\'{\i}a y Astrof\'{\i}sica under grant AYA2016-76682-C3-2-P. AAH 
acknowledges support from the Spanish MINECO and FEDER funding grant AYA2015-6346-C2-1-P. 
AAH, SGB. and AU acknowledge support through grant PGC2018-094671-B-I00 (MCIU/AEI/FEDER,UE). AAH's  work was done under project No. MDM-2017-0737 Unidad de Excelencia "Mar\'{\i}a de Maeztu"- Centro de Astrobiolog\'{\i}a (INTA-CSIC).
AF acknowledges support  from the Spanish MINECO  funding grant  AYA2016-75066-C2-2.
\end{acknowledgements}

\bibliographystyle{aa}
\bibliography{aa-2}

\begin{appendix}

\section{A {\tt $^{3D}$Barolo} model of the NGC~1068 outflow}\label{App1}

   \begin{figure}[b!]
   \centering
    \includegraphics[width=0.32\textwidth]{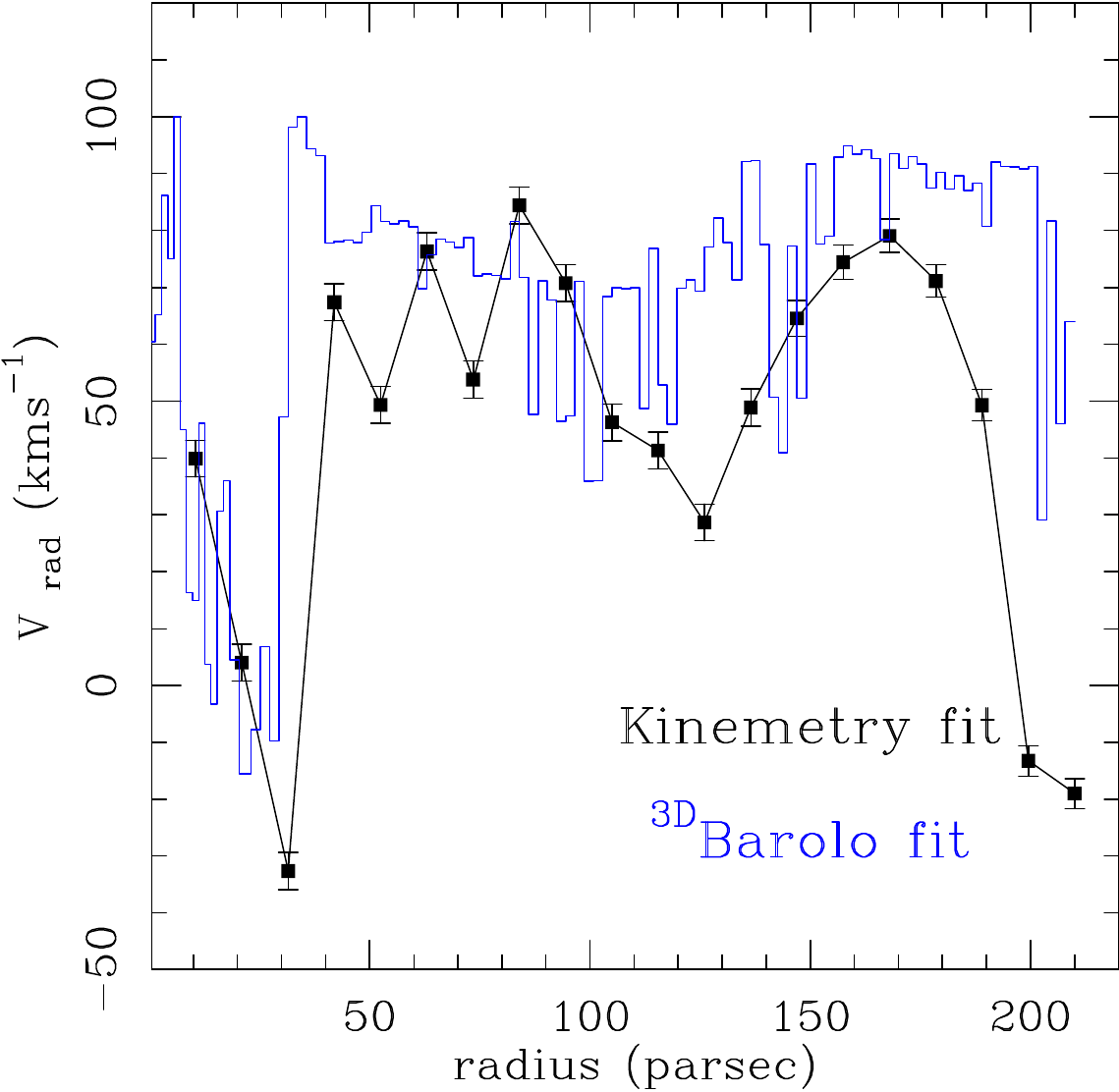}
    \includegraphics[width=0.32\textwidth]{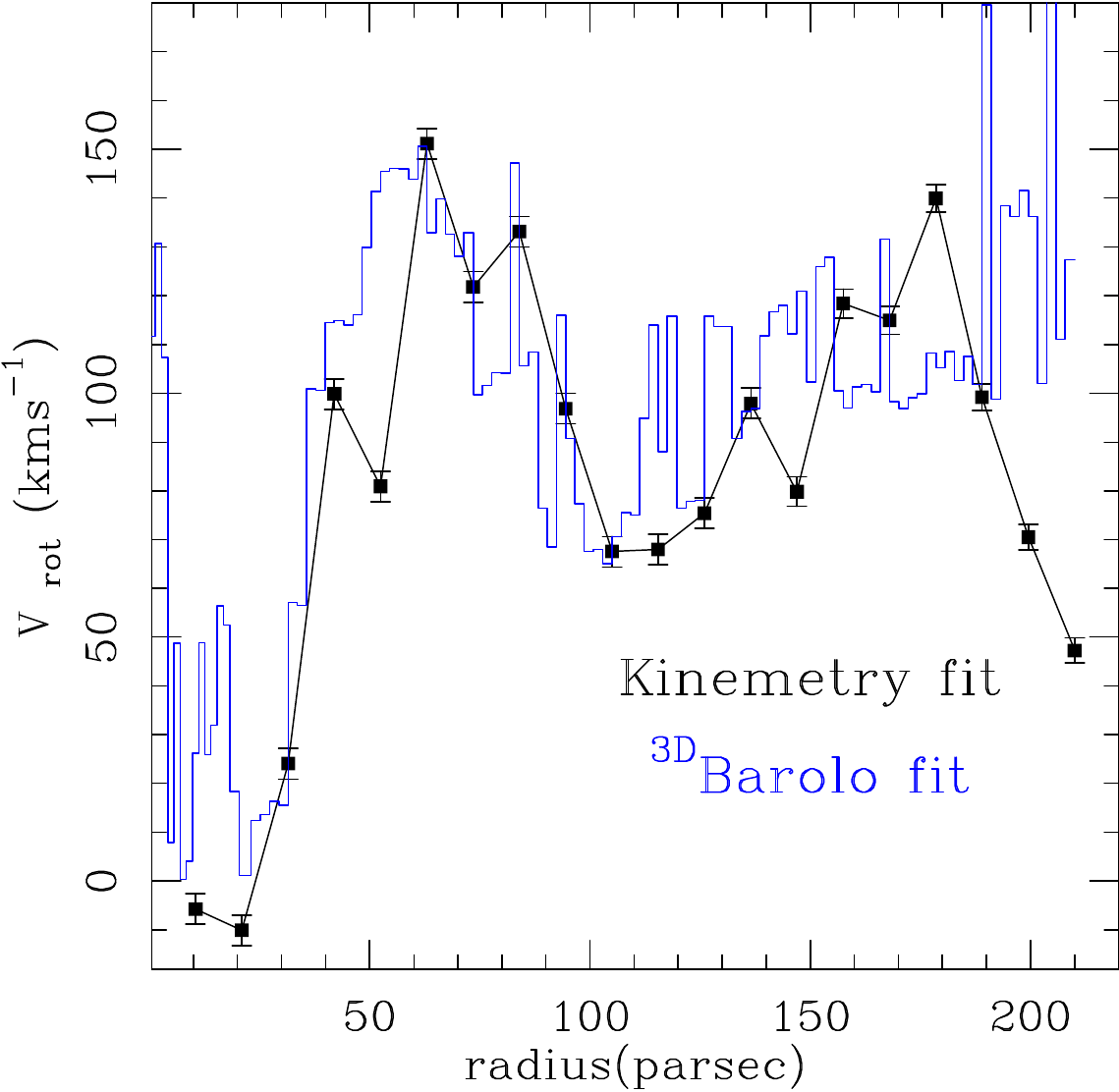}
    \caption{The radial profiles of $v_{\rm rad}$ ({\it upper panel}) and $v_{\rm rot}$ ({\it lower panel}) found by {\tt $^{3D}$Barolo} (blue histograms) and {\tt kinemetry} (black line and markers) to fit the CO(2--1) data of the CND region out to $r\simeq210$~pc.}      
   \label{benchmark}
    \end{figure}


The velocity field of molecular gas reveals strong non-circular motions and a high degree of complexity on a wide range of  spatial scales in the CND of NGC~1068. In the following we attempt to define the  main features of a global {\it model} that best accounts for the main features of the gas kinematics derived from the CO(2--1) data using the {\tt $^{3D}$Barolo} software.

We first benchmarked the solutions found by {\tt $^{3D}$Barolo} against the output provided by {\tt kinemetry} in Sect.~\ref{kinemetry}. For this purpose we  ran {\tt $^{3D}$Barolo} assuming the common set of fixed parameters used in Sect.~\ref{kinemetry} for the AGN position, the disk geometry ($PA$,  $i$), and the systemic velocity ($v_{\rm sys}$), but letting  vary $v_{\rm rot}$, $v_{\rm rad}$, and the velocity dispersion ($\sigma_{\rm gas}$) for a set of 103 radii sampling the CND from $r=0\farcs03~(\simeq2$~pc) to   $r=3\arcsec~(\simeq210$~pc).  We also adopted a fixed Gaussian scale-height for the disk ($H=10$~pc) to reproduce the 2D model of the {\tt kinemetry} disk, which
is by construction collapsed along its vertical axis.

As shown in Fig.~\ref{benchmark}, the $v_{\rm rad}$ and $v_{\rm rot}$ profiles found by {\tt $^{3D}$Barolo} and  {\tt kinemetry} are roughly compatible within the errors.  In particular, taking into account the known orientation of the large-scale disk of NGC~1068, i.e., the NE (SW) side of the disk corresponds  the far (near) side, the positive sign and modulus of the $v_{\rm rad}$  profile ($v_{\rm rad} \simeq (0.4-0.8)\times v_{\rm rot}$ for  30~pc~$<r<210$~pc) similarly found by {\tt $^{3D}$Barolo} and  {\tt kinemetry} identify the presence of a {\em coplanar} outward radial component.

Based on the available observational constraints, for the second {\tt $^{3D}$Barolo} run we imposed two restrictions on the starting range of the initial values for the two geometrical parameters: 1) $PA=PA_{\rm kinemetry}\pm20^{\circ}=289\pm20^{\circ}$ and 2) $i=i_{\rm kinemetry}\pm20^{\circ}=41\pm20^{\circ}$ at all radii. We relax this restriction in the torus region and its connections ($r < 30$~pc), where we gradually change from the known orientation of the highly-inclined torus, by adopting $-60^{\circ}>i>-90^{\circ}$ in the range $r < 20$~pc, towards the orientation of the disk of the galaxy at larger radii. We note that the negative sign of $i$ in the inner region reflects here that the N (S) side of the tilted torus is the near (far) side, as dictated by the favored geometry of the NLR of the galaxy shown in Fig.~\ref{outflow-scheme}.

\begin{figure*}[bth!]
   \centering
    \includegraphics[width=0.8\textwidth]{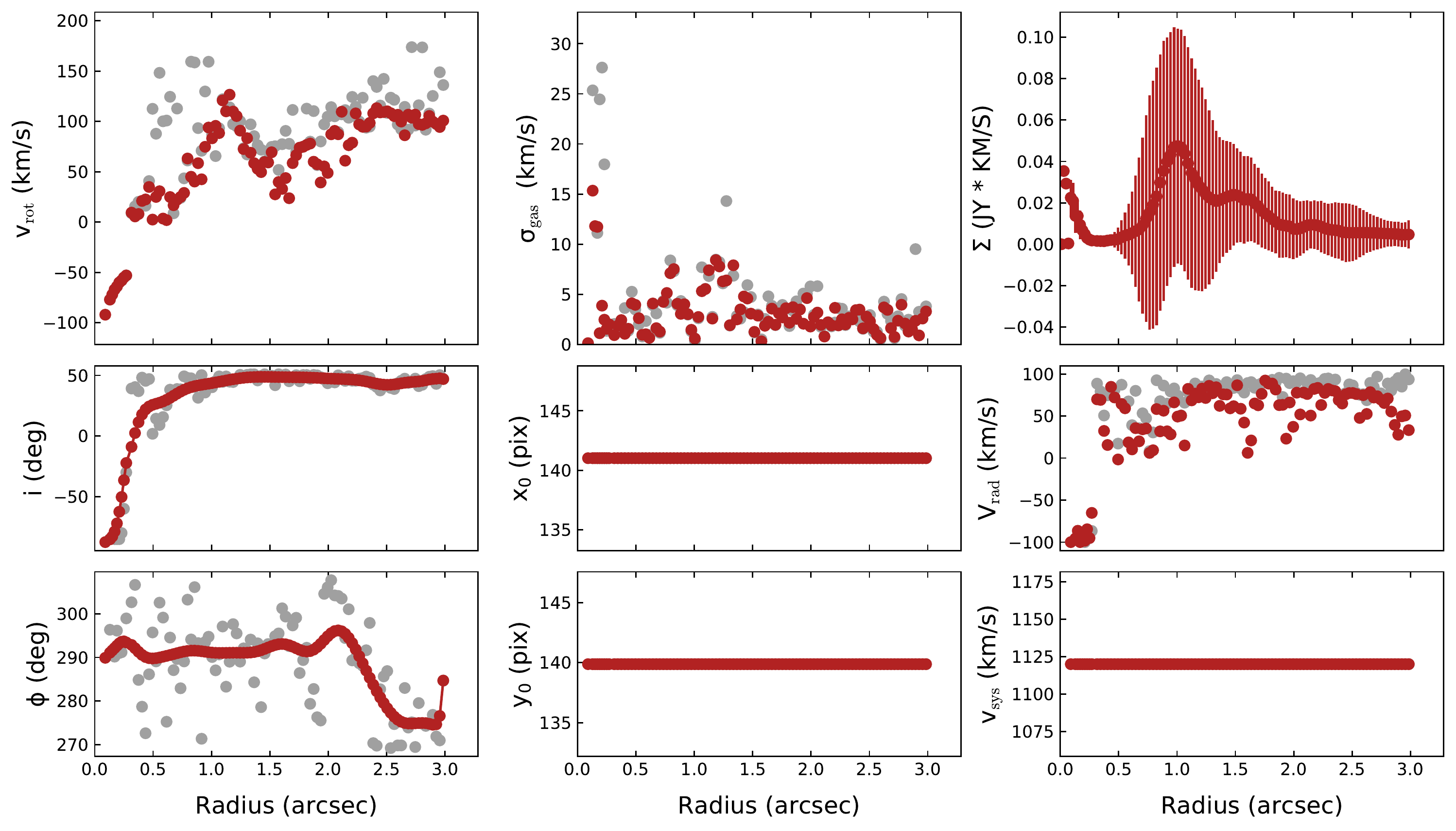}
   \caption{The radial profiles of the best fit found by {\tt $^{3D}$Barolo} for $v_{\rm rot}$, $v_{\rm rad}$, $v_{\rm disp}$ ($\sigma_{\rm gas}$),  $PA$ ($\phi$), and  $i$. The position of the AGN ($x_{\rm o}$ and $y_{\rm o}$ in pixel units) and the systemic velocity ($v_{\rm sys}$) are fixed to the values previously found by {\tt kinemetry}. The size of the vertical bars in the radial profile of integrated intensities ($\Sigma$) reflect the deviations from axisymmetry in the gas distribution.  The grey  and red markers represent the output values of the fitted parameters obtained after the first ({\em step-1}) and second  ({\em step-2}) runs, respectively.}
   \label{Barolo-parameters}
    \end{figure*}

 In order to minimize the occurrence of discontinuities in the derived $v_{\rm rot}$ and $v_{\rm rad}$ profiles, which may appear when the geometrical parameters of the disk  are simultaneously fit with the kinematic profiles,  we adopted the parameter regularization approach described by \citet{Dit15}.  This requires executing the second fitting routine using a two-step process. We first derived a model by fitting all the free parameters simultaneously ({\em step-1}), while in a second iteration we fixed the geometrical parameters  to a  Bezier function that interpolates the values found in the previous iteration for $PA$ and  $i$ with the rest of parameters being let free ({\em step-2}).  The resulting radial profiles of the best-fit parameters 
 found after this second run are shown in  Fig~\ref{Barolo-parameters}. 
 
 We note that $v_{\rm rot}$ and $v_{\rm rad}$ are seen to change sign at $r\simeq20$~pc: both parameters change from negative values  inside this boundary to positive values at larger radii. This change of sign accounts for the change of orientation of the plane of the galaxy relative to the observer, described above. Furthermore, as the $v_{\rm vert}$ parameter is not iteratively fitted by the version of {\tt $^{3D}$Barolo} used in the fit, we chose to explore four different values of $v_{\rm vert}$ (0, 25, 50, 100, and 150~km~s$^{-1}$) and constrained the range of the best-fit solutions to $v_{\rm vert}\simeq100\pm50$~km~s$^{-1}$. In this paper we show the output of the best-fit solution for $v_{\rm vert}\simeq100$~km~s$^{-1}$. The $H$ profile (not shown) shows values going from 7 to 10~pc in the torus region ($r < 20$~pc), to higher values $\simeq20-50$~pc in the outer disk up to $r\simeq210$~pc. The main characteristics of the best-fit solution encapsulated in the radial profiles shown in Fig~\ref{Barolo-parameters} are described in Sect.~\ref{BBarolo}.

\end{appendix}

\end{document}